\pdfoutput=1
\PassOptionsToPackage{table}{xcolor}
\documentclass[pdflatex,sn-mathphys-num]{sn-jnl}

\usepackage{graphicx}%
\usepackage{multirow}%
\usepackage{amsmath,amssymb,amsfonts}%
\usepackage{amsthm}%
\usepackage{mathrsfs}%
\usepackage[title]{appendix}%
\usepackage{xcolor}%
\usepackage{textcomp}%
\usepackage{manyfoot}%
\usepackage{booktabs}%
\usepackage[utf8]{inputenc}
\usepackage{newunicodechar}
\newunicodechar{−}{\textminus}
\usepackage{algorithm}%
\usepackage{algorithmicx}%
\usepackage{algpseudocode}%
\usepackage{listings}%
\usepackage{caption}
\usepackage{subcaption}
\usepackage{overpic}
\usepackage{longtable}
\usepackage{siunitx}
\usepackage{threeparttable}
\usepackage{tabularx}
\usepackage{float}
\usepackage{placeins}

\theoremstyle{thmstyleone}%
\theoremstyle{thmstyletwo}%

\theoremstyle{thmstylethree}%

\begin{document}

\title[PhosTailor]{PhosTailor: structure-to-sequence distillation for kinase discovery in sparse phosphoproteomes}


\author[1]{\fnm{Jingjie} \sur{Zhang}}
\equalcont{These authors contributed equally to this work.}

\author[2]{\fnm{Jun} \sur{Xu}}
\equalcont{These authors contributed equally to this work.}

\author[3]{\fnm{Kewen} \sur{Wang}}
\equalcont{These authors contributed equally to this work.}

\author[4]{\fnm{Yu} \sur{Wang}}

\author[6]{\fnm{Mutian} \sur{He}}

\author[2]{\fnm{Min} \sur{liang}}

\author[1]{\fnm{Jinghao} \sur{Wang}}

\author[1]{\fnm{Hanqun} \sur{Cao}}

\author[1]{\fnm{Zijun} \sur{Gao}}

\author*[5]{\fnm{Ning} \sur{Ma}}\email{eeening.ma@polyu.edu.hk}

\author*[3]{\fnm{Chang-Yu} \sur{Hsieh}}\email{kimhsieh@zju.edu.cn}

\author*[2]{\fnm{Xuhui} \sur{Zeng}}\email{zengxuhui@ntu.edu.cn}

\author*[1]{\fnm{Chunbin} \sur{Gu}}\email{cbgu@cuhk.edu.hk}

\author[1]{\fnm{Pheng-Ann} \sur{Heng}}

\affil[1]{\orgdiv{Department of Computer Science and Engineering}, \orgname{The Chinese University of Hong Kong}, \orgaddress{\state{Hong Kong}, \country{China}}}

\affil[2]{\orgdiv{Institute of Reproductive Medicine, Medical School}, \orgname{Nantong University}, \orgaddress{\city{Nantong}, \postcode{226019}, \state{Jiangsu Province}, \country{China}}}

\affil[3]{\orgdiv{College of Pharmaceutical Sciences}, \orgname{Zhejiang University}, \orgaddress{\city{Hangzhou}, \postcode{310027}, \state{Jiangsu Province}, \country{China}}}

\affil[4]{\orgdiv{School of Computer Science}, \orgname{Peking University}, \orgaddress{\city{Beijing}, \country{China}}}

\affil[5]{\orgdiv{Department of Electrical and Electronic Engineering}, \orgname{The Hong Kong Polytechnic University}, \orgaddress{\state{Hong Kong}, \country{China}}}

\affil[6]{\orgdiv{Faculty of Applied Sciences}, \orgname{Macao Polytechnic University}, \orgaddress{\state{Macao}, \country{China}}}

\abstract{ 
    Phosphoproteomics maps thousands of phosphorylation sites, yet upstream kinase assignment remains difficult in systems where few kinase–substrate pairs have been characterized experimentally. PhosTailor transfers structural and biochemical knowledge into sequence-only predictors, using student-identified hard examples to guide teacher adaptation. PhosTailor outperformed sequence- and structure-aware baselines across five reproductive and cancer benchmarks. For sperm-relevant kinases withheld from reproductive-domain training, its AUROC exceeded the strongest comparator by 13.4 percentage points. The sequence-only student retained near-teacher accuracy with approximately 100-fold higher throughput. Applied to sperm-function proteins, PhosTailor prospectively prioritized TANK-binding kinase 1 (TBK1). Among 410 GSK8612-responsive phosphosites in pooled human sperm, phosphorylation decreased at 68.7\% of top-fifth-ranked sites versus 51.5\% overall. In a donor-paired experiment across seven donors, mean hyperactivation after one hour was 39.7\% with GSK8612 and 26.6\% with vehicle. These findings nominate TBK1 for mechanistic investigation and show how kinase prediction can guide chemical perturbation in sparsely annotated systems.
}


\maketitle

\section{Introduction}\label{intro}

Protein phosphorylation regulates nearly every aspect of cellular physiology and disease, yet most phosphorylation sites identified by quantitative phosphoproteomics remain `orphan' sites without an assigned upstream kinase~\cite{hunter1995protein,cohen2000regulation,cohen2002protein,ubersax2007mechanisms,piersma2024inferring}. Because experimentally curated kinase--substrate relationships are heavily biased towards a small subset of well-studied kinases and proliferating cell lines, this annotation gap is most acute in specialized primary human cells and distinct disease contexts~\cite{canovas2021diversity,johnson2023atlas,maier2025data}. Mature human spermatozoa illustrate this challenge clearly: largely transcriptionally and translationally silent after ejaculation, sperm depend almost exclusively on rapid post-translational phosphorylation signalling to coordinate capacitation, motility and hyperactivation, yet many of the upstream kinases involved remain poorly defined~\cite{ostermeier2002spermatozoal,jodar2013presence,visconti1995capacitation,ficarro2003phosphoproteome,wang2015quantitative,martin2020human,serrano2022sperm,signorelli2012kinases}. A comparable challenge limits the mechanistic interpretation of tumour phosphoproteomes and disease-associated variants, where identifying active upstream kinases is essential for uncovering dysregulated signalling pathways and predicting therapeutic response~\cite{mertins2016proteogenomics,zagorac2018vivo}.

Current computational predictors face substantial limitations when applied to these sparsely annotated biological systems~\cite{luo2019deepphos,chen2023gps,zhou2024using,zhang2025sagephos,zhang2026learning}. Sequence-based models scale readily to proteome-wide datasets~\cite{blom1999sequence,obenauer2003scansite,xue2008gps,luo2019deepphos,chen2023gps,zhou2024using,zhang2026learning}, but do not explicitly capture three-dimensional determinants of kinase recognition, such as residue accessibility, local geometry and spatial context~\cite{ubersax2007mechanisms,johnson2023atlas}. Conversely, structure-aware models incorporate richer biochemical information~\cite{jumper2021highly,tunyasuvunakool2021highly,zhang2022protein,varadi2024alphafold,choo2025sapp,zhang2025sagephos}, yet structural retrieval and graph construction impose considerable computational overhead during large-scale phosphoproteomic screening. More fundamentally, both paradigms are constrained by domain data scarcity: general-purpose models often miss context-specific regulatory relationships, whereas models trained solely on limited domain annotations fail to generalize to understudied kinases or uncharacterized substrates~\cite{maier2025data}. Addressing this limitation requires transferring general structural and biochemical principles of kinase recognition into lightweight, domain-adapted predictors.

Here, we developed PhosTailor, a general-to-specialized learning framework that transfers multimodal structural and biochemical phosphorylation knowledge into fast, sequence-only predictors tailored to sparsely annotated biological domains (Fig.~\ref{fig:framework}a,b). Across five benchmarks spanning male reproductive biology and cancer, PhosTailor consistently outperformed competing sequence- and structure-based methods. Notably, for sperm-relevant kinases withheld from reproductive-domain training, PhosTailor improved the area under the receiver operating characteristic curve (AUROC) by 13.4 percentage points over the strongest comparator. At deployment, the sequence-only predictor matched its multimodal teacher's accuracy while achieving roughly 100-fold higher inference throughput and substantially reducing false-positive predictions. Independent biochemical evaluations confirmed that this structure-free model retains genuine kinase-conditioned specificity rather than generic site propensity. Specifically, its high-confidence predictions recover canonical basophilic, proline-directed and acidophilic motifs, align closely with an experimental Ser/Thr kinase specificity atlas~\cite{johnson2023atlas}, respond selectively to counterfactual kinase substitution, and detect graded motif disruptions caused by clinical variants in the androgen receptor (AR) and cystic fibrosis transmembrane conductance regulator (CFTR) associated with androgen insensitivity and male infertility.

Applying PhosTailor to a curated panel of 60 sperm-function proteins harbouring variants from clinical intrauterine insemination failure prospectively nominated TANK-binding kinase 1 (TBK1), a kinase classically studied in innate immunity and autophagy whose function in mature human spermatozoa remains poorly characterized, alongside mitogen-activated protein kinase kinase 2 (MAP2K2, also known as MEK2) (Fig.~\ref{fig:framework}c). Consistent with a physiological presence in sperm, TBK1 was detected in mature human spermatozoa, and top-ranked candidate substrates mapped to proteins of the axoneme, flagellum and fibrous sheath. Pharmacological perturbation of pooled human sperm with the selective TBK1 inhibitor GSK8612 demonstrated strong concordance between model predictions and phosphoproteomic responses: among 410 responsive phosphosites, phosphorylation decreased at 68.7\% of sites within the top fifth of PhosTailor's TBK1 ranking, compared with 51.5\% across all responsive sites. In paired samples from seven independent donors, computer-assisted sperm analysis revealed that GSK8612 preserved basal and progressive motility while markedly increasing sperm hyperactivation after one hour (39.7\% versus 26.6\% with vehicle), accompanied by elevated curvilinear velocity and lateral head displacement. By contrast, pharmacological inhibition of MEK1/2 suppressed hyperactivation, providing pathway-level contrast for kinase regulation of sperm kinematics. Together, these findings nominate TBK1 as an unexpected candidate regulator of human sperm hyperactivation and establish a generalizable, web-accessible framework for uncovering kinase signalling in sparsely annotated human phosphoproteomes.

\section{Results}\label{results}

\subsection{PhosTailor enables general-to-specialized transfer of multimodal phosphorylation knowledge}

\begin{figure*}
    \centering
    \includegraphics[width=\linewidth]{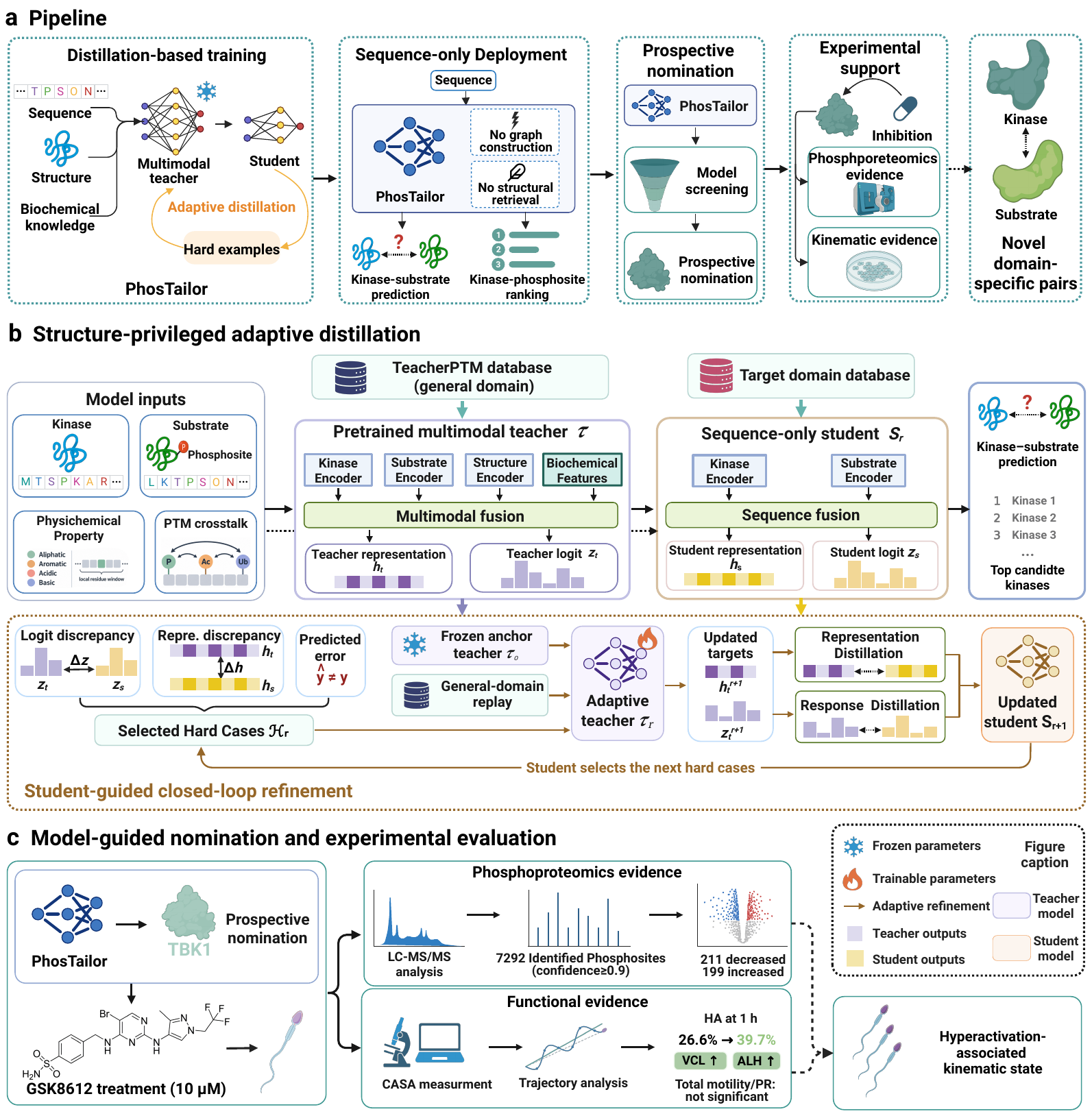}
    \caption{\textbf{PhosTailor enables kinase inference through adaptive multimodal knowledge transfer.}
    \textbf{a}, Overview of privileged training, sequence-only deployment and model-guided experimental investigation, culminating in candidate domain-specific kinase–substrate pairs.
    \textbf{b}, Structure-privileged adaptive distillation transfers sequence, structural and biochemical knowledge from a multimodal teacher to a sequence-only student. Student-guided refinement uses selected hard cases, a frozen anchor teacher and general-domain replay to update teacher targets.
    \textbf{c}, Prospective nomination of TBK1 followed by exploratory pharmacological evaluation with GSK8612. Phosphoproteomics used pooled samples with three technical replicates per condition, whereas computer-assisted sperm analysis (CASA) was performed in a separate donor-paired cohort. After protein-abundance filtering, 410 responsive phosphosites were retained (211 decreased and 199 increased). CASA showed increased hyperactivation (HA) at 1 and 2 h, curvilinear velocity (VCL) at 1 h and amplitude of lateral head displacement (ALH) at 1 and 2 h; no statistically significant differences were detected in total motility or progressive motility (PR).}
    \label{fig:framework}
\end{figure*}

Accurate kinase–substrate prediction in specialized biological contexts requires adapting broadly learned phosphorylation knowledge to domain-specific regulatory patterns while maintaining computational efficiency. To address these challenges, we developed PhosTailor, a general-to-specialized learning framework that transfers multimodal phosphorylation knowledge into lightweight, sequence-only predictors tailored to distinct physiological and disease-associated contexts (Fig.~\ref{fig:framework}b). Rather than relying on a domain-specific architecture, PhosTailor employs a unified knowledge-transfer strategy that can be adapted to different target domains with limited kinase–substrate annotations.

PhosTailor employs a multimodal teacher that integrates kinase and substrate sequence representations with structural and residue-level biochemical features. This knowledge is transferred into a compact sequence-only student through supervised learning, response-level distillation and representation-level alignment. To improve adaptation to specialized domains, an online policy distillation-inspired closed loop identifies difficult target-domain examples based on student uncertainty, prediction errors and teacher–student disagreement. These examples guide selective refinement of an adaptive teacher, while frozen-teacher anchoring and general-domain replay constrain knowledge drift and help retain transferable kinase-recognition information. Updated teacher predictions and representations are subsequently distilled into the student. At deployment, only the domain-adapted student is retained, enabling efficient kinase–substrate–site inference without structural retrieval, graph construction or teacher-side computation.

To support this framework, we curated complementary general and domain-specific phosphorylation resources. TeacherPTM comprises 43,151 enzyme–substrate PTM records integrated from PhosphoSitePlus (PSP), SAGEPhos and OmniPath \cite{hornbeck2015phosphositeplus,zhang2025sagephos,turei2016omnipath}, providing broad phosphorylation knowledge for multimodal teacher training. For reproductive specialization, we constructed three complementary benchmarks: ReproPhos-HPA, comprising 33,489 phosphorylation events associated with proteins expressed in male reproductive tissues according to Human Protein Atlas (HPA) evidence; SpermK-PSP, comprising 16,747 phosphosites involving 202 literature-supported sperm-relevant kinases; and SIGNOR-Phos, derived from the SIGnaling Network Open Resource (SIGNOR) and comprising 3,986 curated causal phosphorylation relationships involving proteins detected in mature human sperm \cite{ponten2008human,uhlen2015tissue,thul2017subcellular,nagaoka2020zglp1,karlsson2021single,uhlen2017pathology,uhlen2019genome,uhlen2019human,castillo2018contribution,lo2026signor}. These resources were harmonized into MReproPhos. To evaluate disease-associated specialization, we additionally constructed a tumour-associated phosphorylation benchmark (Tumor) by integrating kinase–substrate relationships from OmniPath with disease annotations from NCBI PubTator and MeSH, yielding 7,836 positive samples. A cancer-specific subset (Cancer), comprising 2,615 positive samples, was further derived from Tumor. Together, these five benchmarks enabled systematic evaluation of PhosTailor across reproductive and disease-associated phosphorylation contexts. The framework was adapted separately to the reproductive, Tumor and Cancer settings, rather than relying on zero-shot transfer from a reproductive predictor. Detailed curation procedures and dataset statistics are provided in Supplementary Note C.1.

\subsection{PhosTailor improves kinase--substrate prediction across reproductive benchmarks}

We first focused on male reproductive biology, where phosphorylation plays central roles in sperm capacitation, motility and hyperactivation, yet the upstream kinases coordinating these processes remain incompletely characterized. Identifying candidate kinase–substrate relationships in this context could help interpret sperm phosphoproteomic changes and generate testable hypotheses about the signalling mechanisms underlying sperm function.
To systematically evaluate PhosTailor in this setting, we examined three complementary phosphorylation benchmarks: SIGNOR-Phos, comprising curated causal phosphorylation relationships involving proteins detected in mature human sperm; ReproPhos-HPA, covering phosphorylation events associated with proteins expressed in male reproductive tissues; and SpermK-PSP, focusing on phosphorylation events involving literature-supported sperm-relevant kinases. Each benchmark was evaluated under random, kinase cold-start and substrate cold-start splits to assess predictive discrimination and generalization beyond kinases or substrates represented in target-domain training. Across all nine benchmark--split settings, PhosTailor achieved the highest AUROC and AUPRC, with its most pronounced improvements observed for sperm-relevant kinases unseen during target-domain training on SpermK-PSP (Fig.~\ref{fig:perf}).

\subsubsection{PhosTailor generalizes across kinase and substrate shifts on SIGNOR-Phos}
\begin{figure*}[!htp]
   \begingroup
   \centering
   \includegraphics[width=1\linewidth]{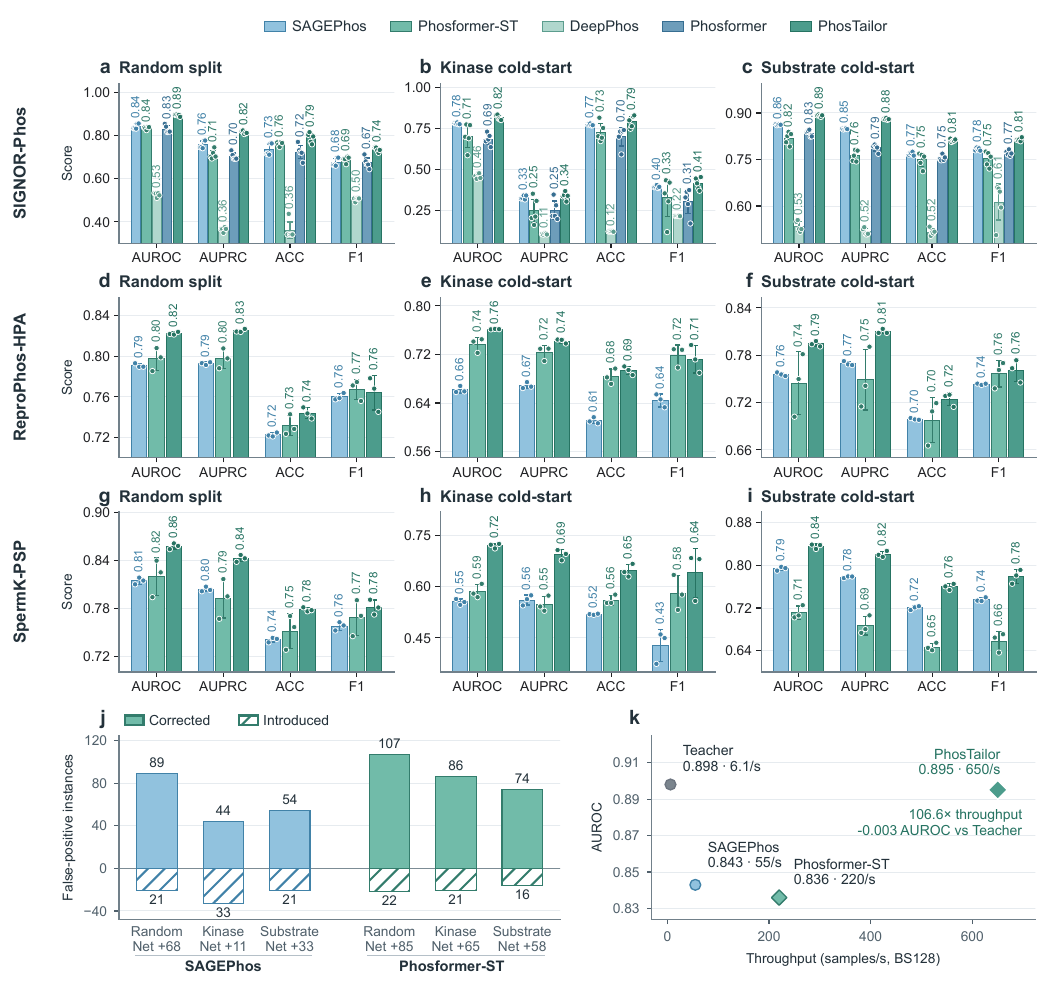}
   \endgroup
   
    \caption{\textbf{PhosTailor generalizes across reproductive phosphorylation landscapes with improved false-positive control and high inference throughput.}
\textbf{a--c}, AUROC, AUPRC, accuracy (ACC) and F1 score on SIGNOR-Phos under random (\textbf{a}), kinase-cold-start (\textbf{b}) and substrate-cold-start (\textbf{c}) splits.
\textbf{d--f}, The same metrics on ReproPhos-HPA under random (\textbf{d}), kinase-cold-start (\textbf{e}) and substrate-cold-start (\textbf{f}) splits.
\textbf{g--i}, The same metrics on SpermK-PSP under random (\textbf{g}), kinase-cold-start (\textbf{h}) and substrate-cold-start (\textbf{i}) splits. Bars show means and error bars denote s.d. across five independent runs for SIGNOR-Phos ($n=5$) and three independent runs for ReproPhos-HPA and SpermK-PSP ($n=3$), with individual runs indicated.
\textbf{j}, Paired comparison of false-positive (FP) instances corrected or introduced by PhosTailor relative to SAGEPhos and Phosformer-ST on SIGNOR-Phos across the three splits, with net reductions indicated.
\textbf{k}, AUROC versus inference throughput on the same GPU (batch size, 128) on SIGNOR-Phos. PhosTailor achieves 106.6$\times$ the teacher's throughput with an AUROC reduction of 0.003.}
    \label{fig:perf}
\end{figure*}

To assess the predictive performance of PhosTailor, we compared it with four representative phosphorylation predictors: DeepPhos \cite{luo2019deepphos}, Phosformer \cite{zhou2023phosformer}, Phosformer-ST \cite{zhou2024using} and SAGEPhos \cite{zhang2025sagephos}. These methods encompass sequence-based and structure-aware prediction approaches. We evaluated all models under random, kinase cold-start and substrate cold-start splits to examine both conventional predictive performance and generalization to kinases or substrates absent from target-domain training. All models were evaluated on matched partitions across five independent seeds (Fig.~\ref{fig:perf}a--c). Dataset construction and evaluation details are provided in Supplementary Information~C.1 and Supplementary Methods~C.1.4.

Under the random split, PhosTailor achieved an AUROC of 0.895 and an AUPRC of 0.817, representing relative improvements of 6.2\% and 7.8\%, respectively, over the strongest baseline, SAGEPhos (Fig.~\ref{fig:perf}a). Improvements were also observed in accuracy and F1 score. These results indicate more reliable discrimination of annotated kinase–substrate phosphorylation relationships under conventional evaluation conditions, providing a basis for prioritizing candidate interactions in reproductive phosphoproteomic studies.

We next evaluated whether PhosTailor could predict phosphorylation relationships involving kinases absent from reproductive-domain training. This kinase cold-start setting is particularly relevant to sperm biology, where many upstream kinases lack experimentally established substrate annotations. Although predictive performance decreased across methods, PhosTailor maintained the highest AUROC (0.818) and AUPRC (0.339), improving over SAGEPhos by 4.7\% and 2.7\%, respectively (Fig.~\ref{fig:perf}b). Its false-positive rate was also 9.1\% lower than that of SAGEPhos (0.179 versus 0.197). These findings suggest that PhosTailor can extend kinase–substrate predictions beyond the kinases represented in reproductive-domain training, supporting the prioritization of candidate upstream regulators with limited sperm-specific annotations.

We further examined generalization to previously unseen substrate proteins, a distinct challenge in sperm phosphoproteomics, where phosphorylation is frequently detected on proteins with limited kinase–substrate annotations. Under substrate cold-start, PhosTailor achieved an AUROC of 0.891 and an AUPRC of 0.881, exceeding SAGEPhos by 3.4\% and 3.8\%, respectively (Fig.~\ref{fig:perf}c). Notably, its false-positive rate decreased to 0.205, a 23.5\% relative reduction compared with the lowest competing rate of 0.268. This performance suggests that PhosTailor can identify candidate kinase–site relationships on proteins not represented during reproductive-domain training, potentially facilitating the investigation of poorly annotated flagellar and other sperm-associated proteins.

We next examined whether these improvements extended to reproductive phosphorylation datasets assembled using complementary tissue-expression and sperm-kinase criteria.

\subsubsection{PhosTailor generalizes across reproductive tissue- and sperm-kinase-focused benchmarks}

To examine whether PhosTailor generalizes across different reproductive phosphorylation datasets, we evaluated it separately on ReproPhos-HPA and SpermK-PSP. ReproPhos-HPA covers phosphorylation events associated with proteins expressed in male reproductive tissues, whereas SpermK-PSP focuses on events involving literature-supported sperm-relevant kinases. These datasets therefore provide complementary assessments of reproductive phosphorylation prediction. We compared PhosTailor with the two strongest baselines, Phosformer-ST and SAGEPhos, under random, substrate cold-start and kinase cold-start splits, with results averaged across three independent runs (Fig.~\ref{fig:perf}d--i).

On ReproPhos-HPA, PhosTailor achieved the highest AUROC and AUPRC across all three splitting settings. Under the random split, it reached an AUROC of 0.823 and an AUPRC of 0.825, exceeding the strongest baseline by 2.5 and 2.7 percentage points, respectively (Fig.~\ref{fig:perf}d). The advantage increased under substrate cold-start, with improvements of 3.9 and 4.1 percentage points in AUROC and AUPRC, respectively (Fig.~\ref{fig:perf}f). These results are relevant to reproductive phosphoproteomics, where newly identified phosphorylation sites often occur on proteins with limited kinase annotations. The performance gains suggest that PhosTailor can extend kinase–site predictions to substrate proteins not encountered during reproductive-domain training.
The advantage was also maintained for previously unseen kinases. Under kinase cold-start, PhosTailor achieved an AUROC of 0.762 and an AUPRC of 0.742, improving over Phosformer-ST by 2.5 and 2.0 percentage points, respectively (Fig.~\ref{fig:perf}e). This result is encouraging because many kinases potentially involved in reproductive signalling have limited experimentally established substrate relationships. The ability to make predictions for kinases excluded from target-domain training may therefore help identify candidate upstream regulators beyond those already well represented in reproductive phosphorylation annotations.

We next evaluated SpermK-PSP, which more specifically represents phosphorylation events involving sperm-relevant kinases. PhosTailor achieved an AUROC of 0.858 and an AUPRC of 0.843 under the random split, exceeding the strongest baselines by 3.8 and 3.9 percentage points, respectively (Fig.~\ref{fig:perf}g). Under substrate cold-start, the corresponding gains were 4.2 and 4.3 percentage points (Fig.~\ref{fig:perf}i). These findings indicate that the predictive advantage extends from the broader reproductive tissue-associated dataset to a more sperm-focused kinase–substrate setting, including when the substrate proteins are absent from target-domain training.
The most pronounced improvement occurred under kinase cold-start on SpermK-PSP. PhosTailor achieved an AUROC of 0.719 and an AUPRC of 0.693, exceeding the strongest comparator by 13.4 and 13.2 percentage points, respectively (Fig.~\ref{fig:perf}h). These differences correspond to relative improvements of approximately 23\% in both metrics. F1 score and accuracy also increased by 6.0 and 8.6 percentage points, respectively. Unlike the random split, this evaluation required predictions for sperm-relevant kinases with no corresponding examples in target-domain training. The substantial gains suggest that PhosTailor is less dependent on reproductive-domain annotations for individual kinases, making it particularly relevant for investigating sperm-associated kinases whose substrate relationships remain poorly characterized.

Together, these results show that the advantages of PhosTailor extend across both broad reproductive tissue-associated data and a more focused sperm-kinase dataset. The improvements under substrate cold-start support predictions on previously unseen reproductive proteins, while the larger gains under kinase cold-start on SpermK-PSP highlight its potential for prioritizing substrates of underannotated sperm-relevant kinases. Combined with structure-free inference, these capabilities support the application of PhosTailor to reproductive phosphoproteomic studies where experimentally established kinase–substrate relationships remain limited.

\subsection{PhosTailor improves prediction across tumour-associated phosphorylation settings}

We next investigated whether the PhosTailor framework could be adapted beyond reproductive biology to tumour-associated phosphorylation. In tumours, aberrant phosphorylation reflects dysregulated signalling networks, and identifying upstream kinases can help uncover altered pathway activity, connect molecular alterations to downstream signalling and prioritize candidate regulatory targets for functional investigation \cite{mertins2016proteogenomics,zagorac2018vivo}. We therefore independently adapted PhosTailor to the tumour-associated benchmark and compared it with SAGEPhos and Phosformer-ST under random, kinase cold-start and substrate cold-start splits (Fig.~\ref{fig:fpr}a--c).

PhosTailor achieved the highest AUROC and AUPRC across all three evaluation settings. Under the random split, it reached an AUROC of 0.86 and an AUPRC of 0.84, outperforming the strongest baselines by approximately 4 percentage points in both metrics (Fig.~\ref{fig:fpr}a). This advantage persisted under kinase cold-start, with an AUROC of 0.78 and an AUPRC of 0.76, exceeding the strongest comparators by approximately 4 percentage points (Fig.~\ref{fig:fpr}b). These results suggest that PhosTailor can transfer biochemical kinase-recognition principles to previously unrepresented kinases, including those absent from both general-domain pretraining and tumour-domain adaptation. This capability is particularly relevant to tumour signalling, where aberrant kinase activities can drive dysregulated pathways despite limited knowledge of their downstream substrates. The largest improvements occurred under substrate cold-start, where PhosTailor achieved an AUROC of 0.77 and an AUPRC of 0.76, corresponding to gains of approximately 7 and 8 percentage points, respectively (Fig.~\ref{fig:fpr}c). This suggests that PhosTailor can extend kinase attribution to previously unrepresented substrate proteins, potentially helping resolve the upstream regulators of newly identified tumour-associated phosphosites and improve the interpretation of dysregulated signalling networks.

\subsection{PhosTailor achieves substantial gains in cancer-associated phosphorylation prediction}

We further investigated whether PhosTailor could capture phosphorylation relationships specifically associated with cancer, beyond the broader tumour-associated setting, by independently adapting the framework to the cancer-specific subset (Fig.~\ref{fig:fpr}d--f). Aberrant phosphorylation is a key feature of oncogenic signalling, influencing pathways that govern cancer cell proliferation, survival and responses to therapy. Identifying the upstream kinases responsible for these events is therefore important for connecting phosphoproteomic alterations to potential oncogenic regulatory networks.

PhosTailor outperformed both baselines across all three evaluation settings, with particularly pronounced gains under random and substrate cold-start evaluations. Under the random split, it achieved an AUROC of 0.86 and an AUPRC of 0.84, exceeding the strongest baselines by approximately 15 and 11 percentage points, respectively (Fig.~\ref{fig:fpr}d). These substantial gains suggest that domain-specific adaptation enables more effective capture of cancer-associated kinase–substrate recognition patterns. Under kinase cold-start, PhosTailor maintained an AUROC and AUPRC of 0.75, corresponding to improvements of approximately 7 and 5 percentage points (Fig.~\ref{fig:fpr}e). This indicates that the transferred biochemical recognition knowledge extends to cancer-associated kinases absent from both general-domain pretraining and cancer-domain adaptation, potentially enabling the prioritization of previously uncharacterized kinases involved in oncogenic signalling. Notably, under substrate cold-start, PhosTailor achieved an AUROC of 0.75 and an AUPRC of 0.74, outperforming the strongest comparators by approximately 11 and 9 percentage points, respectively (Fig.~\ref{fig:fpr}f). This advantage highlights its ability to transfer kinase-recognition patterns to previously unseen cancer-associated substrate proteins, potentially facilitating the identification of upstream regulators for newly characterized cancer phosphosites. Together, these results support the utility of PhosTailor for resolving candidate kinase regulators in cancer-associated phosphoproteomic landscapes. 

\subsection{PhosTailor improves false-positive control while enabling efficient inference}

\begin{figure}[!htp]
    \centering
    \includegraphics[width=1\linewidth]{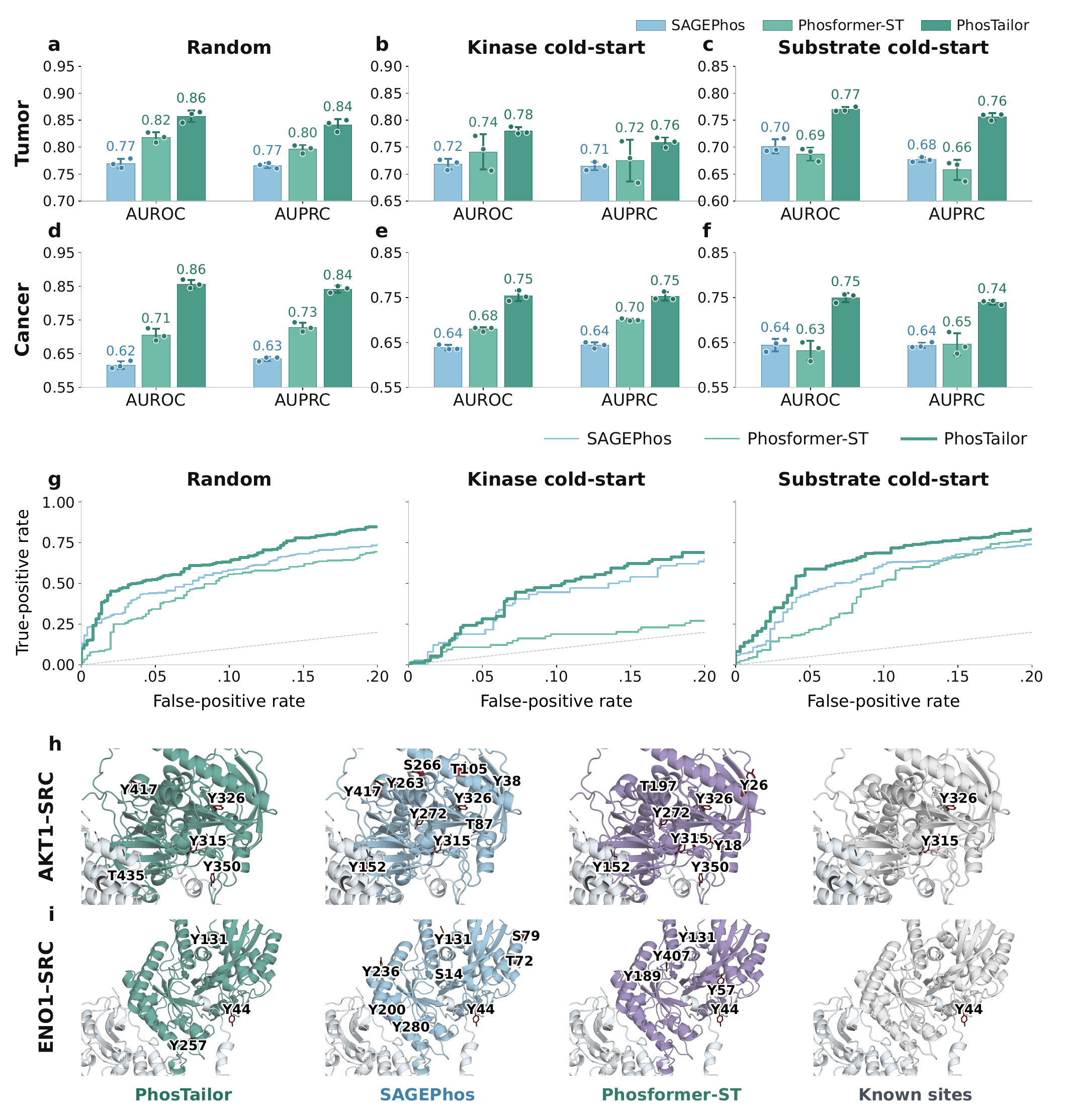}
    \caption{\textbf{PhosTailor enables accurate and efficient disease-associated kinase inference.} \textbf{a–f}, AUROC and AUPRC on the Tumor (a–c) and Cancer (d–f) datasets under random, kinase cold-start and substrate cold-start splits. Bars show mean $\pm$ s.d. across three independent runs ($n=3$), with individual runs indicated. \textbf{g}, ROC curves in the low-false-positive-rate region (FPR $\leq 0.20$); dashed lines indicate random expectation. \textbf{h,i}, Structural mapping of predicted and annotated phosphorylation sites for AKT1–SRC (h) and ENO1–SRC (i), with residue positions labelled.}
    \label{fig:fpr}
\end{figure}

Beyond overall predictive performance, we examined whether PhosTailor could improve kinase--phosphosite prioritization under stringent false-positive constraints. On the reproductive benchmark, PhosTailor generally achieved higher true-positive rates at comparable false-positive rates (FPR $\leq 0.20$) across random, kinase cold-start and substrate cold-start evaluations (Fig.~\ref{fig:fpr}g). The advantages were particularly evident under random and substrate cold-start conditions, indicating improved recovery of annotated phosphorylation relationships while limiting false-positive predictions.

To further characterize these improvements, we performed paired comparisons of false-positive predictions on identical negative-labelled instances (Fig.~\ref{fig:perf}j). Relative to SAGEPhos, PhosTailor corrected 89, 44 and 54 false positives while introducing only 21, 22 and 21 new false positives under random, kinase cold-start and substrate cold-start splits, respectively. Compared with Phosformer-ST, the corresponding numbers were 107, 86 and 74 corrected false positives versus 22, 21 and 16 newly introduced false positives. Across all six comparisons, PhosTailor corrected approximately two to five times as many false positives as it introduced, demonstrating improved screening reliability at the individual prediction level.

To illustrate the improved prediction selectivity underlying PhosTailor's reduced false-positive burden, we examined SRC-associated predictions for two cancer-relevant substrates, AKT1 and ENO1. SRC-family-dependent phosphorylation of AKT1 at Y315 and Y326 has been implicated in AKT activation and cell-survival signalling \cite{chen2001regulation}. ENO1, a glycolytic enzyme associated with cancer progression through FAK–PI3K–AKT signalling, contains a Y44 site previously reported as Src-dependent in alpha-enolase \cite{tanaka1995chicken,fu2015alpha}. We mapped predicted phosphorylation sites onto the protein structures and compared them with reference-annotated sites (Fig.~\ref{fig:fpr}h,i). For AKT1, PhosTailor recovered both annotated sites (Y315 and Y326) among five predicted candidates, compared with ten and eight candidates from SAGEPhos and Phosformer-ST, respectively. Similarly, PhosTailor identified the annotated ENO1 site (Y44) among three candidates, whereas SAGEPhos and Phosformer-ST nominated eight and five sites, respectively. Thus, PhosTailor retained coverage of all reference-annotated sites while substantially reducing additional candidate predictions across both substrates, illustrating improved selectivity and a lower potential false-positive burden for kinase–phosphosite prioritization.

We next assessed whether these predictive advantages could be achieved without the computational burden of multimodal inference (Fig.~\ref{fig:perf}k). The sequence-only PhosTailor student achieved predictive performance comparable to its multimodal teacher (AUROC, 0.895 versus 0.898), while increasing inference throughput from 6.1 to 650 samples per second, a 106.6-fold improvement. PhosTailor also outperformed SAGEPhos (AUROC, 0.843) and Phosformer-ST (AUROC, 0.836), without requiring structural retrieval or graph construction at inference. Together, these results demonstrate that PhosTailor combines improved false-positive control with near-teacher predictive performance and substantially more efficient, structure-free inference.

\subsection{PhosTailor captures kinase-conditioned biochemical specificity}
\begin{figure*}[!htbp]
    \begingroup
    \centering
    \includegraphics[width=1\textwidth]{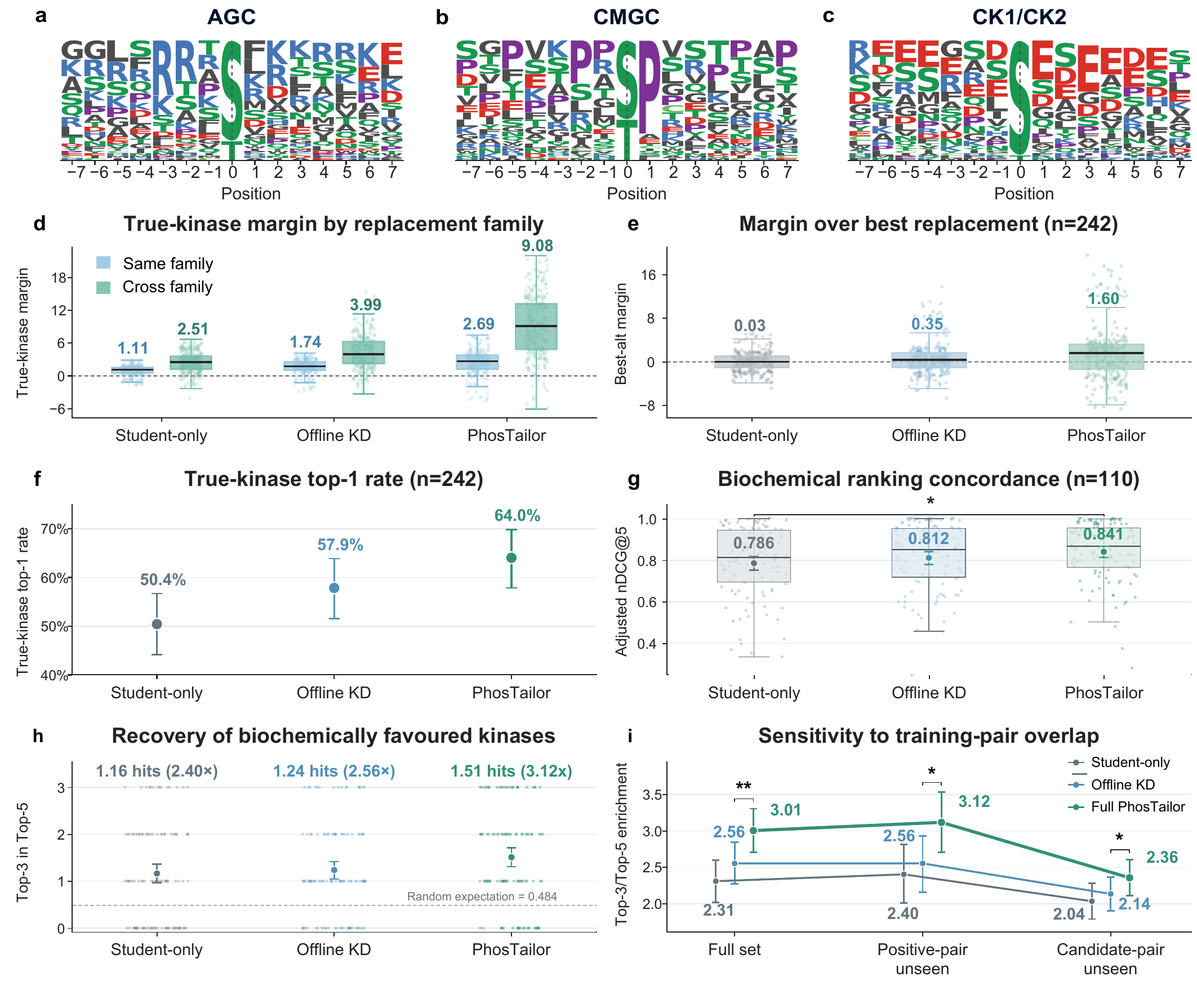}
    \endgroup
    \caption{\textbf{PhosTailor recovers biochemical kinase recognition specificity.}
    (\textbf{a}--\textbf{c}) Sequence logos for high-confidence Ser/Thr-centred predictions for AGC, CMGC and CK1/CK2 kinases. 
    (\textbf{d}--\textbf{f}) Counterfactual substitution of the supported kinase by same- or cross-family alternatives in 242 candidate-pair-unseen kinase\textendash site pairs: logit margins over replacement kinases (\textbf{d}), margin over the highest-scoring replacement (\textbf{e}) and frequency with which the supported kinase ranked first (\textbf{f}). (\textbf{g}) Bias-adjusted nDCG@5 against the Kinase Library atlas across 110 positive-pair-unseen sites. 
    (\textbf{h}) Recovery of atlas-ranked kinases among model top-five predictions; the dashed line marks random expectation. 
    (\textbf{i}) Top-three-in-top-five enrichment across the full, positive-pair-unseen and candidate-pair-unseen sets. Boxes show medians and interquartile ranges; confidence intervals are 95\% where shown. Paired two-sided Wilcoxon tests were adjusted by the Benjamini--Hochberg procedure; asterisks denote adjusted $q < 0.05$, $q < 0.01$ and $q < 0.001$.}

    \label{fig:kinase_family_motifs}
\end{figure*}

Predictive discrimination alone does not establish that a kinase--substrate model has learned kinase specificity rather than generic phosphosite propensity or recurrent substrate-local sequence patterns. We therefore evaluated the biochemical content retained by the final structure-free student using three complementary analyses that interrogate kinase specificity at different levels: recovery of kinase-family substrate motifs from high-confidence predictions, concordance with independently measured intrinsic kinase specificity, and counterfactual sensitivity to kinase identity while holding the substrate and phosphosite fixed.

We first examined whether high-confidence PhosTailor predictions recovered established sequence determinants of kinase--substrate recognition. The frozen model was used to score eligible serine/threonine-centred kinase--site pairs in the held-out test set without using ground-truth labels for candidate selection. Candidates were ranked separately for each kinase and aggregated within kinase groups using a balanced selection strategy, yielding the top 50 unique high-confidence substrate sites for each group. The resulting sequence patterns recapitulated distinct biochemical recognition modes, as shown in Fig.~\ref{fig:kinase_family_motifs}a-c. Predictions for the AGC kinase group showed a pronounced preference for basic residues upstream of the phosphoacceptor, with arginine particularly prominent around positions $-3$ and $-2$; CMGC predictions showed the characteristic proline-directed preference immediately downstream of the phosphoacceptor at position $+1$; and CK1/CK2 predictions were enriched for acidic residues in the local substrate context. Consistent patterns were recovered at finer resolution, including basophilic signatures for protein kinase C (PKC) and calcium/calmodulin-dependent protein kinase II (CAMK2), and a strong $+1$ proline preference for mitogen-activated protein kinase (MAPK) predictions (Supplementary Fig.~A1 and Supplementary Note~C.3). These output-level patterns indicate that the high-confidence predictions preserve family-associated substrate preferences spanning the major basophilic, proline-directed and acidophilic recognition modes of Ser/Thr kinases.

We next asked whether PhosTailor could prioritize kinases for a fixed phosphosite in a manner consistent with experimentally measured intrinsic kinase specificity. We used the Kinase Library Ser/Thr specificity atlas \cite{johnson2023atlas}, which was derived from positional-scanning peptide-library measurements and therefore provides an experimental reference that is independent of the training labels used by PhosTailor. Of the 50 unique kinases represented in the random test set, 31 Ser/Thr kinases could be mapped unambiguously to the atlas and were retained as a common candidate panel. Across 220 eligible phosphosites, each site was paired with all 31 kinases, yielding 6,820 kinase--site combinations. For every phosphosite, the same candidate panel was ranked independently by the model and by the experimental specificity atlas. Unannotated kinase--site combinations were retained as unknown candidates rather than treated as negatives, and model logits were centred separately for each kinase to reduce kinase-wide scoring biases. Ranking concordance was quantified using bias-adjusted normalized discounted cumulative gain at rank five (nDCG@5), together with recovery of the three highest-ranked atlas kinases within the model top five. Detailed candidate mapping, score normalization, ranking metrics, overlap filtering and statistical procedures are provided in Supplementary Note~C.4.

To minimize the contribution of training-pair memorization, the primary analysis excluded phosphosites whose annotated positive kinase--site pair had appeared in either student or teacher training, leaving 110 positive-pair-unseen phosphosites. PhosTailor achieved a mean bias-adjusted nDCG@5 of 0.841, compared with 0.786 for the non-distilled student and 0.812 for fixed-teacher offline distillation (Fig.~\ref{fig:kinase_family_motifs}g). It also recovered an average of 1.51 of the three highest-ranked Kinase Library kinases within its top five predictions, compared with 1.16 and 1.24 for the non-distilled student and offline-distillation models, respectively (Fig.~\ref{fig:kinase_family_motifs}h). These values corresponded to 3.12-fold, 2.40-fold and 2.56-fold enrichment over the random expectation of $3\times5/31=0.484$ recovered kinases per phosphosite (Fig.~\ref{fig:kinase_family_motifs}i). We further tested whether the apparent biochemical concordance depended on any candidate kinase--site combination previously observed during training. Among the 6,820 analysed combinations, 712 (10.44\%) had appeared in either student or teacher training. PhosTailor retained the highest Top3-in-Top5 enrichment on the complete phosphosite set, after excluding previously observed positive pairs, and after removing every candidate kinase--site pair observed during training, with enrichment values of 3.00-fold, 3.12-fold and 2.36-fold, respectively. Under the most stringent candidate-pair-unseen analysis, PhosTailor retained the highest mean enrichment and remained significantly better than fixed-teacher offline distillation after Benjamini--Hochberg correction. Thus, the agreement with the experimental kinase-specificity atlas cannot be explained solely by memorization of previously observed kinase--site relationships.

Finally, we directly tested whether PhosTailor predictions depended on kinase identity through counterfactual kinase substitution. We analysed 242 candidate-pair-unseen positive kinase--site pairs and held the substrate protein, phosphosite and local sequence context fixed while replacing only the supported kinase with sampled same-family or cross-family alternatives. Detailed candidate selection, counterfactual kinase replacement, margin definitions and statistical procedures are provided in Supplementary Note~C.5. Substitution of the supported kinase reduced predicted compatibility across all model variants, with the decrease consistently larger for cross-family than for same-family replacements, consistent with the hierarchical biochemical similarity among kinase alternatives. The magnitude of this response increased progressively with knowledge transfer: median same-family and cross-family logit margins were 1.11 and 2.51 for the non-distilled student, 1.74 and 3.99 after fixed-teacher offline distillation, and 2.69 and 9.08 for full PhosTailor (Fig.~\ref{fig:kinase_family_motifs}d).
The same ordering was observed under the stricter comparison against the highest-scoring replacement kinase, for which the median supported-kinase margin increased from 0.03 to 0.35 and 1.60, respectively (Fig.~\ref{fig:kinase_family_motifs}e).
Accordingly, the supported kinase ranked above all sampled replacement kinases in 64.0\% of candidate-pair-unseen cases for PhosTailor, compared with 57.9\% for offline distillation and 50.4\% for the non-distilled student (Fig.~\ref{fig:kinase_family_motifs}f). Together, motif recovery, independent biochemical ranking concordance and controlled kinase substitution show that the final structure-free student retains kinase-conditioned biochemical information rather than relying primarily on generic substrate-local phosphorylation propensity. This biochemical specificity provides an important basis for using PhosTailor to prioritize upstream kinase programmes in subsequent human sperm phosphoproteomic analyses.

\begin{figure*}[!htbp]
    \begingroup
    \centering
    \includegraphics[width=1\textwidth]{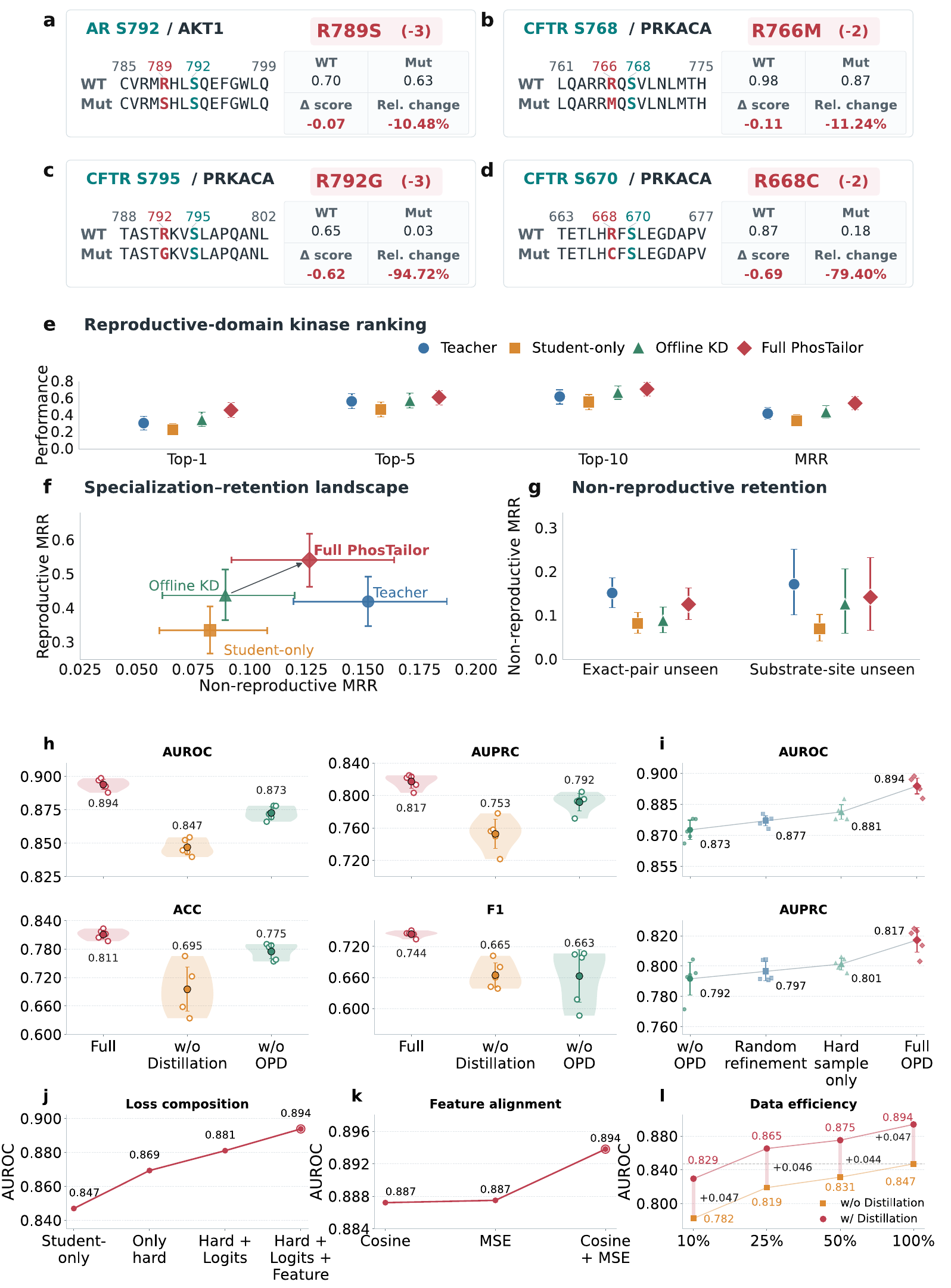}
    \endgroup
    
    \caption{\textbf{Disease-associated variant sensitivity, kinase-ranking retention and ablation analysis of PhosTailor.} (\textbf{a–d}), Predicted kinase–site compatibility for AR R789S (S792) and CFTR R766M (S768), R792G (S795) and R668C (S670), showing wild-type and mutant sequences and prediction scores. (\textbf{e}), Top-1, Top-5 and Top-10 accuracy and mean reciprocal rank (MRR) across 124 reproductive cases. (\textbf{f,g}), Reproductive and non-reproductive kinase-ranking performance (f), and non-reproductive MRR in exact-pair-unseen ($n=240$) and substrate-site-unseen ($n=53$) subsets (g). Error bars in e–g indicate bootstrap 95\% confidence intervals from 20,000 resamples. (\textbf{h}), Module-level ablations assessing the contributions of distillation and online policy distillation (OPD). (\textbf{i}), Comparison of OPD refinement strategies. In h and i, results are from five independent runs, with individual runs and mean $\pm$ s.d. indicated. (\textbf{j}), Stepwise integration of hard-example learning, response distillation and feature alignment. (\textbf{k}), Comparison of cosine, mean-squared-error (MSE) and combined feature-alignment objectives. (\textbf{l}), AUROC with and without distillation across four training-data fractions (10–100\%); connecting segments indicate gains from distillation.}
    \label{fig:case_ab}
\end{figure*}

\subsection{Ablation analyses reveal complementary contributions of knowledge distillation and adaptive refinement}

To determine the contributions of PhosTailor's principal training components, we performed controlled ablation analyses on the reproductive phosphorylation benchmark (Fig.~\ref{fig:case_ab}h--l). Knowledge distillation accounted for the largest performance improvement. Compared with the non-distilled student, the full model increased AUROC from 0.847 to 0.894 and AUPRC from 0.753 to 0.817, while improving accuracy by 16.7\%. 
As the student architecture and evaluation partitions were held constant, these gains indicate that multimodal teacher supervision provides information beyond that learned from target-domain annotations alone.

Student-guided refinement contributed additional improvements beyond conventional fixed-teacher distillation, increasing AUROC from 0.873 to 0.894 (Fig.~\ref{fig:case_ab}i). To examine whether this benefit simply reflected additional teacher optimization or an emphasis on difficult examples, we compared the full framework with random teacher refinement and hard-example prioritization without teacher adaptation. Neither reproduced the performance of the complete procedure, achieving AUROC values of 0.877 and 0.881, respectively. These findings suggest that the advantage of the closed-loop strategy arises from using student-identified difficulties to direct teacher refinement and subsequently transfer the updated supervision back to the student.

Further analysis of the training objectives showed that the different levels of supervision made complementary contributions (Fig.~\ref{fig:case_ab}j,k). Prioritizing difficult examples improved AUROC from 0.847 to 0.869, with response-level distillation and representation alignment progressively increasing performance to 0.894. Combining cosine and mean-squared-error feature alignment also outperformed either objective alone, consistent with the benefit of preserving complementary aspects of the teacher's learned representations. These results indicate that transferring teacher predictions alone does not capture the full benefit of multimodal supervision; aligning intermediate representations provides an additional source of predictive information.

The value of knowledge transfer was particularly evident when target-domain annotations were limited (Fig.~\ref{fig:case_ab}l). Distillation consistently improved AUROC across all evaluated training-data fractions. Notably, with only 25\% of the reproductive training data, the distilled student outperformed the non-distilled model trained on the complete dataset (AUROC, 0.865 versus 0.847). 
This finding suggests that broadly learned phosphorylation knowledge can compensate, in part, for limited domain-specific annotations, reducing the dependence of specialized prediction on extensive target-domain training data. Together, these ablations support the complementary roles of multimodal knowledge transfer and student-guided refinement in developing accurate predictors for sparsely annotated biological contexts. Detailed experimental configurations are provided in Supplementary Note~C.7.

\subsection{PhosTailor balances domain-specific kinase prioritization with broader knowledge retention}

A central question in general-to-specialized learning is whether improved kinase prioritization within a target domain can be achieved without substantially compromising broadly transferable kinase-recognition knowledge. To investigate this trade-off, we evaluated PhosTailor following reproductive-domain adaptation, using reproductive and non-reproductive phosphorylation sites to assess target-domain performance and broader knowledge retention, respectively. We compared Full PhosTailor with the multimodal Teacher, Student-only and Offline KD models under a common kinase-ranking protocol (Fig.~\ref{fig:case_ab}e--g).

The reproductive evaluation cohort was derived from the held-out random test partition of SIGNOR-Phos, yielding 262 unique positive substrate sites after preprocessing and deduplication. To minimize training overlap, we excluded 127 sites with at least one annotated kinase--substrate--site relationship overlapping TeacherPTM or student training, followed by 11 sites with incomplete kinase-panel coverage, inconsistent sequence mappings or unavailable structural inputs. This resulted in 124 reproductive sites with 126 annotated positive kinase--site relationships. For the non-target-domain evaluation, we independently assembled 240 non-reproductive sites from disease-associated phosphorylation records after deduplication and canonicalized training-overlap filtering, including the removal of one previously undetected overlapping case. All models ranked the same panel of 453 human kinases using raw prediction scores. For sites with multiple annotated kinases, the highest-ranked supported kinase was used to calculate Top-1, Top-5, Top-10 and mean reciprocal rank (MRR). No pseudo-negative labels were introduced, allowing kinase prioritization to be assessed directly against annotated positive relationships.

We first examined ranking performance within the specialized target domain. Full PhosTailor achieved the highest point estimates across all four metrics, with Top-1 accuracy reaching 46.0\% compared with 30.6\% for the Teacher, and MRR increasing from 0.420 to 0.542. The improvement in MRR over Student-only remained significant after multiple-testing correction, although differences from the Teacher and Offline KD did not reach statistical significance (adjusted $q=0.057$ and $q=0.056$, respectively). These results indicate that reproductive specialization was associated with improved kinase-ranking point estimates, consistent with more effective prioritization of regulatory relationships within the target domain.

We then examined whether these target-domain gains were accompanied by substantial loss of ranking ability outside the specialized domain. On the 240 non-reproductive sites, Full PhosTailor achieved an MRR of 0.126, retaining approximately 83\% of the Teacher's performance (MRR, 0.152). Notably, it also outperformed conventional Offline KD by approximately 42\% (MRR, 0.126 versus 0.089; paired two-sided Wilcoxon test, Benjamini--Hochberg-adjusted $q=2.42\times10^{-5}$). Thus, although domain adaptation did not fully preserve teacher performance outside the target domain, student-guided refinement was associated with better retention of broader kinase-ranking ability than fixed-teacher distillation.

To assess the robustness of this observation to more stringent training-overlap control, we repeated the analysis after excluding substrate sites encountered during training, irrespective of kinase identity. This retained 53 non-reproductive sites, for which the Full-to-Teacher MRR ratio remained approximately 83\%. However, the difference between Full PhosTailor and Offline KD was no longer statistically significant ($q=0.277$). The corresponding reproductive subset contained only four sites, precluding meaningful statistical inference. Pointwise 95\% percentile confidence intervals were estimated from 20,000 bootstrap resamples of unique substrate sites.

Together, these findings suggest that PhosTailor can improve kinase prioritization within a specialized domain while retaining substantial ranking ability beyond it. The advantage over conventional offline distillation in the primary non-reproductive cohort further supports the role of student-guided adaptation in balancing domain specialization with preservation of transferable phosphorylation knowledge.

\subsection{PhosTailor captures context-dependent disruption of kinase recognition by disease-associated variants}
To assess whether PhosTailor can identify the effects of clinically observed sequence variation on kinase–substrate recognition, we examined four phosphosite-adjacent variants associated with androgen insensitivity syndrome or CFTR-related male infertility. By comparing wild-type and mutant sequences while preserving kinase identity and phosphosite position, we investigated whether the model could resolve mutation-induced changes in kinase compatibility (Fig.~\ref{fig:case_ab}a–d).

PhosTailor predicted a pronounced loss of PRKACA compatibility for the CFTR variant R792G associated with congenital bilateral absence of the vas deferens (CBAVD), with the score decreasing from 0.650 to 0.034 (94.72\%; Fig.~\ref{fig:case_ab}c). This substitution removes the arginine at position −3 relative to S795, disrupting a key basic determinant within the canonical PKA recognition motif. The model captured this disruption through a near-complete loss of predicted compatibility, consistent with previous studies identifying S795 as a PKA-regulated phosphosite and demonstrating reduced chloride-channel activity in R792G-mutant CFTR \cite{ravnik2000two,picciotto1992phosphorylation,vankeerberghen1998characterization}. A similarly strong response was observed for CFTR R668C, for which PhosTailor predicted a reduction in PRKACA–S670 compatibility from 0.868 to 0.179 (79.40\%; Fig.~\ref{fig:case_ab}d). Here, substitution of the −2 arginine removes an important upstream basic residue from the local recognition context, a perturbation captured by the marked decrease in model score. This prediction is mechanistically consistent with independent phosphoproteomic evidence establishing S670 responsiveness to cAMP–PKA signalling, while functional studies have separately shown impaired chloride conductance associated with R668C \cite{schnur2019phosphorylation,el2012cftr}.

Importantly, PhosTailor also distinguished substitutions associated with partial rather than near-complete compatibility loss. For CFTR R766M, the model predicted reduced PRKACA–S768 compatibility (0.977 to 0.867; Fig.~\ref{fig:case_ab}b), capturing the disruption of the −2 arginine within the established PKA recognition context \cite{ravnik2000two,hegedHus2009role}. Despite this substitution, neighbouring upstream arginines remain intact, providing a plausible basis for the high residual compatibility predicted by the model. Notably, the probability-scale reduction corresponded to a logit decrease of approximately 1.88, indicating appreciable sensitivity to this mutation. This response is consistent with biochemical evidence showing that R766M weakens the interaction between phosphorylated CFTR peptides and 14-3-3 proteins, supporting the functional importance of the affected phosphoregulatory region \cite{Stevers2016CFTR}. PhosTailor also detected a reduction in AKT1 compatibility for AR R789S, associated with partial and mild androgen insensitivity syndrome, from 0.700 to 0.626 (Fig.~\ref{fig:case_ab}a). R789S removes the −3 arginine adjacent to the AKT-regulated S792 phosphosite, disrupting a canonical AKT-recognition determinant that was reflected in the model prediction \cite{gottlieb2012androgen,palazzolo2007akt}. The remaining −5 arginine may partially preserve the local recognition context, providing a possible explanation for the more moderate predicted effect.

Together, these cases demonstrate that PhosTailor captures both pronounced and partial disruptions of kinase-recognition motifs, generating context-dependent predictions consistent with independent biochemical and functional evidence.

\subsection{PhosTailor nominates TBK1 as a candidate regulator of human sperm function}

To identify candidate upstream kinases relevant to male reproductive function, we selected 60 proteins encoded by genes harbouring variants identified through whole-exome sequencing of sperm from intrauterine insemination (IUI) failure cases (Supplementary Table~B3a). These proteins were manually prioritized for their reproductive relevance and encompassed key processes involved in sperm signalling, motility, maturation and fertilization. The frozen PhosTailor model was applied to nominate candidate kinases, integrating high-confidence kinase--phosphosite predictions across the protein panel into a cross-protein consensus ranking (Supplementary Note~C.6.1).

From the model-informed candidates, ten kinases were prioritized for biological investigation (Supplementary Table~B3b). Among these, TANK-binding kinase 1 (TBK1) emerged as a candidate of particular interest because its role in mature human sperm remains poorly characterized. We therefore selected TBK1 for subsequent phosphoproteomic and functional investigations. Mitogen-activated protein kinase kinase 2 (MAP2K2, also known as MEK2), the highest-ranked kinase in the biologically prioritized shortlist, was also included in the experimental evaluation.

\subsection{Phosphoproteomic and motility responses to kinase perturbation}

To investigate the model-nominated TBK1 candidate, we used GSK8612, a selective TBK1 inhibitor characterized in published pharmacological studies~\cite{thomson2019discovery}. After 1 h of exposure to 10 $\mu$M GSK8612, 411 of 7,292 quantified phosphosites met the fold-change (FC) criteria ($>1.5$ or $<1/1.5$, equivalently $|\log_2\mathrm{FC}|>\log_2(1.5)$) and the vendor-reported nominal differential-analysis threshold ($P<0.05$), while 79 of 4,507 quantified proteins met the differential-abundance criteria. Excluding one phosphosite associated with altered total-protein abundance yielded 410 filtered responsive sites for subsequent model--experiment concordance analysis.

\begin{figure*}[!htbp]
    \begingroup
    \centering
    \includegraphics[width=1\textwidth]{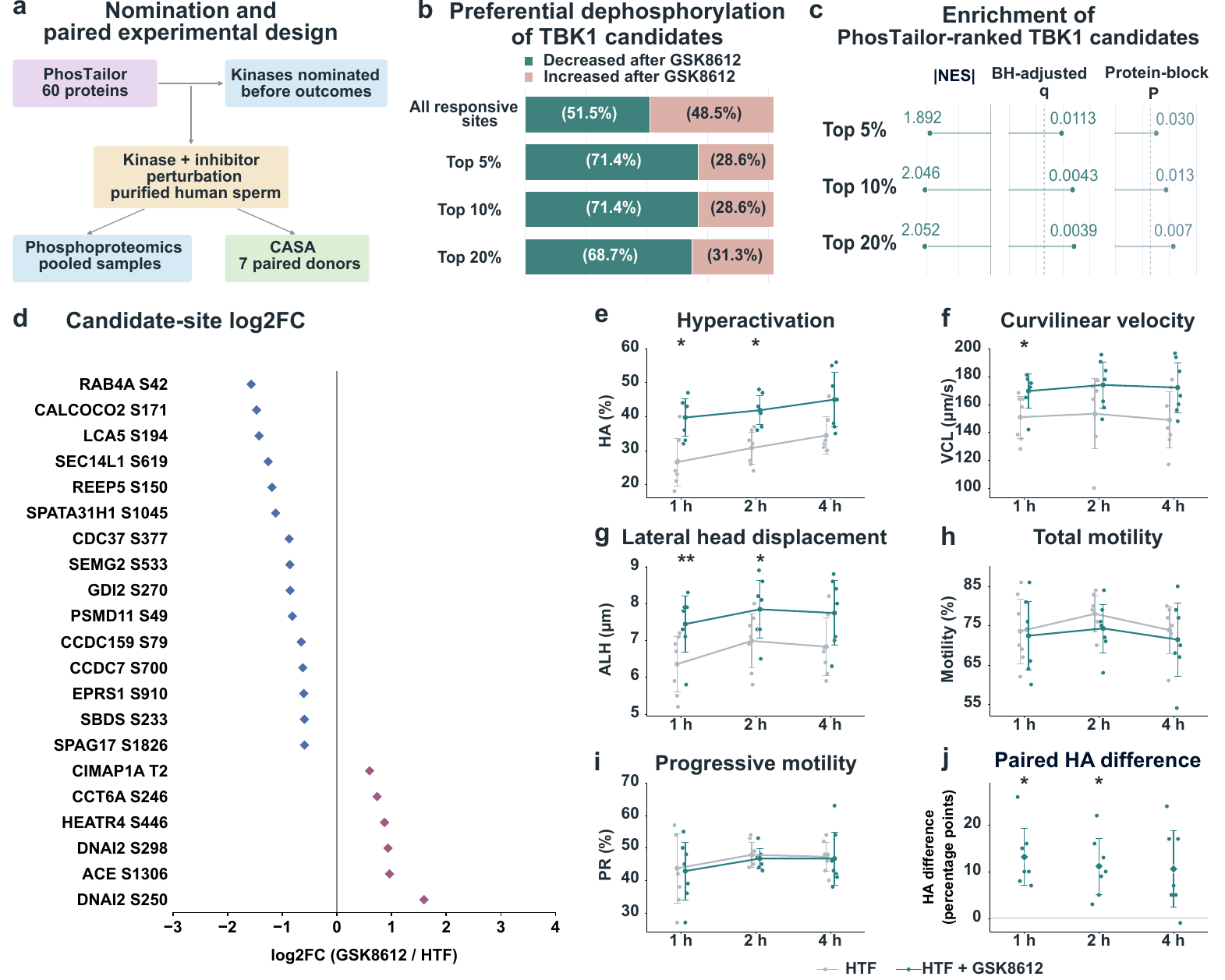}
    \endgroup    
  
    \caption{\textbf{PhosTailor-guided prioritization and experimental evaluation of TBK1-associated phosphorylation.}
    (\textbf{a}) Prospective nomination of TBK1 for perturbation in mature human sperm. (\textbf{b}) Proportions of decreased and increased phosphosites after 1 h of 10 $\mu$M GSK8612 among 410 filtered responsive sites and within the top 5\%, 10\% and 20\% PhosTailor candidate tiers; percentages are shown. (\textbf{c}) Enrichment of PhosTailor-ranked TBK1-compatible sites among phosphosites decreased after perturbation; absolute normalized enrichment scores ($|\mathrm{NES}|$), BH-adjusted $q$ values across the three nested tiers and protein-block-resampling $P$ values are shown. (\textbf{d}) GSK8612-versus-control $\log_2$ fold changes for the Top 5\% candidate phosphosites. Each diamond shows the $\log_2$ ratio of condition-mean intensities; the vertical line marks no change. (\textbf{e}--\textbf{j}) CASA measurements after 1, 2 and 4 h of GSK8612 or vehicle perturbation: hyperactivation (HA; \textbf{e}), curvilinear velocity (VCL; \textbf{f}), amplitude of lateral head displacement (ALH; \textbf{g}), total motility (\textbf{h}), progressive motility (PR; \textbf{i}) and paired-donor HA difference in percentage points (\textbf{j}). Points represent seven independent donors with matched vehicle and inhibitor aliquots; group summaries and paired-difference diamonds show means with pointwise 95\% confidence intervals. Symbols indicate two-sided paired $t$ tests with Holm correction across 15 endpoint-by-time comparisons; panel~\textbf{j} repeats the HA comparisons. *$P<0.05$, **$P<0.01$, ***$P<0.001$. Source values and exact statistics are provided in Supplementary Data 1, 4--6.
}
\label{fig:Model-nominated}
\end{figure*}

To evaluate whether PhosTailor could identify biologically relevant TBK1-associated phosphosites, we used the frozen model to rank 410 GSK8612-responsive phosphosites and selected the Top 5\%, Top 10\% and Top 20\% as candidate sets. We then examined whether these model-prioritized sites were preferentially associated with decreased phosphorylation after GSK8612 treatment (Supplementary Note~C.6.3 and Supplementary Methods~E.7). Whereas only 51.5\% of all responsive sites showed decreased phosphorylation, this proportion increased to 68.7\% among the Top 20\% model-ranked candidates (Fig.~\ref{fig:Model-nominated}b). These candidates were significantly enriched toward the decreased-phosphorylation response across all three thresholds, including Top 20\% (NES = $-2.052$, BH-adjusted $q = 0.0039$; protein-block $P = 0.007$; Fig.~\ref{fig:Model-nominated}c). The corresponding candidate-site responses are shown in Fig.~\ref{fig:Model-nominated}d. Together, these analyses show that PhosTailor prioritized TBK1-compatible candidates enriched for GSK8612-associated decreases within the 410-site response set, providing experimental support for model-guided phosphosite prioritization.

We next investigated functional responses to GSK8612 treatment using computer-assisted sperm analysis (CASA) in a separate cohort of seven independent donors, with matched vehicle and GSK8612 aliquots from each donor (Fig.~\ref{fig:Model-nominated}e--j). After 1 h of GSK8612 exposure, mean hyperactivation (HA) increased from 26.6\% in controls to 39.7\%, corresponding to a paired increase of 13.1 percentage points (pointwise 95\% CI, 7.0--19.2; Holm-adjusted $P = 0.026$ across 15 comparisons). HA was also increased at 2 h (adjusted $P = 0.044$), whereas the difference at 4 h did not remain significant after correction. Curvilinear velocity (VCL) increased at 1 h, and amplitude of lateral head displacement (ALH) increased at 1 and 2 h. Total motility and progressive motility (PR) showed no significant differences. Aligned with the phosphoproteomic concordance, these donor-paired CASA responses provide functional support for investigating the model-nominated TBK1 candidate in hyperactivation-associated sperm movement.

A complementary U0126 perturbation was used to examine the model-nominated MAP2K2 candidate within the MEK1/2 signalling pathway. MEK2 showed a consistently stronger proteomic signal than MAP2K1 (MEK1) across control and inhibitor-treated samples and was identified by 16 versus 9 distinct peptides, supporting examination of MAP2K2-compatible predictions within this pathway. Following 1 h of U0126 exposure, 485 of 7,129 quantitatively compared phosphosites met the nominal differential criteria; after the matched-protein filter, 463 sites were retained, of which 346 (74.7\%) showed decreased phosphorylation, including sites on the sperm movement-associated proteins AKAP4, AKAP3 and DNAH8. Phosphoproteomics comprised three technical LC--MS/MS replicates per condition from pooled material, providing a defined response set for evaluating model-nominated MAP2K2-compatible candidates.

Applying the same analysis to MAP2K2, the Top 5\% PhosTailor-ranked candidates (24 sites) showed nominal enrichment toward sites with decreased signal after U0126 treatment (NES = $-1.601$, empirical $P=0.0315$). This signal did not remain significant after correction across the three candidate tiers (BH-adjusted $q=0.0945$) or protein-block resampling ($P=0.0679$). The model-ranked Top 5\% set provides a focused list of MAP2K2-compatible candidates for follow-up within the U0126-responsive phosphoproteome (Supplementary Note~C.6.5).

In a separate cohort of five independent donors, U0126 treatment was associated at 1~h with lower HA (28.6\% to 16.8\%; paired difference, $-11.8$ percentage points; pointwise 95\% CI, $-15.67$ to $-7.93$; Holm-adjusted $P=0.0149$), VCL (paired difference, $-19.98$~$\mu$m/s; pointwise 95\% CI, $-27.79$ to $-12.17$; Holm-adjusted $P=0.0270$) and ALH (paired difference, $-0.68$~$\mu$m; pointwise 95\% CI, $-1.001$ to $-0.359$; Holm-adjusted $P=0.0461$). VCL was also lower at 2 and 4~h (Holm-adjusted $P=0.0408$ and $0.0016$, respectively). Mean total and PR values were lower after U0126 treatment, but neither endpoint reached significance after Holm correction across all 15 endpoint-by-time comparisons. These changes are directionally consistent with previous human-sperm observations that inhibition of the MEK--extracellular signal-regulated kinase (ERK) pathway reduces motility-associated responses~\cite{almog2008identification}. These donor-paired responses support the involvement of MEK1/2 signalling in hyperactivation-associated sperm movement.

Together, these experiments illustrate a practical route from PhosTailor-guided kinase nomination to phosphoproteomic and donor-level functional investigation. The GSK8612 findings support prioritization of TBK1-compatible phosphosite candidates, with U0126 providing complementary MEK1/2 pathway context for sperm-motility phenotypes.

\section{Discussion}\label{discussion}

Interpreting phosphoproteomic data requires knowing which kinase acts at which site, yet this inference is weakest precisely where it is most needed: in specialized contexts whose kinase–substrate annotations are sparse. PhosTailor addresses this as a general problem of transferring knowledge from a data-rich general domain into annotation-poor specialized ones. Protein structure and residue-level biochemistry are treated as privileged information — available while training the multimodal teacher, discarded at deployment — so that the delivered model is a single sequence-only student requiring no structural retrieval or graph construction. The framework is domain-agnostic by construction: the same procedure was adapted independently to reproductive, tumour-associated and cancer-specific phosphorylation, achieving the highest AUROC and AUPRC across all 15 benchmark–split combinations. Gains were largest where annotation is thinnest, and a distilled student trained on a quarter of the reproductive data outperformed a non-distilled student trained on all of it — transferred general knowledge substitutes for missing domain examples rather than sharpening an already well-supported boundary.

Lightweight deployment is not a convenience but a precondition for biological use. The sequence-only student approached its multimodal teacher's performance while running two orders of magnitude faster, converting proteome-scale kinase screening from an offline batch computation into an interactive query. We have accordingly released PhosTailor as a web server (\url{https://jingjie00-phostailor.hf.space/}) on which experimental biologists can submit substrates or phosphosites and obtain ranked candidate kinases without computational infrastructure, local deployment or structural input, the regime in which most phosphoproteomic datasets are generated. Efficiency did not come at the expense of mechanistic content. Without label access, high-confidence predictions recovered the canonical basophilic, proline-directed and acidophilic recognition modes; ranking against an experimental specificity atlas independent of our training labels recovered top-ranked kinases at over threefold random expectation; and counterfactual substitution reduced predicted compatibility more for cross-family than same-family replacements. Because the stringent analysis tested candidate pairs not observed during training, providing evidence of generalization to new kinase--site combinations. The model also responded to phosphosite-proximal clinical variants in proportion to motif disruption, suggesting a route to interpreting variants of uncertain significance.

An important biological finding of this study was the nomination of TANK-binding kinase 1 (TBK1), a kinase primarily studied in innate immune signalling and autophagy, as a candidate, previously underexplored regulator of mature human sperm function. Nominated prospectively by the frozen PhosTailor model, TBK1 was subsequently detected in human sperm, and model-prioritized candidates included proteins associated with axonemal and flagellar function, the fibrous sheath and CatSper. Within the locked, response-defined 410-site phosphoproteomic universe, the pooled GSK8612 response showed directional concordance with TBK1-compatible predictions. Donor-paired CASA further revealed increased hyperactivation-associated kinematics after GSK8612 treatment. Complementary U0126 treatment reduced hyperactivation-associated kinematics under MEK1/2 pathway inhibition. Together, these findings highlight TBK1 as a candidate for sperm-function research and illustrate how PhosTailor can guide investigation beyond established sperm kinases.

The experiments support model-guided kinase prioritization through pharmacological response profiling. Target-engagement measurements and orthogonal perturbations are needed to resolve kinase-specific mechanisms. The phosphoproteomic measurements were based on one pooled sample per condition with three technical replicates and therefore do not estimate donor-level biological variability. CASA used separate, non-overlapping donor cohorts for GSK8612 ($n=7$) and U0126 ($n=5$). Training resources are also biased toward well-studied kinases. Within these bounds, PhosTailor offers a reusable route from sparsely annotated PTM data to experimentally testable hypotheses, and identifies TBK1 as a candidate for mechanistic investigation in mature human sperm.

\section{Methods}
\label{methods}

\subsection{Model inputs and task formulation}
\label{sec:model_inputs}

We formulated kinase--substrate phosphorylation prediction as a site-level
binary classification task. Each instance comprised an enzyme sequence, a
substrate sequence and a candidate modification site on the substrate, and the
model predicted whether the enzyme modifies the corresponding residue. For
phosphorylation-focused training, candidate sites were restricted to serine,
threonine or tyrosine residues. The substrate was represented by a 15-residue
site-centred sequence window, whereas the enzyme was represented by its
full-length sequence or by a predefined truncated sequence where required.

For sample $i$, we denote the enzyme sequence by $x_i^{(e)}$, the substrate
site-centred sequence by $x_i^{(s)}$, the candidate site index by $c_i$, the
enzyme and substrate structure graphs by $g_i^{(e)}$ and $g_i^{(s)}$, the
sequence-derived physicochemical descriptors by $a_i$, and the binary label by
$y_i\in\{0,1\}$.

The privileged multimodal teacher receives

\begin{equation}
\mathcal{I}_i^{(t)}
=
\left(
x_i^{(e)},
x_i^{(s)},
c_i,
g_i^{(e)},
g_i^{(s)},
a_i
\right),
\label{eq:teacher_input}
\end{equation}

whereas the deployed student receives only structure-free inputs,

\begin{equation}
\mathcal{I}_i^{(s)}
=
\left(
x_i^{(e)},
x_i^{(s)},
c_i,
a_i
\right).
\label{eq:student_input}
\end{equation}

Thus, explicit protein structures constitute privileged information available
during teacher training but are not required during student inference. 
Dataset harmonization and benchmark partitioning are described in Supplementary Note~C.1. Detailed site-centred window construction and input preprocessing are provided in Supplementary Methods~D.1.

\subsection{Structural curation and multimodal encoding}
\label{sec:structural_curation}

For teacher-side structural modelling, experimentally determined structures
from RCSB PDB were prioritized where available, with AlphaFoldDB predictions
used as fallback structures. Records lacking the structural inputs required
after curation were excluded from structure-dependent teacher training.

Available enzyme and substrate structures were converted into residue-level
graphs and encoded using a geometric vector perceptron (GVP)-based structural
encoder. Structural representations were aligned to sequence coordinates
before multimodal fusion. For the substrate, aligned structural features were
restricted to the site-centred window; for the enzyme, sequence and structural
representations were pooled and broadcast to provide global kinase context.
Structure selection and curation are described in Supplementary Note~C.1, whereas structural graph construction and sequence--structure mapping are described in Supplementary Methods~D.2.

\subsection{Privileged teacher and structure-free student}
\label{sec:teacher_student}

The multimodal teacher integrates enzyme and substrate sequence
representations with structural and residue-level physicochemical information.
Sequence representations were generated using an ESM2-based encoder followed
by lightweight projection layers. Substrate sequence and structural features
were fused at residue level, with a learnable embedding marking the candidate
phosphosite. Enzyme sequence and structural representations were fused in
parallel and subsequently combined with the substrate-side representation and
physicochemical descriptors through hierarchical gated fusion. Pooling of the
resulting representation yielded the teacher feature $h_i^{t}$, which was
mapped to the teacher logit $z_i^{t}$.

The student retained the corresponding sequence branches, physicochemical
descriptors and gated-fusion architecture but removed explicit structural
encoders. It therefore required neither structural retrieval nor graph
construction during inference. The student representation $h_i^{s}$ was mapped
to a prediction logit $z_i^{s}$. Additional feature-processing details are provided in Supplementary Methods~D.3, with encoder and fusion configurations summarized in Supplementary Table~B4.

\subsection{Multi-level privileged knowledge distillation}
\label{subsec:kd_objective}

The structure-free student was trained using ground-truth supervision
and response- and representation-level knowledge distillation from
the multimodal teacher. At refinement round $r$, cached teacher
logits $\bar{z}_{i,r}^{\,t}$ and representations
$\bar{h}_{i,r}^{\,t}$ provided supervision for target-domain
training samples. The student objective was

\begin{equation}
\mathcal{L}_{S}
=
\mathcal{L}_{\mathrm{sup}}
+
\alpha\mathcal{L}_{\mathrm{resp}}
+
\beta\mathcal{L}_{\mathrm{repr}},
\label{eq:student_objective}
\end{equation}

where $\mathcal{L}_{\mathrm{sup}}$ is binary cross-entropy against
the observed labels, and $\mathcal{L}_{\mathrm{resp}}$ and
$\mathcal{L}_{\mathrm{repr}}$ denote response- and
representation-level distillation losses, weighted by the non-negative coefficients $\alpha$ and $\beta$, respectively.

For response-level distillation, student and cached teacher logits
were softened using temperature $\tau$,

\begin{equation}
p_i^{s}
=
\sigma\left(\frac{z_i^{s}}{\tau}\right),
\qquad
\bar{p}_{i,r}^{\,t}
=
\sigma\left(\frac{\bar{z}_{i,r}^{\,t}}{\tau}\right),
\label{eq:temperature_scaling}
\end{equation}

where $\sigma(\cdot)$ is the sigmoid function and $\tau>0$
controls prediction softness. Response transfer was optimized as

\begin{equation}
\mathcal{L}_{\mathrm{resp}}
=
\frac{1}{N}
\sum_{i=1}^{N}
a_{i,r}\,
\ell_{\mathrm{resp}}
\left(
p_i^{s},
\bar{p}_{i,r}^{\,t}
\right),
\label{eq:response_loss}
\end{equation}

where $N$ is the minibatch size,
$\ell_{\mathrm{resp}}$ is binary cross-entropy with the
softened teacher probability as the target, and $a_{i,r}$
is a label-aware sample weight that suppresses potentially
misleading teacher supervision.

Representation-level distillation combined cosine similarity
and normalized squared-error alignment,

\begin{equation}
\mathcal{L}_{\mathrm{repr}}
=
\frac{1}{N}
\sum_{i=1}^{N}
\left[
\lambda_{\mathrm{cos}}
\left(
1-
\cos
\left(
g(h_i^{s}),
\bar{h}_{i,r}^{\,t}
\right)
\right)
+
\lambda_{\mathrm{mse}}
\left\|
\widetilde{g(h_i^{s})}
-
\widetilde{\bar{h}_{i,r}^{\,t}}
\right\|_2^2
\right],
\label{eq:representation_loss}
\end{equation}

where $h_i^s$ and $\bar{h}_{i,r}^{\,t}$ are student and
cached teacher representations, $g(\cdot)$ projects student
features into the teacher feature space, and
$\widetilde{(\cdot)}$ denotes feature normalization.
The coefficients $\lambda_{\mathrm{cos}}$ and
$\lambda_{\mathrm{mse}}$ control the relative contributions
of directional and squared-error alignment. Detailed
distillation weighting and normalization procedures are
provided in Supplementary Methods~D.4.

\subsection{Student-guided closed-loop teacher refinement}
\label{sec:closed_loop}

Conventional distillation treats the teacher as a fixed source of supervision. SpermPhos instead maintains a frozen anchor teacher $T_0$, which provides a reference for broadly learned phosphorylation knowledge, and an adaptive teacher $T_r$, which is selectively refined using difficult examples from the target domain, where $r$ denotes the refinement round. This design enables domain-specific adaptation of teacher supervision while constraining deviations from transferable knowledge acquired during general-domain training.

Following an initial warm-up stage, each target-domain training sample was
assigned a hardness score combining student uncertainty, supervised prediction
error and teacher--student disagreement,

\begin{equation}
s_i^{(r)}
=
\lambda_u
\hat{u}_i^{(r)}
+
\lambda_l
\widehat{\ell}_{i}^{\mathrm{sup},(r)}
+
\lambda_d
\hat{d}_i^{(r)}.
\label{eq:hard_score}
\end{equation}

Here, $\hat{u}_i^{(r)}$ denotes normalized student predictive uncertainty,
$\widehat{\ell}_{i}^{\mathrm{sup},(r)}$ denotes normalized sample-level
binary cross-entropy against the ground-truth label, and
$\hat{d}_i^{(r)}$ denotes normalized disagreement between student
and cached teacher predictions. The coefficients $\lambda_u$,
$\lambda_l$ and $\lambda_d$ control the relative contributions of
these three components to sample hardness.

The highest-scoring target-domain examples formed the hard subset
$\mathcal{H}_r$. Validation and test examples were excluded from hard-example
selection, teacher adaptation and target regeneration.

The adaptive teacher was updated using the selected hard subset while being
constrained by general-domain replay and the frozen anchor teacher,

\begin{equation}
\mathcal{L}_{T}^{(r)}
=
\mathcal{L}_{\mathrm{hard}}^{(r)}
+
\lambda_{\mathrm{replay}}
\mathcal{L}_{\mathrm{replay}}^{(r)}
+
\lambda_{\mathrm{anchor}}
\mathcal{L}_{\mathrm{anchor}}^{(r)}.
\label{eq:teacher_total_loss}
\end{equation}

Here, $\mathcal{L}_{\mathrm{hard}}^{(r)}$ adapts the teacher to difficult
target-domain examples, $\mathcal{L}_{\mathrm{replay}}^{(r)}$ retains
performance on TeacherPTM examples, and
$\mathcal{L}_{\mathrm{anchor}}^{(r)}$ regularizes adaptive-teacher responses and
representations towards those of the frozen teacher. The coefficients
$\lambda_{\mathrm{replay}}$ and $\lambda_{\mathrm{anchor}}$ control
the strengths of general-domain replay and frozen-teacher regularization,
respectively, balancing target-domain adaptation against knowledge retention.
Only lightweight adaptation parameters together with higher-level fusion
and prediction modules were updated; the underlying sequence and structural
encoders remained frozen.

After teacher adaptation, updated logits and representations were regenerated
for the selected hard examples and used in the subsequent student-training
stage. The resulting closed-loop transfer can be summarized as

\begin{equation}
S_r
\longrightarrow
\mathcal{H}_r
\longrightarrow
T_{r+1}
\longrightarrow
\mathcal{C}_{r+1}
\longrightarrow
S_{r+1},
\label{eq:opd_loop}
\end{equation}

where $S_r$ and $S_{r+1}$ denote the student states before and after
refinement, $T_{r+1}$ is the updated adaptive teacher, and
$\mathcal{C}_{r+1}$ denotes the supervision cache containing
regenerated teacher targets for selected hard examples while retaining
previously cached targets for other samples. Ground-truth
label supervision remained active throughout all refinement rounds.

For the fixed-teacher offline-distillation baseline, teacher refinement and
iterative target regeneration were disabled; the student was instead trained
against a static cache generated from the frozen anchor teacher. At deployment,
all teacher-side components, structural inputs, replay data and cached targets
were discarded, and prediction was performed exclusively by the final
structure-free student. Hard-example scoring, adaptive-teacher regularization
and selective cache updating are further described in Supplementary
Methods~D.5, with refinement configurations
summarized in Supplementary Table~B4.

\subsection{Training and evaluation}
\label{sec:training_evaluation}

The multimodal teacher was trained on the general-domain TeacherPTM resource, and PhosTailor was subsequently adapted to distinct target domains using the same knowledge-transfer and student-guided refinement framework. For reproductive specialization, training was performed using MReproPhos, with evaluation on we proposed SIGNOR-Phos, ReproPhos-HPA and SpermK-PSP. To assess applicability to disease-associated phosphorylation, PhosTailor was separately adapted and evaluated on the Tumor dataset and its cancer-specific subset, Cancer. These experiments assessed the adaptability of the framework rather than zero-shot transfer from a reproductive-domain predictor.

For each benchmark, predictive performance was evaluated under random, kinase cold-start and substrate cold-start splits. In the cold-start settings, the held-out kinases or substrate proteins were excluded from the corresponding target-domain training partition but were not necessarily absent from general-domain teacher training. Training, validation and test partitions were kept fixed within each evaluation. Hyperparameters and model checkpoints were selected using validation data only, and held-out test data were reserved for final evaluation. Performance was assessed primarily using AUROC and AUPRC, together with F1 score and other threshold-dependent metrics where reported. Data partitioning is described in Supplementary Methods~C.1.4, and the recorded optimization and evaluation configurations are summarized in Supplementary Table~B4.

\subsection{Human sperm sample preparation and proteomic profiling}
\label{sec:Sperm_proteomics}

Human semen samples were obtained at the Affiliated Hospital of Nantong University after informed consent and approval by the Human Research Ethics Committee of the Affiliated Hospital of Nantong University. Samples were screened using study-defined thresholds of total motility $\geq40\%$ and progressive motility (PR) $\geq32\%$. These thresholds correspond to the lower reference limits in the fifth edition (2010) of the World Health Organization (WHO) laboratory manual~\cite{Cooper2010WHOReferenceValues,WHO2010SemenManual}; they were not used alone to diagnose normozoospermia. Spermatozoa were enriched by swim-up purification in saline buffered with 4-(2-hydroxyethyl)-1-piperazineethanesulfonic acid (HEPES) and collected by centrifugation at 300 $\times$ g for 6 min at room temperature. Purified spermatozoa were assigned to global proteomic and phosphoproteomic workflows. On each collection day, purified sperm were divided equally among human tubal fluid (HTF) medium~\cite{Quinn1985HTF} with matched dimethyl sulfoxide (DMSO) vehicle, HTF with 10~$\mu$M GSK8612 and HTF with 10~$\mu$M U0126. Inhibitors were prepared as 10~mM DMSO stocks and stored at $-80\,^\circ$C. Cells were incubated for 1~h at 37~$^\circ$C. After incubation, cells were collected by low-speed centrifugation, snap-frozen in liquid nitrogen and stored at $-80\,^\circ$C. After approximately 10 days of collection, pellets were pooled by treatment condition and divided into three technical LC--MS/MS replicates; thus, $n=3$ denotes technical replicates of pooled material, not independent donors. CASA samples came from separate, non-overlapping donor cohorts. Buffer composition, sample handling and washing procedures are provided in Supplementary Methods.

For global proteomics, proteins were extracted, quantified by bicinchoninic acid assay, quality-controlled by sodium dodecyl sulfate--polyacrylamide gel electrophoresis (SDS--PAGE), reduced, alkylated and digested with trypsin. For phosphoproteomics, proteins were subjected to reduction, alkylation, bead-based cleanup and overnight tryptic digestion, followed by C$_{18}$ desalting and phosphopeptide enrichment. Peptides were separated by reversed-phase ultrahigh-performance liquid chromatography and analyzed on an Orbitrap Astral mass spectrometer in data-independent acquisition (DIA) mode. Detailed extraction, digestion, enrichment, chromatographic and instrument parameters are provided in Supplementary Methods.

\subsection{DIA data processing and differential phosphosite analysis}
\label{sec:DIA_processing}

DIA files from the global proteomic and phosphoproteomic workflows were processed using DIA-NN v2.2 and v1.9, respectively~\cite{Demichev2020DIANN}. Spectral libraries were generated using DIA-NN deep-learning prediction and match-between-runs information. Searches specified trypsin digestion with up to two missed cleavages, and peptide and protein identifications were filtered at a 1\% false discovery rate. The resulting quantitative matrices were used to compare the HTF control and inhibitor-treated groups. For phosphoproteomic analyses, sites with a localization confidence of $\geq 0.9$ were retained. Differential phosphosite calls used the provider-reported nominal two-sided equal-variance $t$-test threshold ($P<0.05$) together with $\mathrm{FC}>1.5$ or $\mathrm{FC}<1/1.5$. These $P$ values were not adjusted for multiple testing, and the resulting site-level calls were treated as exploratory because each condition comprised pooled material measured in three technical replicates.
Additional software settings and quality-control criteria are provided in Supplementary Methods.

\subsection{Prospective kinase nomination and perturbation-concordance analyses}
\label{sec:methods_nomination}

The frozen PhosTailor student was applied to a predefined set of 60 sperm-relevant proteins before proteomic and phosphoproteomic perturbation outcomes were available. Model predictions over eligible phosphosite-centred contexts and a panel of 204 kinases were aggregated into kinase prioritization scores, and TBK1 and MAP2K2 were nominated for pharmacological perturbation following consideration of model support, sperm-proteome detectability and experimental tractability. The model and ranking procedure were kept fixed for all subsequent analyses. Differential phosphosites were filtered against matched global-proteome calls, yielding 410 GSK8612-responsive and 463 U0126-responsive sites after excluding sites assigned to proteins flagged as differentially abundant.
Within each response-defined set, PhosTailor scores defined the Top 5\%, Top 10\% and Top 20\% candidate tiers independently of response direction and magnitude; candidate-set concordance was then evaluated using direction-aware enrichment, permutation testing and protein-block resampling. Candidate-direction comparisons used one-sided Fisher exact tests. Complete selection, ranking and statistical definitions are provided in Supplementary Methods~E.6.

\subsection{Computer-assisted sperm analysis}
\label{sec:methods_casa}

Purified sperm from independent donors were divided within donor into matched DMSO-vehicle and drug aliquots for either GSK8612 (10~$\mu$M; $n=7$ donors) or U0126 (10~$\mu$M; $n=5$ donors); the two drug cohorts did not overlap with each other or with the omics cohort. CASA was performed at 1, 2 and 4 h using a Hamilton Thorne CEROS system and a 20-$\mu$m-deep counting chamber, with at least 200 sperm evaluated per assay. Total motility, PR, hyperactivation (HA), average-path velocity (VAP), straight-line velocity (VSL), curvilinear velocity (VCL) and amplitude of lateral head displacement (ALH) were recorded. For each drug, paired $t$ tests compared treated and matched vehicle aliquots at each time point. The 15-comparison family comprised HA, VCL, ALH, total motility and PR at 1 , 2 and 4 h; Holm correction was applied across this family. Paired mean differences and pointwise 95\% confidence intervals were calculated with donor as the pairing unit. VAP and VSL were recorded as CASA parameters but were not included in the displayed 15-comparison family. Instrument settings and HA thresholds are provided in Supplementary Methods~E.11.

\section*{Data availability}

The phosphorylation datasets used in this study were curated from publicly available resources, including PhosphoSitePlus, SAGEPhos, OmniPath, SIGNOR and the Human Protein Atlas, as described in the Supplementary Information. The curated TeacherPTM and MReproPhos datasets, benchmark partitions and associated annotations will be made publicly available upon acceptance of the manuscript.
Processed phosphosite and protein matrices, protein-filter exclusions, model scores and candidate memberships, enrichment summaries and running curves, CASA donor-level values and statistics, quality-control coordinates, and the curated external HM/LM comparison are provided in Supplementary Data 1--12. 
Raw proteomic and phosphoproteomic repository accessions will be added when the deposition is finalized.

\section*{Code availability}
The source code for PhosTailor, including model training, multimodal teacher--student distillation, student-guided closed-loop refinement and sequence-only inference, is available at \url{https://github.com/ZhangJJ26/PhosTailor}. An interactive, sequence-only PhosTailor web server for kinase--phosphosite prediction is freely accessible at \url{https://jingjie00-phostailor.hf.space/}.

\section*{Ethics statement}

Human semen samples were obtained from donors screened using the study's motility criteria at the Affiliated Hospital of Nantong University. All participants provided written informed consent before sample collection. The study protocol was approved by the Human Research Ethics Committee of the Affiliated Hospital of Nantong University.

\section*{Author contributions}

J.Z., J.X. and K.W. contributed equally to this work. C.G. conceived the original research idea. J.Z. designed the study and developed the PhosTailor framework. J.Z., Y.W. and J.W. contributed to model development, implementation and computational analyses. K.W. curated the phosphorylation datasets and performed computational experiments and statistical analyses. J.X. coordinated human sperm sample collection and conducted pharmacological perturbation, proteomic and phosphoproteomic experiments. M.H. contributed to biological analyses and interpretation. J.Z., J.X. and K.W. drafted the manuscript, with contributions to manuscript preparation and revision from H.C. and Z.G. N.M., C.H., X.Z., C.G. and P.H. supervised the study and secured funding. All authors reviewed and approved the final manuscript.

\section*{Competing interests}

The authors declare no competing interests.

\bibliography{reference}


\FloatBarrier
\clearpage

\begingroup
\centering
{\Large\bfseries Supplementary Information for\\[0.5em]
PhosTailor: structure-to-sequence distillation for kinase discovery in sparse phosphoproteomes\par}
\vspace{1.5em}
\endgroup

\setcounter{figure}{0}
\setcounter{table}{0}
\setcounter{equation}{0}
\renewcommand{\figurename}{Supplementary Fig.}
\renewcommand{\thefigure}{A\arabic{figure}}
\renewcommand{\tablename}{Supplementary Table}
\renewcommand{\thetable}{B\arabic{table}}
\renewcommand{\theequation}{S\arabic{equation}}
\makeatletter
\@ifundefined{theHfigure}{}{\renewcommand{\theHfigure}{A\arabic{figure}}}
\@ifundefined{theHtable}{}{\renewcommand{\theHtable}{B\arabic{table}}}
\@ifundefined{theHequation}{}{\renewcommand{\theHequation}{S\arabic{equation}}}
\makeatother

\begin{appendices}

\section{Supplementary Figures}\label{figs}

\begin{figure}[!htp]
    \centering
    \includegraphics[width=1\linewidth]{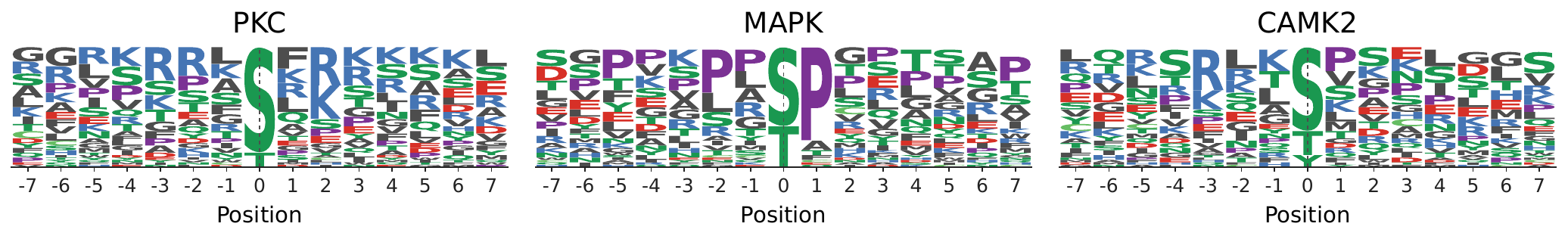}
    \caption{\textbf{Kinase-subfamily motifs recovered from high-confidence PhosTailor predictions.}
    Sequence logos were generated from the top 50 unique serine/threonine-centred predictions for kinase subfamilies with more than 50 eligible candidate sites. Protein kinase C (PKC) and calcium/calmodulin-dependent protein kinase II (CAMK2) predictions showed upstream basic-residue preferences, whereas mitogen-activated protein kinase (MAPK) predictions displayed a pronounced proline preference at position $+1$.}
    \label{fig:subfamily_motifs}
\end{figure}

\begin{figure}[!htp]
    \centering
    \includegraphics[width=1\linewidth]{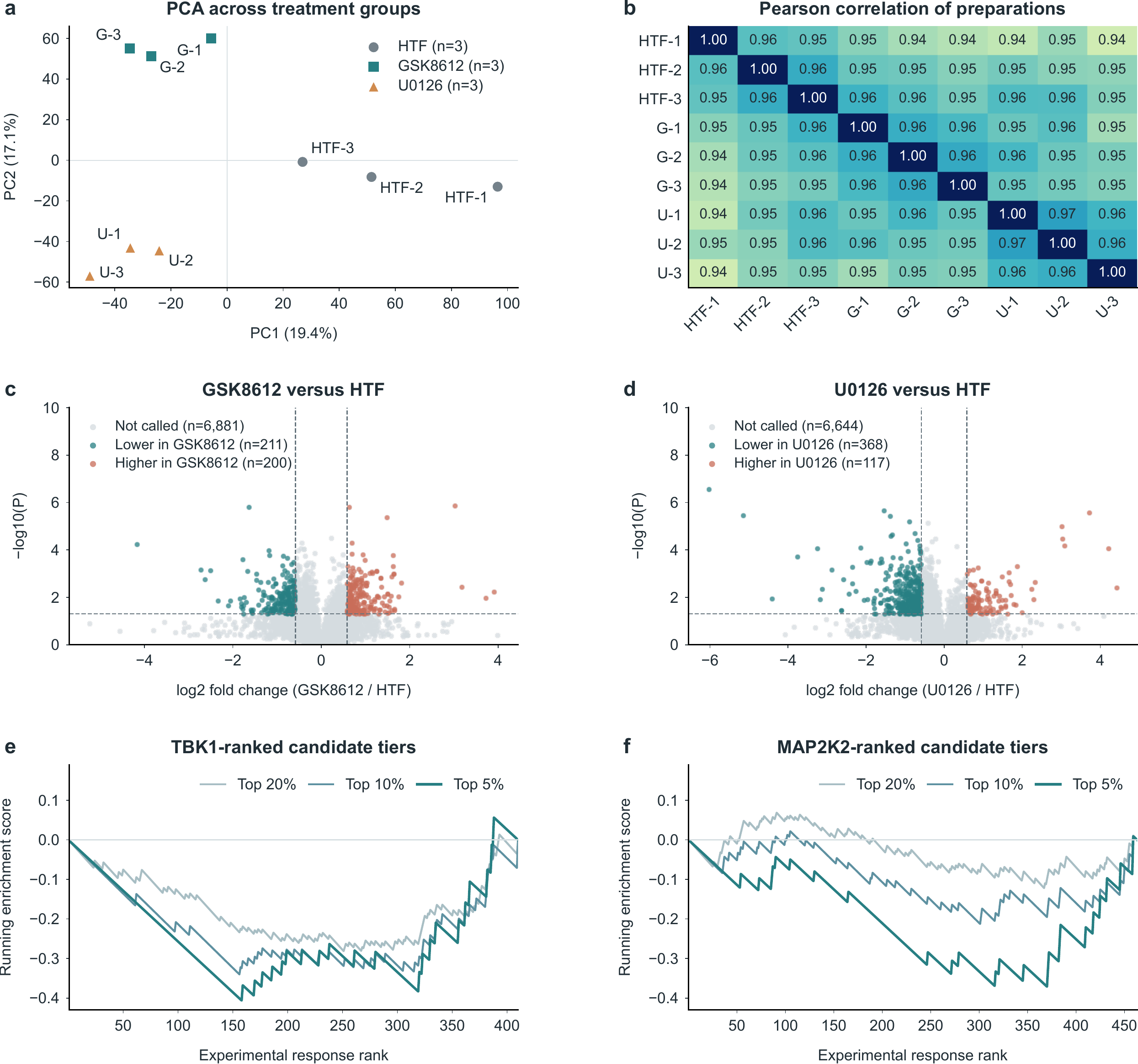}
\caption{\textbf{Phosphoproteomic responses and model-ranked candidate-site enrichment.}
\textbf{a,b}, Principal component analysis and Pearson correlations of nine phosphoproteomic profiles from vehicle-matched human tubal fluid (HTF), GSK8612 and U0126 conditions. Each condition comprised one pool split into three technical replicates, rather than independent biological replicates.
\textbf{c,d}, GSK8612-versus-HTF and U0126-versus-HTF volcano plots. Dashed lines indicate nominal $P=0.05$ and $\log_2$ fold change $=\pm\log_2(1.5)$; differential calls used both thresholds without multiple-testing correction. Counts shown precede protein-abundance filtering.
\textbf{e,f}, Weighted running-enrichment curves for nested Top 5\%, Top 10\% and Top 20\% PhosTailor-ranked TBK1 and MAP2K2 (MEK2) candidate tiers, respectively. Excluding sites assigned to differentially abundant proteins left 410 GSK8612-responsive and 463 U0126-responsive sites. Negative excursions indicate enrichment toward decreased phosphorylation. U0126 responses provide MEK1/2 pathway context for the model-ranked MAP2K2 candidates. Source values and enrichment statistics are provided in Supplementary Data 1, 6, 7, 9 and 10. Experimental procedures and enrichment definitions are provided in Appendix~\ref{sec:supp_experimental_methods}.}

    \label{fig:phosphoproteomic_quality}
\end{figure}

\begin{figure}[!htp]
    \centering
    \includegraphics[width=1\linewidth]{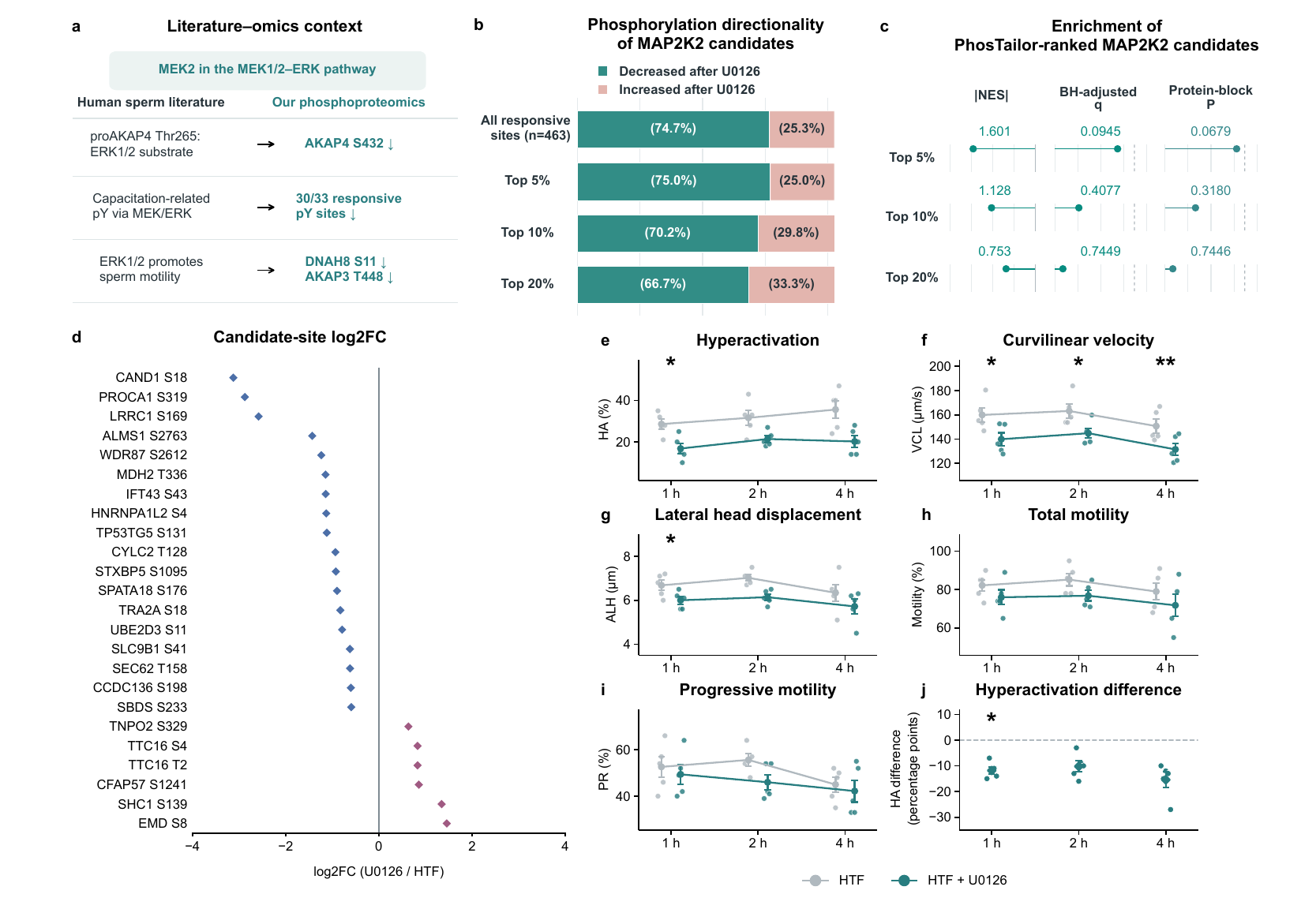}
\caption{\textbf{MAP2K2 (MEK2) candidate responses and sperm-motility changes under U0126 perturbation.}
\textbf{a}, Published sperm mitogen-activated protein kinase kinase (MEK)--extracellular signal-regulated kinase (ERK) pathway findings alongside phosphosite changes observed after U0126 treatment; arrows indicate thematic correspondence~\cite{almog2008identification}. The reported ERK1/2 substrate site is proAKAP4 Thr265~\cite{RahamimBenNavi2016AKAP4}; the AKAP4 site observed here is S432.
\textbf{b}, Response directions among 463 filtered U0126-responsive sites and nested MAP2K2 candidate tiers (Top 5\%, $n=24$; Top 10\%, $n=47$; Top 20\%, $n=93$).
\textbf{c}, Absolute normalized enrichment scores ($|\mathrm{NES}|$), Benjamini--Hochberg-adjusted $q$ values and protein-block $P$ values. Signed NES values were negative; no tier met $q<0.05$ or protein-block $P<0.05$.
\textbf{d}, Top 5\% candidate-site $\log_2$(U0126/HTF) fold changes. Each diamond shows the $\log_2$ ratio of condition-mean intensities; the vertical line marks no change.
\textbf{e--i}, Hyperactivation (HA), curvilinear velocity (VCL), amplitude of lateral head displacement (ALH), total motility and progressive motility (PR) at 1, 2 and 4~h. Points represent five donors with paired vehicle and U0126 aliquots; summaries show mean $\pm$ standard error of the mean (SEM).
\textbf{j}, Within-donor HA differences in percentage points (U0126 minus HTF), shown as mean $\pm$ SEM. Two-sided paired t tests used Holm correction across all 15 endpoint-by-time comparisons for U0126; panel j repeats the HA comparisons. *$P<0.05$, **$P<0.01$, ***$P<0.001$. These paired-donor measurements characterize sperm-motility responses to MEK1/2 pathway inhibition. Source values and exact statistics are provided in Supplementary Data 1, 4--6 and 8. Methods are provided in Section~\ref{sec:supp_methods_casa}.}

    \label{fig:map2k2_u0126_casa}
\end{figure}

\FloatBarrier
\clearpage
\section{Supplementary Tables}\label{tabs}

\begingroup
\definecolor{supprowgray}{RGB}{245,247,249}
\footnotesize
\setlength{\tabcolsep}{5.5pt}
\renewcommand{\arraystretch}{1.10}
\setlength{\LTleft}{0pt}
\setlength{\LTright}{\fill}
\setlength{\LTcapwidth}{\textwidth}
\setlength{\emergencystretch}{2em}

\begin{longtable}{@{}
  >{\raggedright\arraybackslash}p{0.22\linewidth}
  >{\raggedright\arraybackslash}p{0.73\linewidth}@{}}
\caption{\textbf{Major notations.}}\label{tab:major_notations}\\
\toprule
\textbf{Notation} & \textbf{Definition} \\
\midrule
\endfirsthead
\multicolumn{2}{@{}l}{\textit{Supplementary Table~\ref{tab:major_notations} (continued)}} \\
\toprule
\textbf{Notation} & \textbf{Definition} \\
\midrule
\endhead
\midrule
\multicolumn{2}{r@{}}{\textit{Continued on next page}} \\
\endfoot
\bottomrule
\endlastfoot
\rowcolor{supprowgray} PhosTailor & General-to-specialized phosphorylation prediction framework. \\
OPD & Online policy distillation; adapted closed-loop training. \\
\rowcolor{supprowgray} MReproPhos & Integrated male reproductive kinase–substrate–phosphosite dataset. \\
ReproPhos-HPA & Phosphorylation dataset filtered using protein-expression evidence from male reproductive tissues. \\
\rowcolor{supprowgray} SpermK-PSP & Phosphorylation dataset involving literature-supported sperm-relevant kinases. \\
SIGNOR-Phos & SIGNOR-derived causal phosphorylation relationships restricted to proteins detected in mature human sperm. \\
\rowcolor{supprowgray} TeacherPTM & General-domain enzyme–substrate PTM dataset used for teacher training. \\
CASA & Computer-assisted sperm analysis. \\
\rowcolor{supprowgray} HA & Hyperactivation. \\
VCL & Curvilinear velocity. \\
\rowcolor{supprowgray} ALH & Amplitude of lateral head displacement. \\
VAP & Average path velocity. \\
\rowcolor{supprowgray} VSL & Straight-line velocity. \\
KPS & Kinase prioritization score. \\
\rowcolor{supprowgray} Tumor & Tumour-associated phosphorylation benchmark. \\
Cancer & Cancer-specific subset of the Tumor benchmark. \\
\rowcolor{supprowgray} LIN & Linearity. \\
PR & Progressive motility. \\
\rowcolor{supprowgray} TBK1 & TANK-binding kinase 1. \\
MAP2K1 (MEK1) & Mitogen-activated protein kinase kinase 1. \\
\rowcolor{supprowgray} MAP2K2 (MEK2) & Mitogen-activated protein kinase kinase 2. \\
MEK1/2 & Mitogen-activated protein kinase kinases 1 and 2. \\
\rowcolor{supprowgray} ERK & Extracellular signal-regulated kinase. \\
PSP & PhosphoSitePlus. \\
\rowcolor{supprowgray} AGC & Protein kinase A, G and C family group. \\
NES & Normalized enrichment score. \\
\rowcolor{supprowgray} MSE & Mean squared error. \\
CI & Confidence interval. \\
\rowcolor{supprowgray} s.d. & Standard deviation. \\
$q$ value & Multiple-testing-adjusted P value. \\
\rowcolor{supprowgray} NCBI & National Center for Biotechnology Information. \\
MeSH & Medical Subject Headings. \\
\rowcolor{supprowgray} PCA & Principal component analysis. \\
nDCG@5 & Normalized discounted cumulative gain at rank five. \\
\rowcolor{supprowgray} PTM & Post-translational modification. \\
DIA & Data-independent acquisition. \\
\rowcolor{supprowgray} DIA-NN & Software for DIA mass-spectrometry data analysis. \\
LC--MS/MS & Liquid chromatography--tandem mass spectrometry. \\
\rowcolor{supprowgray} UHPLC & Ultra-high-performance liquid chromatography. \\
SDS--PAGE & Sodium dodecyl sulfate--polyacrylamide gel electrophoresis. \\
\rowcolor{supprowgray} FDR & False discovery rate. \\
BH procedure & Benjamini--Hochberg multiple-testing correction. \\
\rowcolor{supprowgray} Offline KD & Fixed-teacher offline knowledge distillation. \\
HTF & Human tubal fluid medium. \\
\rowcolor{supprowgray} ES & Enrichment score. \\
FC & Fold change. \\
\rowcolor{supprowgray} GVP & Geometric vector perceptron. \\
PDB & Protein Data Bank. \\
\rowcolor{supprowgray} AlphaFoldDB & AlphaFold structure-prediction database. \\
HPA & Human Protein Atlas. \\
\rowcolor{supprowgray} SIGNOR & SIGnaling Network Open Resource. \\
OmniPath & Signalling pathway and molecular-interaction resource. \\
\rowcolor{supprowgray} HM/LM & High-motility/low-motility sperm fractions. \\
PKA & Protein kinase A. \\
\rowcolor{supprowgray} PKC & Protein kinase C. \\
CAMK2 & Calcium/calmodulin-dependent protein kinase II. \\
\rowcolor{supprowgray} MAPK & Mitogen-activated protein kinase. \\
CMGC & CDK, MAPK, GSK3 and CLK kinase group. \\
\rowcolor{supprowgray} CK1/CK2 & Casein kinase 1 and casein kinase 2. \\
cAMP & Cyclic adenosine monophosphate. \\
\rowcolor{supprowgray} BCA & Bicinchoninic acid assay. \\
WHO & World Health Organization. \\
\rowcolor{supprowgray} GSK8612 & Selective TBK1 inhibitor used for pharmacological perturbation. \\
AUROC & Area under the receiver operating characteristic curve. \\
\rowcolor{supprowgray} AUPRC & Area under the precision--recall curve. \\
Accuracy (ACC) & Proportion of correctly classified observations. \\
\rowcolor{supprowgray} F1 score & Harmonic mean of precision and recall. \\

\end{longtable}

\begin{longtable}{@{}
  >{\centering\arraybackslash}p{0.09\linewidth}
  >{\raggedright\arraybackslash}p{0.34\linewidth}
  >{\raggedright\arraybackslash}p{0.49\linewidth}@{}}
\caption{\textbf{Mapping of test-set kinases to the Kinase Library specificity atlas.}
The table lists the 31 kinases represented in the held-out random test split that could
be unambiguously mapped to Ser/Thr specificity matrices in Kinase Library v0.1.0.
The candidate panel was defined independently of model predictions and biochemical
concordance results.}\label{tab:kinase_library_panel}\\
\toprule
\textbf{Index} & \textbf{Kinase Library label} & \textbf{Standardized kinase identifier} \\
\midrule
\endfirsthead
\multicolumn{3}{@{}l}{\textit{Supplementary Table~\ref{tab:kinase_library_panel} (continued)}} \\
\toprule
\textbf{Index} & \textbf{Kinase Library label} & \textbf{Standardized kinase identifier} \\
\midrule
\endhead
\midrule
\multicolumn{3}{r@{}}{\textit{Continued on next page}} \\
\endfoot
\bottomrule
\endlastfoot
\rowcolor{supprowgray} 1  & AKT1   & AKT1 \\
2  & AKT2   & AKT2 \\
\rowcolor{supprowgray} 3  & AMPKA1 & AMPKA1 \\
4  & AMPKA2 & AMPKA2 \\
\rowcolor{supprowgray} 5  & CAMK1G & CAMK1G \\
6  & CAMK2A & CAMK2A \\
\rowcolor{supprowgray} 7  & CAMK2B & CAMK2B \\
8  & CAMK2D & CAMK2D \\
\rowcolor{supprowgray} 9  & CAMK2G & CAMK2G \\
10 & CAMK4  & CAMK4 \\
\rowcolor{supprowgray} 11 & CAMKK1 & CAMKK1 \\
12 & CAMKK2 & CAMKK2 \\
\rowcolor{supprowgray} 13 & CK1A   & CSNK1A1 \\
14 & CK2A1  & CSNK2A1 \\
\rowcolor{supprowgray} 15 & ERK2   & MAPK1 \\
16 & P38A   & MAPK14 \\
\rowcolor{supprowgray} 17 & ERK1   & MAPK3 \\
18 & PKACA  & PRKACA \\
\rowcolor{supprowgray} 19 & PKACB  & PRKACB \\
20 & PKACG  & PRKACG \\
\rowcolor{supprowgray} 21 & PKCA   & PRKCA \\
22 & PKCB   & PRKCB \\
\rowcolor{supprowgray} 23 & PKCD   & PRKCD \\
24 & PKCE   & PRKCE \\
\rowcolor{supprowgray} 25 & PKCG   & PRKCG \\
26 & PKCH   & PRKCH \\
\rowcolor{supprowgray} 27 & PKCI   & PRKCI \\
28 & PKCT   & PRKCQ \\
\rowcolor{supprowgray} 29 & PKCZ   & PRKCZ \\
30 & ROCK1  & ROCK1 \\
\rowcolor{supprowgray} 31 & ROCK2  & ROCK2 \\

\end{longtable}

\begin{longtable}{@{}
  >{\centering\arraybackslash}p{0.085\linewidth}
  >{\raggedright\arraybackslash}p{0.21\linewidth}
  >{\raggedright\arraybackslash}p{0.165\linewidth}
  >{\raggedright\arraybackslash}p{0.46\linewidth}@{}}
\caption{\textbf{Reproductive protein panel and top 10 candidate kinases.}}\label{tab:reproductive_proteins_kinases}\\
\toprule
\multicolumn{4}{@{}l}{\textbf{a, Reproductive-associated substrate proteins ($n=60$)}} \\
\midrule
\textbf{No.} & \textbf{Gene} & \textbf{UniProt} & \textbf{Reproductive or functional context} \\
\midrule
\endfirsthead
\multicolumn{4}{@{}l}{\textit{Supplementary Table~\ref{tab:reproductive_proteins_kinases} (continued)}} \\
\toprule
\textbf{No.} & \textbf{Gene / kinase} & \textbf{UniProt} & \textbf{Reproductive or functional context} \\
\midrule
\endhead
\midrule
\multicolumn{4}{r@{}}{\textit{Continued on next page}} \\
\endfoot
\bottomrule
\endlastfoot
\rowcolor{supprowgray} 1 & NECTIN3 & Q9NQS3 & Cell adhesion; male reproductive literature \\
2 & MROH7 & Q68CQ1 & Testis-enriched candidate; knockout studies \\
\rowcolor{supprowgray} 3 & CATSPERD & Q86XM0 & CatSper Ca$^{2+}$ channel accessory subunit \\
4 & NUP210L & Q5VU65 & Sperm nuclear-envelope organization and head morphology \\
\rowcolor{supprowgray} 5 & PRSS21 & Q9Y6M0 & Sperm maturation and fertilization-associated serine protease \\
6 & ADAM21 & Q9UKJ8 & Testis-associated ADAM-family protein \\
\rowcolor{supprowgray} 7 & MYCBPAP & Q8TBZ2 & Sperm tail and head morphogenesis \\
8 & ARRDC5 & A6NEK1 & Spermatogenesis and sperm morphogenesis \\
\rowcolor{supprowgray} 9 & INSL6 & Q9Y581 & Testis-associated factor; spermatogenic function \\
10 & SUN3 & Q8TAQ9 & Spermatid nuclear envelope and sperm head shaping \\
\rowcolor{supprowgray} 11 & CCDC136 & Q96JN2 & Acrosome formation and fertilization \\
12 & DPEP3 & Q9H4B8 & Testis-associated peptidase; limited fertility phenotype \\
\rowcolor{supprowgray} 13 & IRGC & Q6NXR0 & Germ-cell-associated GTPase \\
14 & SUN5 & Q8TC36 & Sperm head--tail coupling and male infertility \\
\rowcolor{supprowgray} 15 & ADAM29 & Q9UKF5 & Testis-associated ADAM-family candidate \\
16 & CATSPER2 & Q96P56 & CatSper Ca$^{2+}$ channel pore-forming subunit \\
\rowcolor{supprowgray} 17 & SLC9B1 & Q4ZJI4 & Solute carrier; sperm ion-transport candidate \\
18 & TCP11 & Q8WWU5 & Testis-associated protein implicated in sperm function \\
\rowcolor{supprowgray} 19 & MMEL1 & Q495T6 & Neprilysin-family protein; reproductive association \\
20 & TMEM232 & C9JQI7 & Sperm morphology and male fertility \\
\rowcolor{supprowgray} 21 & CATSPERG & Q6ZRH7 & CatSper Ca$^{2+}$ channel accessory subunit \\
22 & ADAM20 & O43506 & Sperm--egg fusion-associated candidate \\
\rowcolor{supprowgray} 23 & ATP1A4 & Q13733 & Sperm-associated Na$^+$/K$^+$-ATPase \\
24 & ATP8B3 & O60423 & Testis-associated phospholipid transporter \\
\rowcolor{supprowgray} 25 & CCDC188 & H7C350 & Sperm head--neck connection \\
26 & ARMC12 & Q5T9G4 & Sperm mitochondrial-sheath organization \\
\rowcolor{supprowgray} 27 & KPNA2 & P52292 & Nuclear import; germ-cell-associated factor \\
28 & TMPRSS12 & Q86WS5 & Sperm motility and reproductive tract migration \\
\rowcolor{supprowgray} 29 & GLIPR1L1 & Q6UWM5 & Sperm acrosomal function and gamete interaction \\
30 & EQTN & Q9NQ60 & Sperm--egg adhesion \\
\rowcolor{supprowgray} 31 & TEX101 & Q9BY14 & Testis and sperm maturation-associated protein \\
32 & DDX20 & Q9UHI6 & Prospermatogonial cell-cycle regulation \\
\rowcolor{supprowgray} 33 & IZUMO1 & Q8IYV9 & Sperm--egg membrane fusion \\
34 & SSMEM1 & Q8WWF3 & Sperm head and acrosome formation \\
\rowcolor{supprowgray} 35 & SPEM1 & Q8N4L4 & Spermatid cytoplasm removal and sperm morphogenesis \\
36 & SPATA3 & Q8NHX4 & Spermatogenesis-associated candidate \\
\rowcolor{supprowgray} 37 & TEX38 & Q6PEX7 & Sperm morphogenesis and migration \\
38 & TMEM249 (CATSPERQ) & Q2WGJ8 & Sperm hyperactivation-associated membrane protein \\
\rowcolor{supprowgray} 39 & BAG5 & Q9UL15 & Chaperone-associated regulation of spermatogenesis \\
40 & CFAP61 & Q8NHU2 & Flagellar assembly and sperm motility \\
\rowcolor{supprowgray} 41 & CFAP65 & Q6ZU64 & Flagellar structure and male infertility \\
42 & CEP41 & Q9BYV8 & Centrosome and cilia-associated protein \\
\rowcolor{supprowgray} 43 & ADAM2 & Q99965 & Sperm adhesion and gamete interaction \\
44 & RPL10L & Q96L21 & Testis-associated ribosomal protein \\
\rowcolor{supprowgray} 45 & ZPBP2 & Q6X784 & Acrosomal protein associated with gamete interaction \\
46 & CATSPER3 & Q86XQ3 & CatSper Ca$^{2+}$ channel pore-forming subunit \\
\rowcolor{supprowgray} 47 & CATSPER4 & Q7RTX7 & CatSper Ca$^{2+}$ channel pore-forming subunit \\
48 & LRRC52 & Q8N7C0 & Regulatory subunit of sperm K$^+$ channels \\
\rowcolor{supprowgray} 49 & IFT122 & Q9HBG6 & Intraflagellar transport and ciliary assembly \\
50 & CATSPER1 & Q8NEC5 & CatSper Ca$^{2+}$ channel pore-forming subunit \\
\rowcolor{supprowgray} 51 & DPY19L2 & Q6NUT2 & Sperm head shaping and globozoospermia \\
52 & KCNU1 & A8MYU2 & Sperm K$^+$ channel and membrane potential \\
\rowcolor{supprowgray} 53 & TRIM36 & Q9NQ86 & Ubiquitin-ligase-related spermatogenic regulation \\
54 & LRRC37A & A6NMS7 & Acrosomal integrity and sperm head morphology \\
\rowcolor{supprowgray} 55 & SPEM3 & A0A1B0GUW6 & Spermatid-associated protein; limited functional evidence \\
56 & ADAM30 & Q9UKF2 & Testis-associated ADAM-family candidate \\
\rowcolor{supprowgray} 57 & SLC22A16 & Q86VW1 & Carnitine/organic-cation transporter family \\
58 & AKAP3 & O75969 & Fibrous-sheath assembly and sperm motility \\
\rowcolor{supprowgray} 59 & AKAP4 & Q5JQC9 & Fibrous-sheath integrity and flagellar motility \\
60 & CABYR & O75952 & Fibrous-sheath organization and sperm motility \\
\addlinespace[9pt]
\multicolumn{4}{@{}l}{\textbf{b, Top 10 kinase candidates ($n=10$)}} \\
\midrule
Priority & Kinase & UniProt & Biological context \\
\midrule
\rowcolor{supprowgray} 1 & MAP2K2 (MEK2) & P36507 & MEK2--ERK signalling associated with sperm function \\
2 & LIMK1 & P53667 & Actin--cofilin regulation in capacitation and acrosomal signalling \\
\rowcolor{supprowgray} 3 & DCAF1 & Q9Y4B6 & Spermatid-enriched protein with atypical Ser/Thr kinase activity; direct sperm function uncharacterized \\
4 & EEF2K & O00418 & Translation elongation control; reported involvement in spermatogenic regulation \\
\rowcolor{supprowgray} 5 & TBK1 & Q9UHD2 & Underexplored in mature human sperm; selected for experimental investigation \\
6 & PASK & Q96RG2 & Metabolic protein kinase with testis-associated expression \\
\rowcolor{supprowgray} 7 & CAMKK2 & Q96RR4 & Ca$^{2+}$/calmodulin-associated signalling; limited direct human sperm evidence \\
8 & TRPM7 & Q96QT4 & Divalent-cation channel with an intrinsic protein kinase domain \\
\rowcolor{supprowgray} 9 & NEK9 & Q8TD19 & Microtubule and cell-division regulation; direct sperm-specific evidence limited \\
10 & NEK7 & Q8TDX7 & Microtubule and cell-cycle regulation; direct sperm-specific evidence limited \\

\end{longtable}
\endgroup

\begingroup
\footnotesize
\setlength{\tabcolsep}{4pt}
\renewcommand{\arraystretch}{1.13}
\setlength{\LTleft}{0pt}
\setlength{\LTright}{0pt}
\begin{longtable}{@{}p{0.23\linewidth}p{0.25\linewidth}p{0.485\linewidth}@{}}
\caption{\textbf{Model architecture, optimization, adaptive distillation and evaluation configurations of PhosTailor.} }\label{tab:supp_phostailor_config}\\
\toprule
\textbf{Component} & \textbf{Parameter} & \textbf{Configuration / reported value}\\
\midrule
\endfirsthead
\multicolumn{3}{@{}l}{\textit{Supplementary Table~\ref{tab:supp_phostailor_config} (continued)}}\\
\toprule
\textbf{Component} & \textbf{Parameter} & \textbf{Configuration / reported value}\\
\midrule
\endhead
\midrule
\multicolumn{3}{r@{}}{\textit{Continued on next page}}\\
\endfoot
\bottomrule
\endlastfoot

\multicolumn{3}{@{}l}{\textbf{A. Model architecture and input processing}}\\
\addlinespace[2pt]
Teacher kinase encoder & Backbone & ESM2-650M (\texttt{esm2\_t33\_650M\_UR50D}); 33 layers, 1,280-dimensional hidden states, 20 attention heads.\\
Teacher substrate encoder & Backbone & COMPASS-PTM based on ESM2-650M; combines 1,280-dimensional ESM2 features, 320-dimensional chemical representations and four physicochemical features.\\
Student encoder & Backbone & Shared ESM2-8M (\texttt{esm2\_t6\_8M\_UR50D}); six layers, 320-dimensional hidden states, 20 attention heads.\\
Sequence encoder adaptation & LoRA & Teacher kinase and student sequence encoders: rank 32, scaling factor 64, dropout 0.1; query/key/value/dense targets, frozen base backbone. COMPASS-PTM internal LoRA: rank 16, scaling factor 16.\\
Structural encoder & GVP & Shared two-layer GVP; 512 scalar and eight vector channels; 21-dimensional node and 14-dimensional edge input features; 512-dimensional structural output.\\
Structural graph & Residues and edges & C$\alpha$ residue nodes; sequence-neighbour edges, spatial edges within 10~\AA{}, and up to ten nearest-neighbour edges per residue.\\
Sequence inputs & Length policy & Kinase sequence cap: 1,020 residues; student substrate input: 15-residue site-centred window. Teacher kinase and substrate sequence caps: 1,020 residues.\\
Structure limits & Stage-dependent maximum & Original teacher pretraining: 1,020 structural nodes; inspected OPD/downstream configuration: 512 nodes.\\
Residue-to-sequence mapping & Stage-dependent method & Original teacher pretraining: Needleman--Wunsch hard alignment (match $+2$; mismatch/gap $-1$), with interpolation fallback. Inspected OPD/cache configurations: linear interpolation (\texttt{align\_corners=False}).\\
Multimodal fusion & Representations & Hierarchical gated fusion; teacher feature size 1,280 and student feature size 320.\\
\addlinespace[4pt]

\multicolumn{3}{@{}l}{\textbf{B. Model optimization and checkpoint selection}}\\
\addlinespace[2pt]
Teacher pretraining & Optimizer; learning rate & Adam; $1\times10^{-4}$.\\
Teacher pretraining & Seed; batch size & Seed 26; batch size 8; 32 gradient-accumulation steps (effective batch size 256).\\
Teacher pretraining & Training duration & Maximum configured epochs: 500.\\
Student, random split & Optimization & Adam; learning rate $1\times10^{-4}$; batch size 64; seed 42.\\
Student, kinase cold-start & Optimization & Adam; learning rate $2\times10^{-4}$; batch size 64; seed 26.\\
Student, substrate cold-start & Optimization & Adam; learning rate $2\times10^{-4}$; batch size 64; seed 26.\\
Student training & Epoch and stopping rules & Maximum configured epochs: 1,000; validation-AUROC-based early stopping with patience 12 and minimum improvement 0.001.\\
Checkpoint selection & Metric & Teacher: save when validation AUROC strictly increases; student: save when validation AUROC improves by more than 0.001.\\
\addlinespace[4pt]

\multicolumn{3}{@{}l}{\textbf{C. Knowledge distillation and student-guided refinement}}\\
\addlinespace[2pt]
Student objective & Relative weights & Supervised loss: 1; response-level distillation: 0.35; representation-level distillation: 0.8.\\
Response distillation & Temperature & $T=2.8$; logit-distillation term includes $T^2$ scaling.\\
Feature distillation & Alignment objective & $0.5\times$ cosine loss $+\;0.5\times$ MSE; LayerNorm applied before the MSE term.\\
Label-aware distillation & Reverse suppression & Enabled; suppression weight 0.8, shaping exponent $\gamma=1$; reverse suppression disabled during the first three epochs.\\
Additional KD weighting & Uncertainty-weighted KD & Disabled in the audited student configuration.\\
Refinement schedule & Warm-up; interval & Five warm-up epochs, followed by refinement every three epochs (first refinement after epoch 5).\\
Hard-example mining & Subset size & Top 10\% of target-domain training samples, subject to a minimum of 256 unique sample identifiers.\\
Hardness score & Component weights & Normalized uncertainty: 0.3; label BCE: 0.3; teacher--student disagreement: 0.4.\\
Hard-example resampling & Sampling weights & Relative weights of hard versus other samples: 3 versus 1; resampling with replacement.\\
Adaptive-teacher update & Optimizer and steps & Adam; learning rate $5\times10^{-5}$; eight optimizer steps per refinement round.\\
Frozen-teacher anchoring & Regularization & Frozen anchor $T_0$; temperature 2.0 and anchor-loss coefficient 1.0; response and representation constraints.\\
General-domain replay & Loss coefficient & Replay-loss coefficient 0.5; replay examples resampled from the general-domain training pool during refinement.\\
\addlinespace[4pt]

\multicolumn{3}{@{}l}{\textbf{D. Evaluation protocol}}\\
\addlinespace[2pt]
Classification metrics & Decision threshold & Probability threshold 0.5 for threshold-dependent metrics; AUROC and AUPRC calculated from continuous model scores.\\
Evaluation implementation & Batch size and precision & Student evaluation: batch size 64, FP32, strict loading; saved teacher random-split evaluation: batch size 8.\\
\addlinespace[4pt]

\multicolumn{3}{@{}l}{\textbf{E. Hardware, software}}\\
\addlinespace[2pt]
Software & Python / PyTorch & Python 3.12.3; PyTorch 2.5.1 for the efficiency benchmark and in archived training requirements.\\
Archived dependencies & Principal packages & Transformers 4.48.1; PEFT 0.17.1; Gemmi 0.7.5; NumPy 2.3.3; pandas 2.3.3; scikit-learn 1.7.2; tokenizers 0.21.4.\\
Inference benchmark hardware & Runtime & PyTorch CUDA runtime 12.1; FP32 inference, \texttt{eval()} mode, no autocast.\\

\end{longtable}
\endgroup

\clearpage
\section{Supplementary Notes}\label{Notes}
\subsection{Datasets}
\label{sec:supp_datasets}

\subsubsection{Preliminaries: source resources}
We curated enzyme--substrate datasets from multiple public and literature-derived resources. These resources provided complementary information for constructing the teacher dataset and the male-reproduction-focused student datasets.

OmniPath~\cite{turei2016omnipath} was used as a curated source of enzyme--post-translational modification (PTM) relationships. OmniPath integrates curated knowledge from more than one hundred primary resources into interconnected modules covering protein--protein and gene-regulatory interactions, enzyme--PTM relationships, protein complexes, functional protein annotations and intercellular communication. In this study, OmniPath contributed curated enzyme--substrate PTM records to the unified dataset.

PhosphoSitePlus (PSP)~\cite{hornbeck2015phosphositeplus} is a curated mammalian PTM knowledgebase containing experimentally supported phosphorylation, acetylation, ubiquitylation, methylation and other modification sites. Its annotations are largely derived from mass-spectrometry evidence and include regulatory PTM sites linked to cellular processes, molecular functions and protein--protein interactions, as well as PTMVar records connecting PTMs with missense variants. PSP was used as a major source of curated enzyme--substrate PTM relationships and modification-site annotations.

The SAGEPhos dataset~\cite{zhang2025sagephos} integrates phosphorylation sites from major repositories, including Phospho.ELM, PhosphoNetworks and PhosphoSitePlus, together with structural information derived from AlphaFoldDB. After quality control and redundancy removal, the dataset contains 18,360 high-confidence positive phosphorylation sites. We included SAGEPhos as an additional source of curated phosphorylation-site information for constructing the teacher dataset.

UniProtKB~\cite{boutet2016uniprotkb} is a widely used protein knowledgebase that provides standardized protein sequences and functional annotations through manually curated Swiss-Prot records and computationally annotated TrEMBL records. In this study, UniProtKB was used to retrieve full-length protein sequences for enzymes and substrates and to harmonize protein identifiers across resources. RCSB Protein Data Bank (RCSB PDB)~\cite{burley2019rcsb} and AlphaFold Protein Structure Database (AlphaFoldDB)~\cite{varadi2022alphafold} were used as structural resources for enzymes and substrates. RCSB PDB provides experimentally determined macromolecular structures, whereas AlphaFoldDB provides predicted protein structures generated by AlphaFold with residue-level confidence estimates.

The Human Protein Atlas (HPA), compiled by Uhlén et al.~\cite{ponten2008human,uhlen2015tissue,thul2017subcellular,nagaoka2020zglp1,karlsson2021single,uhlen2017pathology,uhlen2019genome,uhlen2019human}, was used to define the male reproductive protein space. HPA is a comprehensive multi-omics resource that combines antibody-based imaging, mass-spectrometry-driven proteomics, transcriptomics and systems-biology analyses to characterize the spatial distribution of human proteins across cells, tissues and organs. We used HPA tissue-expression evidence from testis, epididymis, prostate and seminal vesicle to define a broad male reproductive protein background for student-dataset construction.

We also incorporated a literature-derived ejaculated-sperm proteome compiled by Castillo et al.~\cite{castillo2018contribution}. This resource integrates sperm proteomic evidence from multiple independent studies and was used to identify proteins and kinases present in mature spermatozoa. Together with manual literature curation of sperm-related functional evidence, this resource supported the construction of a sperm-function-focused student dataset.

SIGNOR 3.0~\cite{lo2023signor} was used as an additional standardized knowledgebase of experimentally supported causal interactions. SIGNOR curates causal relationships between proteins and diverse biological entities, including stimuli, phenotypes, enzyme inhibitors, complexes and protein families. Each record is linked to experimental evidence and contains metadata describing the direction of the regulatory effect and the underlying molecular mechanism. Because phosphorylation is the most abundant annotation type in SIGNOR, this resource provides a high-confidence source of phosphorylation-related causal interactions with defined regulatory mechanisms and, when available, functional consequences.

Together, these resources provided complementary evidence for constructing the distillation datasets. OmniPath, PSP and SAGEPhos contributed broad enzyme--substrate PTM and phosphorylation-site records for the teacher dataset. UniProtKB provided full-length enzyme and substrate sequences. RCSB PDB and AlphaFoldDB provided structural inputs for structure-level modelling. HPA defined a broad male reproductive protein background. The literature-derived sperm proteome and manual functional curation identified sperm-relevant kinases. SIGNOR contributed standardized causal phosphorylation evidence for a high-confidence student subset.

\subsubsection{Teacher dataset}
The teacher dataset was constructed as a broad enzyme--substrate post-translational modification (PTM) dataset by integrating records from PhosphoSitePlus \cite{hornbeck2015phosphositeplus}, SAGEPhos \cite{zhang2025sagephos} and OmniPath \cite{turei2016omnipath}. These resources provide complementary coverage of curated enzyme--substrate relationships across diverse biological contexts.

To ensure a unified sequence-level representation, the full-length protein sequences of enzymes and substrates were retrieved from UniProt and linked to the corresponding enzyme--substrate records. After harmonizing enzyme identifiers, substrate identifiers and modification-site annotations across the three resources, duplicate records were removed at the enzyme--substrate--site level. The final teacher dataset contained 43,151 enzyme--substrate PTM records. This dataset was not restricted to specific-domain biology and was used to train the teacher model to capture general enzyme--substrate association patterns. 

We selected one representative structure per accession using a source-prioritized ranking framework. Candidate structures were collected from RCSB PDB(experimental) \cite{burley2019rcsb} and AlphaFoldDB (computed) \cite{varadi2022alphafold}. Ranking combined source provenance, coverage, and file availability, with higher base preference assigned to RCSB PDB candidates and AlphaFoldDB used as fallback when experimental structures were absent or inferior. For each accession, the top-ranked candidate was retained as the final structure for downstream analysis. This strategy balances reliability and coverage by explicitly prioritizing experimentally determined structures while preserving samples through computed models when needed.

The resulting teacher model provided broad multimodal PTM knowledge that was subsequently transferred to the specific-domain-focused student model through knowledge distillation.

\subsubsection{Student datasets}

The student model was designed to specialize the general enzyme--substrate knowledge learned by the teacher model toward distinct physiological and disease-associated phosphorylation contexts. For male reproductive specialization, we constructed three complementary datasets using biological filters based on reproductive tissue expression, literature-supported sperm kinase relevance and experimentally curated causal phosphorylation relationships. These datasets were termed ReproPhos-HPA, SpermK-PSP and SIGNOR-Phos, respectively, and were subsequently integrated into MReproPhos.

To further investigate the applicability of the general-to-specialized learning framework beyond reproductive biology, we constructed two additional disease-associated phosphorylation datasets, termed Tumor and Cancer. These datasets were derived by integrating curated kinase--substrate phosphorylation relationships with disease annotations, providing complementary target domains for evaluating phosphorylation prediction in broader tumour-associated and cancer-specific contexts. The framework was adapted separately to the reproductive, Tumor and Cancer domains using the same knowledge-transfer strategy, rather than directly applying a reproductive-domain predictor to other disease contexts.

\paragraph{ReproPhos-HPA: a broad HPA-guided male reproductive phosphorylation dataset.}
The first student dataset, termed ReproPhos-HPA, was constructed through a coarse screening strategy based on the male reproductive proteome defined from HPA. 

Within the male reproductive system, we considered four major anatomical components: testis, epididymis, prostate and seminal vesicle. These tissues represent complementary stages and environments of male reproductive biology.
Transcriptome analysis in HPA showed that 77\% of human proteins were expressed in testis (15,481/20,162), including 1,992 testis-enriched proteins. The corresponding proportions were 65\% in epididymis (13,069/20,162; 315 enriched), 71\% in prostate (14,287/20,162; 126 enriched), and 65\% in seminal vesicle (13,204/20,162; 73 enriched). After de-duplication across the four tissues, we obtained 16,887 non-redundant proteins expressed in at least one male reproductive tissue, among which 2,506 showed higher expression in reproductive tissues than in any other organ. This male reproductive protein set accounted for 83.76\% of the HPA protein-coding proteome considered in this study (16,887/20,162).

Using this HPA-derived male reproductive protein set as a biological filter, we extracted kinase--substrate phosphorylation records from the teacher dataset. A teacher-dataset record was retained when the substrate protein was included in the male reproductive protein set. For each retained event, we preserved the kinase, substrate, modified residue, phosphosite-centered motif, full-length protein sequences and source annotations when available. After filtering for unique kinase--substrate phosphorylation relationships and harmonizing the site-level annotations, we assembled 33,489 distinct phosphorylation events. ReproPhos-HPA therefore represents a broad male-reproductive phosphorylation subset derived from the teacher dataset and captures phosphorylation events potentially related to spermatogenesis, sperm development and sperm maturation.

\paragraph{SpermK-PSP: a literature-supported sperm kinase and PhosphoSitePlus dataset.}
The second student dataset, termed SpermK-PSP, was constructed through a fine screening strategy focused on sperm-relevant kinases. We started from the ejaculated-sperm proteome compiled by Castillo et al. \cite{castillo2018contribution}, which consolidated proteomic evidence from 46 independent studies that met rigorous technical quality thresholds. This resource produced a non-redundant list of 6,871 proteins detected in ejaculated human sperm.

To identify kinases present in mature spermatozoa, we cross-referenced this sperm proteome with the kinase compendium in PhosphoSitePlus. This step identified 505 kinases that were present in mature spermatozoa. To further determine which of these kinases were functionally relevant to sperm biology, we performed a systematic PubMed search using the query string ``(kinase name) AND (sperm function OR reproduct* OR fertility)''. Kinases were retained only when peer-reviewed studies reported measurable effects on sperm physiology, fertility or reproductive processes. Kinases with evidence of no effect or without functional data were excluded. This literature-validation procedure yielded 202 sperm-relevant kinases. Loss or aberrant activity of these kinases can compromise male reproductive capacity to varying extents, and in the most severe cases leads to complete male infertility. Systematic investigation of the implicated kinases enables dissection of their associated signalling pathways, providing a comprehensive framework for understanding the role of phosphorylation events in male reproductive function.

Using these 202 literature-supported sperm kinases as a biological filter, we extracted corresponding enzyme--substrate phosphorylation records from the teacher dataset. A teacher-dataset record was retained when its kinase belonged to the literature-supported sperm kinase set. After filtering for unique kinase--substrate phosphorylation relationships and harmonizing the site-level annotations, we assembled 16,747 phosphosites. For each retained event, we preserved the catalyzing kinase, substrate protein, modified serine, threonine or tyrosine residue, phosphosite-centered sequence window, full-length protein sequences and source annotations when available. Compared with ReproPhos-HPA, SpermK-PSP provides a more functionally specific phosphorylation dataset focused on mature sperm biology. It supports the study of phosphorylation-mediated regulation in processes such as capacitation, motility regulation, hyperactivation and fertility-associated sperm function.

\paragraph{SIGNOR-Phos: a standardized causal phosphorylation dataset.}
The third dataset, termed SIGNOR-Phos, was constructed by integrating phosphorylation relationships from SIGNOR 4.0 with an experimentally derived human sperm proteome. SIGNOR (SIGnaling Network Open Resource) \cite{lo2026signor} is a manually curated database of causal relationships between proteins and other biological entities, with each interaction linked to its supporting experimental literature. In addition to identifying the upstream regulator and downstream target, SIGNOR annotates the underlying molecular mechanism, the direction of the regulatory effect, and, when available, the modified residue or site. Given the extensive representation of phosphorylation-mediated interactions in SIGNOR, we used the database as a standardized source of causal kinase–substrate relationships for which the responsible kinase, substrate, phosphorylation site, and functional consequence, such as activation or inhibition, had been experimentally characterized.

To restrict these relationships to proteins with evidence of occurrence in mature human spermatozoa, we used the ejaculated human sperm proteome compiled by Castillo et al. as the protein background set~\cite{castillo2018contribution}. This resource integrated 46 independent proteomic studies that satisfied predefined technical quality criteria. Protein identifiers reported across the included studies were standardized and deduplicated, producing a non-redundant catalogue of 6,871 proteins detected in ejaculated human spermatozoa. This experimentally derived protein set defined the molecular background within which potentially sperm-relevant phosphorylation relationships were identified.

We extracted human protein--protein interactions from SIGNOR 4.0 annotated with a phosphorylation regulatory mechanism. Protein identifiers were harmonized to standardized human gene symbols and UniProt accessions, and interactions were retained only when both the kinase and substrate were present in the sperm proteome. For each retained relationship, we preserved the kinase and substrate identifiers, phosphorylation-site annotations where available, regulatory effects and supporting references. The resulting SIGNOR-Phos dataset comprised 3,986 curated phosphorylation records, providing a sperm-proteome-constrained resource of experimentally supported causal phosphorylation relationships.

\paragraph{MReproPhos: an integrated male reproductive phosphorylation dataset}
Finally, we integrated ReproPhos-HPA, SpermK-PSP and SIGNOR-Phos to construct a comprehensive student-side dataset for male reproductive phosphorylation modelling, called MReproPhos. These three datasets represent complementary evidence levels. ReproPhos-HPA provides broad coverage of phosphorylation events associated with male reproductive tissues. SpermK-PSP provides a functionally refined set of phosphorylation events linked to literature-supported sperm kinases. SIGNOR-Phos provides standardized causal phosphorylation records with experimentally supported regulatory effects.

Records from the three datasets were harmonized at the kinase--substrate--site level. Duplicate entries across resources were merged while preserving source annotations. When available, each retained entry contained the kinase, substrate, modified residue, phosphosite-centered motif, full-length protein sequence and evidence source. The integrated student dataset was used to specialize the student model toward male reproductive phosphorylation biology during distillation, allowing it to inherit general enzyme--substrate knowledge from the teacher model while emphasizing reproductive and sperm-function-related phosphorylation events.

\paragraph{Tumor: a disease-annotated tumour-associated phosphorylation dataset.}

To extend kinase--substrate phosphorylation modelling beyond male reproductive biology, we constructed a tumour-associated dataset, termed Tumor, by integrating curated phosphorylation relationships from OmniPath \cite{turei2016omnipath} with disease annotations from NCBI PubTator and Medical Subject Headings (MeSH).

OmniPath provides literature-curated molecular interactions, including enzyme--substrate relationships involving protein phosphorylation. NCBI PubTator provides biomedical literature annotations linking molecular entities to disease-related information, whereas MeSH provides a standardized vocabulary for identifying and organizing disease concepts. We used these complementary resources to identify kinase--substrate phosphorylation records associated with tumour-related biological contexts.

Following integration and data curation, the Tumor dataset comprised 7,836 positive phosphorylation samples. Unlike the reproductive datasets, which were constructed using tissue-expression or sperm-specific functional evidence, Tumor was defined through disease-associated annotations. It therefore provided a distinct target domain for investigating whether broadly learned phosphorylation knowledge could be adapted to tumour-associated kinase--substrate relationships.

The PhosTailor framework was independently adapted to this dataset, with predictive performance evaluated under random, kinase cold-start and substrate cold-start settings. This evaluation was designed to assess domain-specific predictive discrimination and generalization to kinases or substrate proteins absent from the corresponding target-domain training partition.

\paragraph{Cancer: a cancer-specific phosphorylation dataset.}

To further examine domain specialization within the broader tumour-associated phosphorylation context, we constructed a cancer-specific dataset, termed Cancer, as a subset of Tumor. Cancer-associated phosphorylation records were identified using the disease annotations incorporated during Tumor dataset construction, restricting the target domain to phosphorylation relationships associated with cancer-related disease concepts.

The resulting Cancer dataset contained 2,615 positive phosphorylation samples. As a subset of Tumor, Cancer represents a more narrowly defined disease-associated phosphorylation context rather than an independent source of kinase--substrate annotations. This distinction allowed us to examine whether the general-to-specialized learning framework could also support adaptation to a more focused disease-associated distribution.

We independently adapted PhosTailor to the Cancer dataset using the same multimodal teacher--student knowledge-transfer framework and student-guided refinement strategy. As for Tumor, performance was evaluated under random, kinase cold-start and substrate cold-start settings.

\subsubsection{Random and cold-start splits}
\label{sec:supp_data_partitioning}

To evaluate predictive performance and generalization across distinct kinase--substrate distributions, we constructed three complementary benchmark partitions: random split, kinase cold-start and substrate cold-start. These partitioning strategies were applied to the reproductive phosphorylation benchmarks and the additional disease-associated datasets, including Tumor and Cancer.

\paragraph{Random split.}
The random split provided a conventional evaluation of predictive performance on held-out kinase--substrate--phosphosite records. Following identifier harmonization and record-level deduplication, samples were assigned to training, validation and test partitions. Kinase and substrate identities could recur across partitions, whereas the held-out test records were excluded from target-domain training. This setting assessed the ability of PhosTailor to discriminate annotated phosphorylation relationships under a conventional evaluation distribution.

\paragraph{Kinase cold-start split.}
To evaluate generalization to kinases without corresponding target-domain training examples, the kinase cold-start split excluded all test-set kinase identities from the target-domain training partition. Consequently, the student was required to predict kinase--substrate--phosphosite relationships involving kinases not represented during target-domain training. This setting assessed whether the framework transferred kinase-conditioned information beyond the kinase identities explicitly observed in the specialized training data.

\paragraph{Substrate cold-start split.}
The substrate cold-start split evaluated generalization to previously unseen substrate proteins. All substrate identities assigned to the test partition were excluded from the target-domain training partition, requiring the model to predict phosphorylation relationships on substrate proteins absent from target-domain training. This setting assessed whether the learned kinase--substrate compatibility representations remained informative across changes in substrate identity and sequence context.

\paragraph{Partition consistency and evaluation protocol.}
For each benchmark and partitioning strategy, training, validation and test partitions were established before model optimization and remained fixed throughout the corresponding evaluation. Models were compared using matched data partitions, with hyperparameters and checkpoints selected exclusively from validation performance. Held-out test records were excluded from student optimization, hard-example mining, adaptive-teacher refinement and supervision-cache regeneration.

Importantly, the random, kinase cold-start and substrate cold-start partitioning strategies were consistently applied to both general-domain teacher pretraining and target-domain student training. For each setting, the multimodal teacher was independently trained on the corresponding TeacherPTM partition, yielding a split-specific checkpoint. The student was subsequently trained using the corresponding teacher checkpoint and strictly aligned target-domain data partitions. In the kinase and substrate cold-start settings, held-out test kinase or substrate identities were excluded from the training data at both stages. This coordinated partitioning protocol maintained consistent training--test separation throughout the teacher--student knowledge-transfer process and prevented information leakage across the corresponding evaluation partitions.

\subsection{Evaluation metrics}
\label{sec:supp_evaluation}

Model performance was evaluated as a site-level binary classification task. For each enzyme--substrate--site record \(i\), the model outputs a logit \(z_i\), which was converted into a predicted probability by the sigmoid function:
\[
p_i=\sigma(z_i).
\]
For threshold-based metrics, the binary prediction was defined as
\[
\hat{y}_i=
\mathbb{I}(p_i\ge \tau),
\]
where \(\tau\) is the decision threshold. Unless otherwise specified, \(\tau=0.5\). When threshold selection was required, the threshold was selected on the validation set and fixed before test-set evaluation.

We reported accuracy (ACC), area under the receiver operating characteristic curve (AUROC), area under the precision--recall curve (AUPRC) and F1 score. Let TP, TN, FP and FN denote the numbers of true positives, true negatives, false positives and false negatives, respectively. Accuracy was computed as
\[
\mathrm{ACC}
=
\frac{\mathrm{TP}+\mathrm{TN}}
{\mathrm{TP}+\mathrm{TN}+\mathrm{FP}+\mathrm{FN}}.
\]

Precision and recall were defined as
\[
\mathrm{Precision}
=
\frac{\mathrm{TP}}
{\mathrm{TP}+\mathrm{FP}},
\qquad
\mathrm{Recall}
=
\frac{\mathrm{TP}}
{\mathrm{TP}+\mathrm{FN}}.
\]
The F1 score was computed as the harmonic mean of precision and recall:
\[
\mathrm{F1}
=
\frac{
2\cdot \mathrm{Precision}\cdot \mathrm{Recall}
}{
\mathrm{Precision}+\mathrm{Recall}
}.
\]

AUROC was computed from the receiver operating characteristic curve by varying the decision threshold over the predicted probabilities. It summarizes the trade-off between the true positive rate and false positive rate:
\[
\mathrm{TPR}
=
\frac{\mathrm{TP}}{\mathrm{TP}+\mathrm{FN}},
\qquad
\mathrm{FPR}
=
\frac{\mathrm{FP}}{\mathrm{FP}+\mathrm{TN}}.
\]
AUPRC was computed from the precision--recall curve obtained by varying the decision threshold over the predicted probabilities. Because enzyme--substrate modification datasets are often class-imbalanced, AUPRC was included as a class-imbalance-sensitive threshold-independent metric.

For student-model evaluation, metrics were computed on MReproPhos and, when applicable, on its constituent subsets, including ReproPhos-HPA, SpermK-PSP and SIGNOR-Phos. This evaluation assessed whether the structure-free student model preserved teacher-derived enzyme--substrate knowledge while specializing toward male reproductive phosphorylation biology.

\subsection{Subfamily-level recovery of kinase substrate motifs}
\label{sec:subfamily_motifs}

To determine whether the sequence preferences observed at the broad kinase-family level remained detectable at a finer classification resolution, we further examined kinase subfamilies with more than 50 eligible candidate sites in the held-out test set. For each qualifying subfamily, candidate kinase--site pairs were ranked according to the frozen PhosTailor prediction scores, and sequence logos were constructed from the top 50 unique serine/threonine-centred substrate sites. This analysis was performed without using ground-truth labels for candidate selection and therefore characterizes the sequence patterns preferentially prioritized by the model.

Protein kinase C (PKC) predictions exhibited a pronounced basophilic sequence pattern. Arginine was particularly prominent at upstream positions near $-3$ and $-2$, while additional lysine and arginine preferences were observed on both sides of the candidate phosphoacceptor (Supplementary Fig.~\ref{fig:subfamily_motifs}). This pattern was consistent with the upstream basic-residue preference recovered for the broader AGC kinase group, indicating that the family-level biochemical signature remained recognizable after restricting the analysis to a more specific kinase subfamily.

Mitogen-activated protein kinase (MAPK) predictions displayed the clearest position-specific motif among the examined subfamilies. Proline was strongly preferred immediately downstream of the phosphoacceptor at position $+1$, forming the characteristic serine/threonine--proline pattern associated with proline-directed phosphorylation. The recovery of this motif was consistent with the corresponding CMGC-family logo and showed that the dominant sequence determinant remained evident at the MAPK subfamily level.

Calcium/calmodulin-dependent protein kinase II (CAMK2) predictions also showed a preference for basic residues upstream of the candidate site. In particular, arginine was prominent around position $-3$, accompanied by additional lysine and arginine signals across the upstream sequence context. Although this preference was distributed across several positions rather than being dominated by a single residue, the resulting logo remained consistent with a basophilic substrate-recognition mode.

Collectively, the PKC, MAPK and CAMK2 analyses showed that the major sequence patterns identified at the broader kinase-family level were retained in subfamilies with sufficient test-set coverage. The agreement between the AGC--PKC and CMGC--MAPK patterns further indicates that the kinase-conditioned substrate preferences recovered by PhosTailor are coherent across hierarchical levels of kinase classification. Together with the family-level analysis, these results support the ability of PhosTailor to prioritize substrate sites carrying biologically recognizable kinase-specific sequence contexts.

\subsection{Independent biochemical specificity concordance analysis}
\label{sec:supp_kinase_library}

\paragraph{Kinase specificity atlas and candidate panel.}
We evaluated whether PhosTailor preserved intrinsic kinase--substrate recognition preferences by comparing model-derived kinase rankings with experimentally measured specificity profiles from Kinase Library v0.1.0 \cite{johnson2023atlas}. This atlas contains positional-scanning peptide-library specificity matrices for 303 human serine/threonine (Ser/Thr) kinases.

The candidate kinase panel was defined independently of model predictions. Of the 50 unique kinases represented in the held-out random test split, 31 Ser/Thr kinases could be unambiguously mapped to available Kinase Library specificity matrices. These 31 kinases were retained as a common candidate panel for all models. Kinases without an available specificity matrix or an unambiguous identifier mapping were excluded. The correspondence between standardized kinase identifiers and Kinase Library labels is provided in Supplementary Table~\ref{tab:kinase_library_panel}.

\paragraph{Phosphosite selection and candidate construction.}
We identified 220 unique substrate--phosphosite combinations centred on serine or threonine residues from the held-out random test set. Each retained phosphosite had at least one curated positive kinase--substrate--site annotation involving a kinase in the common candidate panel.

For each phosphosite, all 31 candidate kinases were paired with the same substrate and site context, yielding $220 \times 31 = 6,820$ candidate kinase--substrate--site combinations.

These combinations were independently evaluated using the non-distilled student (Student-only), fixed-teacher offline distillation (Offline KD) and full PhosTailor models. Unannotated kinase--site combinations were retained as unknown candidates rather than treated as experimentally established negatives. Accordingly, the analysis assessed agreement with intrinsic biochemical specificity rather than binary kinase--substrate classification accuracy.

\paragraph{Model inference and kinase-wise score adjustment.}
All models were evaluated using their selected checkpoints and the same sequence preprocessing pipeline as in the main experiments. Inference was performed in deterministic evaluation mode, with dropout and stochastic augmentation disabled. Raw prediction logits were used to construct kinase rankings.

To reduce kinase-specific scoring biases, logits were centred independently for each model and candidate kinase. For kinase $k$ and phosphosite $i$, the adjusted logit was defined as
\begin{equation}
\tilde{z}_{ik}
=
z_{ik}
-
\frac{1}{|\mathcal{I}_k|}
\sum_{i' \in \mathcal{I}_k} z_{i'k},
\label{eq:supp_kinase_adjustment}
\end{equation}
where $z_{ik}$ denotes the original model logit and $\mathcal{I}_k$ denotes the set of phosphosites retained for kinase $k$ in the corresponding analysis.

For each phosphosite, candidate kinases were ranked according to their adjusted logits $\tilde{z}_{ik}$. In overlap-filtered sensitivity analyses, kinase-wise centring was recalculated after applying the corresponding candidate-pair exclusions.

\paragraph{Biochemical ranking concordance.}
Kinase Library percentile scores were used as continuous measures of intrinsic biochemical relevance. For phosphosite $i$, let $\pi_i(r)$ denote the kinase ranked at position $r$ by the model, and let $g_{i,\pi_i(r)}$ denote its corresponding Kinase Library relevance score.

Discounted cumulative gain at rank five was calculated as
\begin{equation}
\mathrm{DCG@5}_i
=
\sum_{r=1}^{\min(5,K_i)}
\frac{g_{i,\pi_i(r)}}{\log_2(r+1)},
\label{eq:supp_kinase_dcg}
\end{equation}
where $K_i$ is the number of eligible candidate kinases for phosphosite $i$.

The ideal discounted cumulative gain, $\mathrm{IDCG@5}_i$, was obtained by ranking the same candidate kinases in descending order of their Kinase Library relevance scores. Normalized discounted cumulative gain was then defined as
\begin{equation}
\mathrm{nDCG@5}_i
=
\frac{\mathrm{DCG@5}_i}
{\mathrm{IDCG@5}_i}.
\label{eq:supp_kinase_ndcg}
\end{equation}

An nDCG@5 value of one indicates perfect agreement between the model-derived ranking and the biochemical reference ranking within the first five candidate positions. The bias-adjusted nDCG@5 used kinase-wise centred model logits for candidate ranking.

\paragraph{Recovery of biochemically favoured kinases.}
To assess whether models preferentially recovered kinases with high intrinsic biochemical compatibility, we measured the overlap between the three highest-ranked Kinase Library kinases and the five highest-ranked model predictions.

For phosphosite $i$, let $\mathcal{A}_i^{(3)}$ denote the atlas top-three kinase set and $\mathcal{M}_i^{(5)}$ denote the model top-five kinase set. The number of recovered kinases was defined as
\begin{equation}
H_i
=
\left|
\mathcal{A}_i^{(3)}
\cap
\mathcal{M}_i^{(5)}
\right|.
\label{eq:supp_kinase_recovery}
\end{equation}

Under random ranking, the expected number of recovered kinases is
\begin{equation}
H_i^{\mathrm{random}}
=
\frac{3 \times 5}{K_i}.
\label{eq:supp_kinase_random}
\end{equation}

The site-level enrichment over random expectation was therefore calculated as
\begin{equation}
E_i
=
\frac{H_i}{3 \times 5/K_i}.
\label{eq:supp_kinase_enrichment}
\end{equation}

For the complete 31-kinase candidate panel, random selection would recover an expected
\[
\frac{3 \times 5}{31}
=
0.484
\]
kinases per phosphosite.

Cohort-level recovery and enrichment were summarized as the means of the corresponding site-level values:
\begin{equation}
\overline{H}
=
\frac{1}{N}\sum_{i=1}^{N}H_i,
\qquad
\overline{E}
=
\frac{1}{N}\sum_{i=1}^{N}E_i,
\label{eq:supp_kinase_cohort}
\end{equation}
where $N$ denotes the number of phosphosites retained in the corresponding analysis.

For the candidate-pair-unseen analysis, $K_i$ and the corresponding random expectation were recalculated separately for each phosphosite after removing training-overlapping candidate pairs.

\paragraph{Training-overlap sensitivity analyses.}
To assess whether biochemical ranking concordance could be explained by memorization of previously observed phosphorylation relationships, we performed three complementary evaluations with increasingly stringent training-overlap control.

Exact kinase--substrate--site overlap was defined using the kinase identifier, substrate identifier, phosphoacceptor residue type and residue position. Overlap was assessed against both TeacherPTM pretraining records and the MReproPhos student-training partition. Validation and test records were not considered training overlaps.

First, the full analysis included all 220 eligible Ser/Thr phosphosites with the common panel of 31 candidate kinases.

Second, the positive-pair-unseen analysis excluded phosphosites whose annotated positive kinase--substrate--site relationships overlapped either teacher or student training. This filtering retained 110 phosphosites and reduced the potential contribution of direct memorization of annotated positive relationships.

Third, the candidate-pair-unseen analysis applied a more stringent exclusion rule by removing every candidate kinase--substrate--site combination previously observed during teacher or student training, irrespective of whether the combination represented the annotated positive kinase for that phosphosite.

Among the original 6,820 candidate combinations, 712 (10.44\%) overlapped teacher or student training and were excluded in this sensitivity analysis. The number of eligible kinases, $K_i$, was recalculated for each phosphosite after filtering, and sites with fewer than ten remaining candidate kinases were excluded.

For all three evaluation settings, model and biochemical-reference rankings were constructed over the corresponding eligible candidate sets. These analyses distinguish direct exposure to annotated positive kinase--site relationships from broader overlap within the candidate kinase space.

\paragraph{Statistical analysis.}
Phosphosites were treated as the independent statistical units. Mean nDCG@5, Top3-in-Top5 recovery and enrichment were calculated across the retained phosphosites.

Non-parametric bootstrap resampling over phosphosites was used to estimate 95\% confidence intervals. Pairwise comparisons between Student-only, Offline KD and PhosTailor were performed on matched phosphosites using two-sided Wilcoxon signed-rank tests.

For candidate-pair-unseen comparisons, all evaluated models were restricted to the same eligible candidate sets. Multiple comparisons within each analysis setting were corrected using the Benjamini--Hochberg procedure, and adjusted $P$ values were reported as $q$ values. Statistical significance was indicated by $q < 0.05$, $q < 0.01$ and $q < 0.001$.

\paragraph{Scope of interpretation.}
Kinase Library measures intrinsic Ser/Thr kinase preferences for local peptide sequence contexts under positional-scanning peptide-library conditions. Agreement between PhosTailor rankings and the experimental atlas therefore provides independent evidence that the model preserves kinase-conditioned biochemical specificity.

\subsection{Counterfactual kinase-substitution analysis}
\label{sec:supp_counterfactual}

\paragraph{Selection of evaluated kinase--site pairs.}
The counterfactual analysis was performed using positive kinase--site pairs from the held-out test set. To minimize the contribution of memorized kinase--site associations, we retained only candidate-pair-unseen examples, defined as test pairs for which the exact combination of kinase identity, substrate identity, residue type and residue position was absent from the training set. This procedure resulted in 242 kinase--site pairs for the final analysis.

\paragraph{Construction of counterfactual kinase replacements.}
For each retained pair, the substrate protein, phosphosite position, site-centred peptide window and all substrate-side input features were held fixed. Only the kinase identity and kinase sequence were replaced. Kinase-family annotations were obtained from the kinase-family mapping used for the motif analysis. For each supported kinase--site pair, we sampled up to five kinases from the same family and five kinases from different families, using a fixed random seed of 42 and sampling without replacement. The supported kinase itself, kinases already annotated as regulators of the same phosphosite in the available training, validation or test annotations, kinases without valid sequences, and kinases that could not be processed by the original model input pipeline were excluded. This generated 533 same-family replacements and 1,210 cross-family replacements. At least one eligible same-family replacement was available for 198 of the 242 pairs, whereas all 242 pairs were retained for the cross-family analysis.

\paragraph{Model inference.}
The non-distilled student, offline knowledge distillation model (Offline KD) and Full PhosTailor were evaluated using the same counterfactual pairs and the same preprocessing pipeline as in the main experiments. All models were evaluated using their selected checkpoints in deterministic evaluation mode, with dropout and stochastic augmentation disabled. Raw prediction logits were used for the primary analysis to avoid compression of score differences caused by sigmoid saturation.

\paragraph{Counterfactual metrics.}
For a phosphosite $s_i$, let $z_i^{\mathrm{true}}$ denote the logit assigned to the supported kinase. Let $\mathcal{K}_{i}^{\mathrm{same}}$ and $\mathcal{K}_{i}^{\mathrm{cross}}$ denote the eligible same-family and cross-family replacement sets, respectively. The same-family and cross-family margins were defined as

\begin{equation}
\Delta_{i}^{\mathrm{same}}
=
z_i^{\mathrm{true}}
-
\operatorname{median}_{k \in \mathcal{K}_{i}^{\mathrm{same}}}
z_i^{k},
\end{equation}

and

\begin{equation}
\Delta_{i}^{\mathrm{cross}}
=
z_i^{\mathrm{true}}
-
\operatorname{median}_{k \in \mathcal{K}_{i}^{\mathrm{cross}}}
z_i^{k}.
\end{equation}

To provide a more stringent comparison, the margin against the highest-scoring replacement was defined as

\begin{equation}
\Delta_{i}^{\mathrm{best}}
=
z_i^{\mathrm{true}}
-
\max_{k \in \mathcal{K}_{i}^{\mathrm{same}}
\cup
\mathcal{K}_{i}^{\mathrm{cross}}}
z_i^{k}.
\end{equation}

The replacement-set top-1 indicator was defined as one when the supported kinase received a higher logit than every sampled replacement kinase, and zero otherwise. Ties were not counted as top-1.

\paragraph{Statistical analysis.}
Within each model, supported-kinase logits were compared with the corresponding median same-family and cross-family replacement logits using paired two-sided Wilcoxon signed-rank tests. Counterfactual margins from Full PhosTailor were also compared with those from the non-distilled student and Offline KD using paired two-sided Wilcoxon signed-rank tests on the common set of kinase--site pairs. Multiple comparisons were controlled using the Benjamini--Hochberg procedure. For replacement-set top-1 rates, 95\% Wilson confidence intervals were calculated.

\paragraph{Interpretation.}
The sampled replacement kinases were used as counterfactual alternatives rather than definitive biological negative regulators. The analysis therefore evaluates whether model predictions are sensitive to kinase identity and relative kinase--site compatibility, but does not establish that every lower-scoring replacement kinase is biologically incapable of phosphorylating the corresponding site.

\subsection{Model-guided wet-lab evaluation}
\label{supp:wetlab_methods}
\label{supp:wetlab_results}

This section reports the wet-lab observations and model--experiment concordance analyses. Experimental procedures and statistical definitions are consolidated in Appendix~\ref{sec:supp_experimental_methods}.

\subsubsection{Prospective kinase nomination}
\label{supp:kinase_selection}

The nomination panel comprised 60 proteins manually prioritized for reproductive relevance from genes harbouring variants identified in intrauterine insemination (IUI) failure cases. These variants informed panel selection; their association with IUI failure has not been established.

The panel yielded 6,690 complete-window serine, threonine and tyrosine contexts, evaluated against 204 candidate kinases. MAP2K2 ranked first and TBK1 fifth in the cross-protein consensus list (Supplementary Table~\ref{tab:reproductive_proteins_kinases}). Both kinase selections and the checkpoint were recorded as fixed before perturbation outcomes were assessed. The aggregation procedure is specified in Section~\ref{sec:supp_kinase_nomination}.

\subsubsection{Proteomic detection and perturbation-associated responses}

Global proteomic and phosphoproteomic measurements compared vehicle-matched HTF, GSK8612 and U0126 conditions after 1~h of treatment. Each condition represented one pool collected over approximately 10 days and split into three technical replicates. These measurements therefore describe the pooled material and do not estimate donor-level biological variability. Supplementary Fig.~\ref{fig:phosphoproteomic_quality}a,b shows the sample-level quality-control profiles.

The GSK8612 and U0126 comparisons identified 79 and 118 differentially abundant proteins, respectively, and 411 and 485 differential phosphosites. The phosphosite calls used fold-change and nominal $P$-value thresholds without multiple-testing correction (Supplementary Fig.~\ref{fig:phosphoproteomic_quality}c,d).

TBK1 (Q9UHD2), MAP2K1 (MEK1; Q02750) and MAP2K2 (MEK2; P36507) had non-missing protein intensities in all nine measurements. Mean non-imputed MAP2K2 and MAP2K1 signals were $3.260\times10^6$ and $0.836\times10^6$, respectively. Their signal ratio was 3.90 across conditions and 3.68 in HTF alone. MAP2K2 and MAP2K1 were supported by 16 and nine distinct peptide sequences (21 and 12 precursor-charge records), respectively. No sequence was shared between the assigned peptide sets. Sequence-distinct peptide support documents detection of both MEK proteins in the pooled sperm proteome. The intensity comparisons describe signals under the acquisition conditions.

\subsubsection{Locked response sets and candidate coverage}
\label{sup:phos_filter}

Excluding sites assigned to differentially abundant proteins removed one of 411 GSK8612-responsive sites and 22 of 485 U0126-responsive sites. The resulting analysis sets contained 410 TBK1-oriented and 463 MAP2K2-oriented sites, respectively (Section~\ref{sec:supp_locked_universes}). Here, filtered responsive sites means that no protein-abundance change met the applied differential criteria; it does not demonstrate equivalent abundance.

The GSK8612 set contained 211 decreased and 199 increased sites, whereas the U0126 set contained 346 decreased and 117 increased sites. Model-defined Top 5\%, Top 10\% and Top 20\% tiers contained 21, 42 and 83 TBK1 candidates and 24, 47 and 93 MAP2K2 candidates. These nested sets assessed sensitivity to candidate coverage and were not independent replications.

Candidate membership depended on frozen model scores within the experimentally selected differential-site backgrounds. No parameters were updated using the perturbation data. This design evaluates conditional rank concordance; it does not establish prospective discrimination across all measurable phosphosites.

\subsubsection{TBK1-oriented concordance and directional checks}

TBK1 candidate tiers were enriched toward the GSK8612-decreased end of the experimental ranking (Supplementary Fig.~\ref{fig:phosphoproteomic_quality}e). For Top 5\%, Top 10\% and Top 20\%, ES values were $-0.406$, $-0.340$ and $-0.281$, and NES values were $-1.892$, $-2.046$ and $-2.052$. Corresponding empirical $P$ values were 0.0113, 0.0029 and 0.0013; BH-adjusted $q$ values were 0.0113, 0.0043 and 0.0039.

Protein-block NES values were $-1.797$, $-1.907$ and $-1.932$, with $P=0.0298$, 0.0128 and 0.0070, respectively. Decreased phosphorylation occurred at 15/21 (71.4\%), 30/42 (71.4\%) and 57/83 (68.7\%) candidates, compared with 211/410 (51.5\%) in the background. Exploratory one-sided Fisher tests against the remaining sites yielded odds ratios of 2.46, 2.58 and 2.46, with $P=0.0475$, 0.00464 and $3.11\times10^{-4}$. These site-based tests do not account for within-protein dependence and complement the protein-block analysis rather than replacing it.

Direction-aware leading edges contained 21, 39 and 47 sites. Alternative $\log_2\mathrm{FC}$ ranking retained negative NES values ($-1.310$, $-1.200$ and $-1.374$), and ten sites were shared between the two Top 5\% leading edges. Direction was therefore consistent across rankings of the same data; statistical support under the protein-block null was established for the direction-aware analysis.

\subsubsection{MAP2K2-oriented concordance under U0126 perturbation}
\label{sup:enrich_map2k2}

For MAP2K2, Top 5\%, Top 10\% and Top 20\% tiers yielded ES values of $-0.371$, $-0.213$ and $-0.122$ and NES values of $-1.601$, $-1.128$ and $-0.753$. Empirical $P$ values were 0.0315, 0.2718 and 0.7449; BH-adjusted $q$ values were 0.0945, 0.4077 and 0.7449. Protein-block NES values were $-1.513$, $-1.061$ and $-0.718$, with $P=0.0679$, 0.3180 and 0.7446. The nominal Top 5\% signal did not remain significant after tier-wise correction or protein-block resampling (Supplementary Figs.~\ref{fig:phosphoproteomic_quality}f and~\ref{fig:map2k2_u0126_casa}c).

Decreased phosphorylation occurred at 18/24 (75.0\%), 33/47 (70.2\%) and 62/93 (66.7\%) candidates, compared with 346/463 (74.7\%) in the background. These proportions did not consistently exceed the background proportion (Supplementary Fig.~\ref{fig:map2k2_u0126_casa}b). The nominal Top 5\% enrichment concerned weighted response ranks, rather than an excess proportion of decreased sites.

Direction-aware leading edges contained 10, 20 and 20 sites. Alternative effect-size ranking yielded Top 5\% NES $=-1.468$ and empirical $P=0.0628$. Seven sites from seven proteins occurred in both Top 5\% leading edges. These candidates were stable across two rankings of the same experiment. The 24 Top 5\% candidate-site response profiles are shown in Supplementary Fig.~\ref{fig:map2k2_u0126_casa}d. This rank-stable subset supports focused follow-up of MAP2K2-compatible candidates within the MEK1/2 pathway context.

\subsubsection{Exploratory external comparison and functional annotations}
\label{sec:supp_external_results}

Table S1 of Martin-Hidalgo \textit{et al.}~\cite{martin2020human} contains 210 differential phosphopeptide records. Re-parsing the row-wise $\log_2(\mathrm{HM}/\mathrm{LM})$ values yielded 108 positive and 102 negative values; these signs conflict with the aggregate direction counts in the source-table footer. Among the seven previously reported site correspondences, six had a singly localized phosphoresidue compatible with a retained GSK8612-responsive site: SYPL1 S2 in the study sequence (corresponding to S20 in canonical SYPL1), PSMB6 S203, SUGT1 S11, KCNU1 S1098, NT5C1B S598 and AK9 S510. All six had matching ratio signs in the two datasets. The GARIN3 record contains both S532 and S536 on the same phosphopeptide, so its ratio was excluded from the single-site direction count. Adjacent AK9 S511 records were also not assigned to S510. 
The source records and mapping audit are provided in Supplementary Data 11 and 12. 
This descriptive, post hoc comparison provides cross-dataset directional context for the GSK8612-responsive candidates.

TBK1-compatible candidates included axonemal/dynein, cilia/flagella, fibrous-sheath and CatSper-associated components. Representative proteins included DNAI1, DNAI2, HYDIN, SPAG17, CFAP91, CFAP221, CFAP54, CFAP44, FSIP2, CABYR, CATSPER4 and C2CD6/CATSPERT. These descriptive annotations place model-prioritized candidates in sperm structures and signalling systems relevant to motility.

\subsubsection{Donor-level sperm-motility observations}
\label{sec:supp_casa_results}

CASA used separate paired-donor cohorts: seven donors for GSK8612 and five for U0126. These cohorts did not overlap with each other or with the pooled omics material. The U0126 measurements showed lower mean HA, VCL and ALH at 1, 2 and 4~h (Supplementary Fig.~\ref{fig:map2k2_u0126_casa}e--j). Mean within-donor HA differences were $-11.8$, $-10.2$ and $-15.4$ percentage points, respectively.

Across the 15 endpoint-by-time comparisons, Holm-adjusted $P$ values for HA at 1, 2 and 4~h were 0.0149, 0.0753 and 0.0621, respectively. VCL was lower at all three time points (adjusted $P=0.0270$, $0.0408$ and $0.0016$), and ALH was lower at 1~h (adjusted $P=0.0461$). Total motility and PR did not meet the $P<0.05$ threshold. Panel~j repeats the HA comparisons (Section~\ref{sec:supp_methods_casa}).

Together with the phosphoproteomic analyses, these donor-paired observations support functional investigation of model-nominated candidates in sperm movement. Candidate memberships and leading-edge annotations are provided in Supplementary Data 1 and 8; enrichment summaries and donor-level measurements and statistics are provided in Supplementary Data 4--6.

\subsection{Experimental configurations for ablation analyses}
\label{sec:supp_ablation}

To assess the contributions of individual components of PhosTailor, we conducted controlled ablation experiments using the random split of the reproductive phosphorylation benchmark (Fig.~5h--l). Unless otherwise specified, all configurations used the same target-domain data partitions and sequence-only student architecture as the full model, with each experiment evaluated across five independent runs. The analyses examined four aspects of the framework: multimodal knowledge distillation, student-guided adaptive refinement, distillation objective composition and data efficiency.

\paragraph{Module-level ablations.}
We first distinguished the contributions of teacher-guided knowledge transfer and adaptive teacher refinement (Fig.~5h). Three configurations were compared. The full PhosTailor model incorporated multimodal knowledge distillation and student-guided closed-loop refinement. The non-distilled student retained supervised training but removed response-level and representation-level teacher supervision. The fixed-teacher offline-distillation variant retained teacher-guided supervision while disabling adaptive teacher refinement and iterative target regeneration. These comparisons isolated the contributions of privileged knowledge transfer and subsequent adaptation of teacher supervision.

\paragraph{Refinement-strategy ablations.}
To further investigate the contribution of student-guided refinement (Fig.~5i), we compared four configurations: fixed-teacher distillation, random refinement, hard-sample-only learning and full PhosTailor. In random refinement, the adaptive teacher was updated using randomly selected target-domain examples instead of student-identified difficult examples. In the hard-sample-only configuration, difficult examples identified by the student received preferential learning emphasis, but the teacher remained unchanged and teacher targets were not regenerated. In the full refinement procedure, student uncertainty, prediction errors and teacher--student disagreement jointly informed hard-example selection, guiding constrained adaptive-teacher updates. Frozen-teacher anchoring and general-domain replay were retained to constrain knowledge drift.

\paragraph{Loss-composition and representation-alignment ablations.}
We next examined the contributions of different training objectives by progressively introducing hard-example learning, response-level logit distillation and representation-level feature alignment into the non-distilled student (Fig.~5j). To further investigate representation alignment, we compared cosine-only, mean-squared-error (MSE)-only and combined cosine--MSE objectives (Fig.~5k). These experiments assessed the complementary contributions of response-level and representation-level knowledge transfer.

\paragraph{Data-efficiency analysis.}
To assess the data efficiency of knowledge distillation under limited target-domain supervision, we constructed nested training subsets containing 10\%, 25\%, 50\% and 100\% of the reproductive-domain training data (Fig.~5l). Subsets were generated by stratified random sampling without replacement. Positive and negative training instances were independently shuffled using a fixed sampling seed, and the first $\lfloor fN_c \rfloor$ instances from each class were selected at each fraction $f$, where $N_c$ denotes the number of training instances in class $c$. The 100\% setting used the complete training partition. Using the same class-specific random permutations ensured nested subsets, $D_{10\%}\subseteq D_{25\%}\subseteq D_{50\%}\subseteq D_{100\%}$, while approximately preserving the original class distribution.
Validation and test instances were excluded from all training and refinement procedures.

\section{Supplementary Computational Methods}
\label{sec:supp_computational_methods}

\subsection{Task formulation and model inputs}
\label{sec:supp_methods_data}

We formulated kinase--substrate phosphorylation prediction as a site-level binary classification task. Each instance comprised a kinase, a substrate protein and a candidate phosphorylation site on the substrate. Given the kinase sequence and the substrate sequence context surrounding a candidate residue, the model predicted the compatibility between the kinase and the corresponding phosphorylation site. Candidate phosphoacceptor residues were restricted to serine, threonine and tyrosine.

For each instance $i$, let $x_i^{(k)}$ denote the kinase amino-acid sequence, $x_i^{(s)}$ the substrate sequence, and $c_i$ the candidate phosphosite position. The substrate sequence context was represented by a 15-residue window centred on the candidate phosphoacceptor. Kinase inputs comprised the corresponding full-length amino-acid sequences, subject to a predefined deterministic truncation policy when the sequence exceeded the model's maximum supported length.

Each instance was associated with a binary label $y_i \in \{0,1\}$, where $y_i=1$ indicated an annotated kinase--substrate--phosphosite relationship and $y_i=0$ denoted a negative-labelled instance generated according to the dataset-specific sampling protocol. Negative-labelled instances were used for model training and benchmark evaluation and were not necessarily experimentally established non-substrate relationships. Known positive interactions and ambiguous annotated kinase--site relationships were excluded from the negative-sampling pool. Detailed dataset construction, identifier harmonization, negative sampling and benchmark partitioning are described in Supplementary Note~\ref{sec:supp_datasets}.

The multimodal teacher incorporated kinase and substrate sequence representations together with structural and sequence-derived biochemical information. In contrast, the student used only sequence-derived inputs, including kinase sequences, substrate-site sequence contexts and lightweight physicochemical descriptors. Explicit protein structures and structural graphs were therefore available as privileged information during teacher training but were not required for student inference. The corresponding teacher and student architectures, structural feature processing and multimodal fusion procedures are described in the subsequent sections.

For each kinase--substrate--phosphosite instance, the model generated a compatibility logit $z_i$, which was transformed into a predicted probability using the sigmoid function:
\begin{equation}
p_i = \sigma(z_i).
\end{equation}
The resulting probability was used for phosphorylation classification and kinase--substrate--site prioritization. Unless otherwise specified, a probability threshold of 0.5 was used for threshold-dependent classification metrics, whereas threshold-independent metrics were calculated from continuous prediction scores.

\subsection{Privileged multimodal teacher architecture}
\label{sec:supp_methods_teacher}

\paragraph{Sequence and biochemical representations.}
The multimodal teacher integrated kinase and substrate sequence representations with structural and physicochemical information. Kinase sequences were encoded using ESM2-650M, comprising 33 Transformer layers with 1,280-dimensional hidden representations. The substrate sequence encoder incorporated COMPASS-PTM~\cite{zhang2026learning}, built upon ESM2-650M, to integrate sequence-derived representations with 320-dimensional chemical features and four residue-level physicochemical descriptors. These complementary representations provided sequence-contextual and biochemical information for kinase--substrate--phosphosite compatibility modelling.

\paragraph{Structural graph construction and encoding.}
Protein structures were obtained from experimentally determined RCSB PDB entries when available, with AlphaFoldDB predictions used as fallback structural sources. Structures were parsed using Gemmi, retaining amino-acid residues with valid C$\alpha$ coordinates. Residue identities were resolved using chain identifiers, residue numbering and insertion codes.

Each protein structure was converted into a residue-level graph, with amino-acid residues represented as nodes and structural relationships represented as edges. Graph connectivity combined sequence-neighbour relationships, spatial contacts within 10~\AA{} and up to ten nearest neighbours per residue. Node features encoded residue identities, whereas edge features captured inter-residue distances, relative displacement vectors, local geometric orientations and edge types. The resulting graphs were processed using a shared two-layer geometric vector perceptron (GVP) encoder to generate residue-level structural representations for kinase and substrate proteins. The corresponding graph feature dimensions, structural encoder configuration and residue limits are summarized in Supplementary Table~\ref{tab:supp_phostailor_config}.

\paragraph{Sequence--structure correspondence.}
Structural representations were aligned with the corresponding sequence positions before multimodal fusion. In the original teacher-pretraining configuration, residue correspondence was established using Needleman--Wunsch global sequence alignment. Structural features were assigned to their aligned sequence positions, with unmatched positions zero-filled and linear interpolation used when hard alignment failed.

In the inspected adaptive-refinement and teacher-cache configurations, structural representations were instead resized to the corresponding sequence length using linear interpolation. Missing or unparseable structures were handled using a zero-feature fallback at the model-input level where applicable. Sequence and structural length limits differed between the original teacher-pretraining and subsequent refinement configurations, as detailed in Supplementary Table~\ref{tab:supp_phostailor_config}.

\paragraph{Hierarchical multimodal fusion and prediction.}
The teacher integrated sequence, structural and biochemical representations through a hierarchical gated-fusion architecture. Substrate sequence and aligned structural features were first combined to construct residue-level multimodal representations, which were subsequently restricted to the 15-residue window centred on the candidate phosphosite. A learnable phosphosite embedding was introduced to identify the central phosphoacceptor position.

In parallel, kinase sequence and structural representations were fused to characterize the kinase context. The resulting kinase representation was pooled and broadcast to match the length of the substrate-site window. Kinase-side and substrate-side representations were then integrated with residue-level physicochemical descriptors through a second gated-fusion stage.

The resulting fused representations were aggregated across the substrate-site window to obtain the teacher representation $h_i^t$. A linear prediction head subsequently mapped $h_i^t$ to the kinase--substrate--phosphosite compatibility logit $z_i^t$. The teacher was pretrained on TeacherPTM using binary cross-entropy supervision, providing multimodal representations for subsequent knowledge distillation.

\subsection{Structure-free student architecture}
\label{sec:supp_methods_student}

The student model was designed to retain kinase-conditioned phosphorylation prediction capabilities while eliminating the requirement for explicit protein structures during inference. It preserved the sequence-based kinase and substrate branches, lightweight biochemical descriptors and hierarchical gated-fusion interface of the teacher, but replaced the high-capacity sequence backbone with a compact ESM2-8M encoder, comprising six Transformer layers with 320-dimensional hidden representations.

Kinase and substrate sequences were encoded using the shared lightweight sequence backbone. The substrate input consisted of a 15-residue window centred on the candidate serine, threonine or tyrosine residue. The kinase sequence representation was pooled to obtain a global kinase embedding, which was subsequently broadcast to match the substrate-window length. Sequence-derived physicochemical descriptors were retained to provide local biochemical context.

Unlike the multimodal teacher, the student did not use experimentally determined or predicted protein structures. The sequence representations and lightweight biochemical descriptors were integrated through the student fusion module to generate a kinase-conditioned substrate-site representation.

The fused features were aggregated across the substrate-site window to obtain the student representation $h_i^s$, which was mapped to the compatibility logit $z_i^s$ through a linear prediction head. The student therefore performed phosphorylation prediction using sequence-derived information alone, without structural retrieval, graph construction or structural encoding.

Although explicit structural information was unavailable to the student, its representations and predictions were guided during training by the multimodal teacher through response-level and representation-level knowledge distillation, as described in the following sections. The final deployed model retained only the structure-free student architecture.

\subsection{Multi-level knowledge distillation}
\label{sec:supp_methods_distillation}

\paragraph{Teacher supervision and cache initialization.}
Knowledge distillation transferred multimodal phosphorylation knowledge from the pretrained teacher to the structure-free student through complementary response-level and representation-level supervision. The pretrained multimodal teacher checkpoint was retained as a frozen anchor teacher $T_0$, while an adaptive teacher $T_r$ was initialized from the same parameters for subsequent refinement. Before student training, $T_0$ was evaluated deterministically on the target-domain training samples to generate an initial supervision cache,
\begin{equation}
\mathcal{C}_0 =
\left\{
\left(\bar{z}_{i,0}^{t}, \bar{h}_{i,0}^{t}\right)
\right\}_{i=1}^{N},
\end{equation}
where $\bar{z}_{i,0}^{t}$ and $\bar{h}_{i,0}^{t}$ denote the cached teacher logit and intermediate representation for sample $i$, respectively. The cache subsequently provided teacher supervision during student optimization and was selectively updated through the closed-loop refinement procedure described below.

\paragraph{Student training objective.}
The structure-free student was optimized using a combination of ground-truth supervision, response-level distillation and representation-level alignment:
\begin{equation}
\mathcal{L}_{S}
=
\mathcal{L}_{\mathrm{sup}}
+
\alpha \mathcal{L}_{\mathrm{resp}}
+
\beta \mathcal{L}_{\mathrm{repr}},
\end{equation}
where $\mathcal{L}_{\mathrm{sup}}$ denotes binary cross-entropy supervision, $\mathcal{L}_{\mathrm{resp}}$ transfers teacher predictive responses and $\mathcal{L}_{\mathrm{repr}}$ aligns student and teacher representations. The corresponding loss weights were $\alpha=0.35$ and $\beta=0.8$.

\paragraph{Response-level and label-aware distillation.}
Response-level distillation aligned student predictions with temperature-scaled teacher predictions stored in the supervision cache. For sample $i$ at refinement state $r$, the softened student and teacher probabilities were defined as
\begin{equation}
p_i^s = \sigma\left(\frac{z_i^s}{T}\right),
\qquad
\bar{p}_{i,r}^{t}
=
\sigma\left(\frac{\bar{z}_{i,r}^{t}}{T}\right),
\end{equation}
where $T=2.8$ denotes the distillation temperature.

To reduce the influence of potentially inaccurate teacher supervision, response distillation incorporated a sample-specific label-aware weighting mechanism. In particular, a reverse-suppression function downweighted teacher predictions that were inconsistent with the observed target-domain labels. The resulting sample weight was expressed as
\begin{equation}
a_{i,r}
=
w_i^{\mathrm{LA}}
\phi_{\mathrm{rev}}
\left(
\bar{p}_{i,r}^{t},y_i;
\gamma,\lambda_{\mathrm{rev}},e
\right),
\end{equation}
where $w_i^{\mathrm{LA}}$ denotes the label-aware weight, $\phi_{\mathrm{rev}}$ the reverse-suppression function, $\gamma$ the shaping exponent, $\lambda_{\mathrm{rev}}$ the suppression strength and $e$ the training epoch. Here, $\gamma=1$ and $\lambda_{\mathrm{rev}}=0.8$, with reverse suppression disabled during the first three training epochs. The response-distillation objective used temperature-scaled soft-target binary cross-entropy with the configured $T^2$ scaling.

\paragraph{Representation-level alignment.}
In addition to transferring predictive responses, the student was encouraged to preserve the teacher's intermediate feature structure. Representation alignment combined cosine similarity and mean-squared-error (MSE) objectives, with equal relative weights of 0.5. Cosine alignment encouraged agreement in representation direction, whereas the MSE component constrained feature-level differences.

Because the teacher and student representations had different dimensions, student features were projected from 320 to 1,280 dimensions before alignment. Feature normalization using LayerNorm was applied before the MSE component, as specified in Supplementary Table~\ref{tab:supp_phostailor_config}. Together, response-level and representation-level distillation enabled the student to learn from teacher predictions and multimodal representations while retaining sequence-only inference.

\subsection{Student-guided closed-loop refinement}
\label{sec:supp_methods_refinement}

\paragraph{Student-induced hard-example identification.}
Following an initial student warm-up stage, PhosTailor employed a student-guided closed-loop refinement strategy to adapt multimodal teacher supervision to the reproductive domain. At each refinement round $r$, difficult target-domain training examples were identified using three complementary signals: student predictive uncertainty, supervised classification error and teacher--student disagreement.

For each training example $i$, student predictive uncertainty was defined as
\begin{equation}
u_i^{(r)}
=
1-2\left|
\sigma\left(z_i^{s,(r)}\right)-0.5
\right|,
\end{equation}
where $z_i^{s,(r)}$ denotes the current student logit. Teacher--student disagreement was calculated using the most recently cached teacher prediction:
\begin{equation}
d_i^{(r)}
=
\left|
\sigma\left(\bar{z}_{i,r}^{t}\right)
-
\sigma\left(z_i^{s,(r)}\right)
\right|.
\end{equation}

Supervised classification error was measured using the per-sample binary cross-entropy loss. The three components were normalized over the target-domain training partition and combined into a composite hardness score:
\begin{equation}
s_i^{(r)}
=
\lambda_u\widehat{u}_i^{(r)}
+
\lambda_l\widehat{\ell}_{i,\mathrm{sup}}^{(r)}
+
\lambda_d\widehat{d}_i^{(r)}.
\end{equation}

Training examples were ranked according to their hardness scores, and the highest-scoring samples were selected to form the hard-example subset $\mathcal{H}_r$. The warm-up duration, refinement interval, component weights and hard-example selection criteria are summarized in Supplementary Table~\ref{tab:supp_phostailor_config}. Validation and test samples were excluded from hard-example mining and all subsequent refinement procedures.

\paragraph{Constrained adaptive-teacher refinement.}
At each refinement round, the adaptive teacher was updated using the student-selected hard examples while preserving broadly transferable phosphorylation knowledge through general-domain replay and frozen-teacher anchoring. The teacher optimization objective was defined as
\begin{equation}
\mathcal{L}_{T}^{(r)}
=
\mathcal{L}_{\mathrm{hard}}^{(r)}
+
\lambda_{\mathrm{replay}}
\mathcal{L}_{\mathrm{replay}}^{(r)}
+
\lambda_{\mathrm{anchor}}
\mathcal{L}_{\mathrm{anchor}}^{(r)}.
\end{equation}

The hard-example term applied binary cross-entropy supervision to the selected target-domain examples:
\begin{equation}
\mathcal{L}_{\mathrm{hard}}^{(r)}
=
\frac{1}{|\mathcal{H}_r|}
\sum_{i\in\mathcal{H}_r}
\ell_{\mathrm{BCE}}
\left(z_i^{t,(r)},y_i\right).
\end{equation}

General-domain replay incorporated additional samples from the TeacherPTM training partition:
\begin{equation}
\mathcal{L}_{\mathrm{replay}}^{(r)}
=
\frac{1}{|\mathcal{B}_r^{\mathrm{replay}}|}
\sum_{j\in\mathcal{B}_r^{\mathrm{replay}}}
\ell_{\mathrm{BCE}}
\left(z_j^{t,(r)},y_j\right),
\end{equation}
where $\mathcal{B}_r^{\mathrm{replay}}$ denotes the replay batch sampled at refinement round $r$.

Frozen-teacher anchoring further constrained the adaptive teacher relative to the pretrained teacher $T_0$. The anchor objective combined temperature-scaled KL divergence between teacher predictive distributions with cosine-based alignment of intermediate representations. Together, replay and anchoring limited drift in both predictive behaviour and representation structure during target-domain adaptation.

Teacher updates were restricted to the configured parameter-efficient adaptation and higher-level fusion and prediction modules, while pretrained sequence and structural backbone weights remained frozen. Optimization settings, regularization coefficients, trainable modules and split-specific update configurations are provided in Supplementary Table~\ref{tab:supp_phostailor_config}.

\paragraph{Selective teacher-target regeneration and cache updating.}
Following each refinement round, the updated teacher $T_{r+1}$ was switched to deterministic evaluation mode to regenerate response- and representation-level targets for the selected hard examples:
\begin{equation}
\left(
z_i^{t,\mathrm{new}},
h_i^{t,\mathrm{new}}
\right)
=
T_{r+1}\left(\mathcal{I}_i^{(t)}\right),
\qquad i\in\mathcal{H}_r,
\end{equation}
where $\mathcal{I}_i^{(t)}$ denotes the corresponding multimodal teacher input.

The supervision cache was updated selectively:
\begin{equation}
\left(
\bar{z}_{i,r+1}^{t},
\bar{h}_{i,r+1}^{t}
\right)
=
\begin{cases}
\left(
z_i^{t,\mathrm{new}},
h_i^{t,\mathrm{new}}
\right),
& i\in\mathcal{H}_r,\\[4pt]
\left(
\bar{z}_{i,r}^{t},
\bar{h}_{i,r}^{t}
\right),
& i\notin\mathcal{H}_r.
\end{cases}
\end{equation}

Thus, $\mathcal{C}_{r+1}$ contained updated supervision for selected hard examples, while the remaining samples retained their previously cached teacher targets. This selective regeneration strategy concentrated teacher-side computation on student-identified difficulties without requiring a full refresh of the target-domain supervision cache.

\paragraph{Label-anchored student updating.}
After teacher adaptation and cache regeneration, student optimization continued from the current state $S_r$ using ground-truth labels and the updated cached supervision $\mathcal{C}_{r+1}$, following the composite distillation objective defined in Supplementary Methods~\ref{sec:supp_methods_distillation}.

Selected hard examples received increased sampling probabilities during subsequent student training, allowing the updated teacher supervision to exert greater influence on difficult target-domain instances. The sampling procedure and relative weights are provided in Supplementary Table~\ref{tab:supp_phostailor_config}.

The resulting closed-loop refinement process was summarized as
\begin{equation}
S_r
\longrightarrow
\mathcal{H}_r
\longrightarrow
T_{r+1}
\longrightarrow
\mathcal{C}_{r+1}
\longrightarrow
S_{r+1}.
\end{equation}

Ground-truth supervision remained active throughout refinement, and validation and test samples were excluded from teacher adaptation and target regeneration. In contrast, fixed-teacher offline distillation disabled adaptive-teacher updates and iterative cache regeneration, using static supervision from the frozen teacher throughout student training.

\subsection{Implementation and inference benchmarking}
\label{sec:supp_methods_training}

Training and evaluation configurations, including optimizers, learning rates, batch sizes, random seeds, checkpoint-selection criteria and software environments, are summarized in Supplementary Table~\ref{tab:supp_phostailor_config}. Architecture-specific and refinement-specific configurations were recorded separately for teacher pretraining, student adaptation and adaptive-teacher updates.

Inference throughput was measured using frozen teacher and student checkpoints on a common set of eligible test examples. Models were evaluated in FP32 using deterministic evaluation mode without gradient computation or automatic mixed precision. Measurements were performed after GPU warm-up, with CUDA synchronization applied to ensure accurate timing. Forward-only measurements excluded model loading, input preprocessing, batch collation and CPU--GPU transfers. Latency and throughput were evaluated across multiple batch sizes using repeated measurements. Complete hardware specifications and timing configurations are provided in Supplementary Table~\ref{tab:supp_phostailor_config}.

Detailed data-partitioning procedures and ablation configurations are provided in Supplementary Methods~\ref{sec:supp_methods_data} and Supplementary Note~\ref{sec:supp_ablation}, respectively.

\section{Supplementary Experimental Methods}
\label{sec:supp_experimental_methods}

\subsection{Human sperm donors, purification and pharmacological treatment}
\label{sec:supp_methods_human}
\label{sec:supp_human_sperm}

The collections and experiments of human sperm samples were approved by the Human Research Ethics Committee of the Affiliated Hospital of Nantong University. Samples were screened using study-defined thresholds of total motility $\geq40\%$ and progressive motility (PR) $\geq32\%$, corresponding to the lower reference limits in the fifth edition of the World Health Organization (WHO) semen manual \cite{Cooper2010WHOReferenceValues,WHO2010SemenManual}. These thresholds alone were not used to diagnose normozoospermia.

After liquefaction, motile spermatozoa were enriched by swim-up in HEPES-buffered saline (HS), pH~7.4. HS contained 135~mM NaCl, 5~mM KCl, 1~mM MgSO$_4$, 2~mM CaCl$_2$, 20~mM 4-(2-hydroxyethyl)-1-piperazineethanesulfonic acid (HEPES), 5~mM glucose, 10~mM lactic acid and 1~mM sodium pyruvate. Motile sperm were collected at $300\times g$ for 6~min at room temperature.

For omics experiments, daily collections were allocated to HTF with matched dimethyl sulfoxide (DMSO) vehicle, HTF with 10~$\mu$M GSK8612 or HTF with 10~$\mu$M U0126. Compounds were obtained from MedChemExpress (GSK8612, HY-111941; U0126, HY-12031A), prepared as 10-mM DMSO stocks and stored at $-80~^\circ$C. A human tubal fluid (HTF) medium~\cite{Quinn1985HTF} containing bovine serum albumin and NaHCO$_3$ was purchased from Aibei Biotechnology (Nanjing, China). Vehicle controls contained 0.1\% DMSO. Samples were incubated for 1~h at $37~^\circ$C, pelleted, snap-frozen in liquid nitrogen and stored at $-80~^\circ$C.

Collections spanned approximately 10 days. Pellets were pooled separately by condition before the service provider divided each pool into three technical replicates. Each condition therefore contained one biological pool; the three measurements were not independent donors. CASA samples were collected separately.

\subsection{Protein extraction, digestion and phosphopeptide enrichment}
\label{sec:supp_sample_preparation}

For global proteomics, proteins were extracted from sperm pellets, quantified by BCA assay and assessed by SDS--PAGE. Aliquots were denatured, reduced and alkylated before tryptic digestion at $37~^\circ$C for 2~h. Peptides were desalted using C$_{18}$ columns, concentrated at $45~^\circ$C and resuspended for analysis.

For phosphoproteomics, extracts were quantified by BCA assay. SDS lysis buffer and dithiothreitol were added, followed by 1~h at room temperature and iodoacetamide alkylation in the dark. Magnetic-bead processing preceded tryptic digestion at an enzyme-to-protein mass ratio of 1:25 for 16~h at $37~^\circ$C. Digestion was terminated by addition of 10\% formic acid; peptides were C$_{18}$-desalted, concentrated at $45~^\circ$C and subjected to phosphopeptide enrichment.

\subsection{LC--MS/MS acquisition}
\label{sec:supp_methods_lcms}
\label{sec:supp_lcms}

Peptides were separated on a Vanquish Neo UHPLC system (Thermo Scientific). Mobile phase A was 0.1\% formic acid in water; mobile phase B was 0.1\% formic acid in 80\% acetonitrile. The phosphoproteomic workflow used a PepMap Neo 5~$\mu$m C$_{18}$ trap column ($300~\mu$m $\times$ 5~mm) and a $\mu$PAC Neo High Throughput analytical column. Initial equilibration used 96\% mobile phase A.

DIA acquisition used an Orbitrap Astral mass spectrometer (Thermo Scientific), positive-ion electrospray and 1.9~kV. Reported MS1 settings were 380--980 $m/z$, resolution 240,000, AGC target 500\% and maximum injection time 3~ms. Reported MS2 settings were resolution 80,000, AGC target 500\%, maximum injection time 3~ms, RF lens 40\%, HCD activation, 2~Th isolation width, normalized collision energy 25\% and cycle time 0.6~s.

\subsection{DIA processing, differential calls and sample-level quality control}
\label{sec:supp_methods_dia}
\label{sec:supp_dia}

Global-proteomic and phosphoproteomic DIA data were processed using DIA-NN v2.2 and v1.9, respectively~\cite{Demichev2020DIANN}. Libraries used DIA-NN deep-learning predictions and match-between-runs information. Trypsin digestion allowed up to two missed cleavages; precursor and protein identifications were filtered at 1\% false-discovery rate. Phosphosites with localization confidence $\geq0.9$ were retained. Identification FDR and localization confidence are distinct from differential-analysis multiplicity control.

Differential phosphosites met fold change $>1.5$ or $<2/3$ and the provider-reported t-test threshold $P<0.05$. These $P$ values were nominal and unadjusted. Tests across technical replicates do not estimate donor-level variability; the differential calls and downstream concordance analyses are exploratory.

The PCA visualization used 7,949 phosphosites quantified in at least six of nine profiles. Reported preprocessing comprised log$_2$ transformation and within-sample median imputation. The current figure imports provider-generated PCA coordinates and Pearson correlations.

\subsection{Protein-abundance filtering and locked analysis universes}
\label{sec:supp_locked_universes}

Phosphosites assigned to proteins meeting the differential-abundance criteria in the matched inhibitor-versus-vehicle comparison were excluded before candidate ranking. This was an exclusion-based filter, rather than division of phosphosite intensity by protein abundance or a quantitative correction model. The retained sites defined the fixed backgrounds for the downstream analyses reported in Section~\ref{sup:phos_filter}.

The differential-protein list was obtained from the quantitative comparison table using the provider's positive/negative differential flags. Missing-value-only protein lists were not combined with this list.

\subsection{Prospective model-guided kinase nomination}
\label{sec:supp_methods_nomination}
\label{sec:supp_kinase_nomination}

The frozen checkpoint evaluated complete 15-residue windows centred on eligible serine, threonine and tyrosine residues in the predefined protein panel. For each protein and kinase, the kinase prioritization score (KPS) summed site-level prediction scores greater than 0.5. The 50 highest-ranked kinases were retained per protein, and their KPS values were aggregated across proteins into a consensus ranking. The ten highest-ranked kinases were considered for experimental selection.

Follow-up candidate scoring used the frozen PhosTailor checkpoint and did not use measured fold changes, differential $P$ values or treatment labels as model inputs.

\subsection{Candidate tiers, response ranking and enrichment}
\label{sec:supp_enrichment}

Within each locked background, candidate sites were ordered by decreasing model score, with deterministic site-label ordering for ties. For rank $r_i=1$ at the highest score and background size $N$, rank percentile was
\begin{equation}
\pi_i=1-\frac{r_i-1}{N}.
\label{eq:supp_site_percentile}
\end{equation}
Boundary-inclusive thresholds $\pi_i\geq0.95$, $\pi_i\geq0.90$ and $\pi_i\geq0.80$ defined the nested candidate tiers. Experimental responses were ordered separately using
\begin{equation}
R_i=\operatorname{sign}(\log_2\mathrm{FC}_i)
\left[-\log_{10}\left(\max(P_i,10^{-300})\right)\right],
\label{eq:supp_direction_score}
\end{equation}
where fold change and nominal $P_i$ refer to inhibitor versus vehicle-matched HTF. Descending order placed increased sites first and decreased sites last. An alternative sensitivity analysis ordered sites by $\log_2\mathrm{FC}$ alone.

A weighted running sum increased at candidate positions in proportion to the absolute ranking score, normalized by the total candidate weight. Non-candidate positions contributed a uniform decrement. ES was the signed value at the largest absolute deviation from zero. For negative ES, the leading edge comprised candidate sites at or beyond the peak; for positive ES, it comprised sites at or before the peak.

Site-level nulls used 10,000 random candidate sets of matched size and seed 2026. NES divided the observed ES by the mean absolute null ES of the same sign. Two-sided empirical $P$ values compared absolute ES with the permutation distribution; BH correction was applied across three tiers separately for each kinase.

Protein-block resampling grouped all sites assigned to a protein, permuted blocks and selected complete blocks until the observed tier size was reached or exceeded. The running-sum statistic was then recalculated using 10,000 permutations and seed 2026. Realized null sizes could therefore exceed the nominal size. This procedure assesses within-protein clustering, not donor-to-donor reproducibility.

\subsection{Complementary TBK1 candidate-direction analysis}
\label{sec:supp_tbk1_direction}

Sites were classified by the sign of GSK8612-versus-HTF $\log_2\mathrm{FC}$. For each candidate tier, a $2\times2$ table compared decreased/increased counts in the tier with those among the remaining sites. One-sided Fisher exact tests evaluated excess decreased phosphorylation; sample odds ratios summarize the tables. These exploratory tests used sites as units and did not model within-protein dependence. The nested tiers were sensitivity checks, not independent biological replications.

The accompanying reanalysis provides raw $P$ values and BH-adjusted values across the three directional tests. It reproduces the recorded candidate memberships without re-estimating model scores. Primary rank-based conclusions remain tied to the direction-aware enrichment and protein-block analyses.

\subsection{External HM/LM comparison and site mapping}
\label{sec:supp_methods_external}
\label{sec:supp_hmlm}

The external comparison used 210 differential phosphopeptide records reported by Martin-Hidalgo \textit{et al.}~\cite{martin2020human}. Direction was assigned from each row's $\log_2(\mathrm{HM}/\mathrm{LM})$ value. The source analysis used row-wise values rather than the conflicting aggregate direction description in the original table footer. Legacy identifiers were mapped conservatively, including FAM71B/FA71B to GARIN3.

An exact-site match required concordant protein identity, alignment of the localized external phosphoresidue to the central model-window residue and exact overlapping sequence compatibility. At least seven overlapping amino acids were required; sequence-external padding was ignored. Duplicate external measurements mapping to one locked TBK1 site were collapsed.

Ratio-sign concordance was assessed by comparing the signs of the external $\log_2(\mathrm{HM}/\mathrm{LM})$ values and the study $\log_2(\mathrm{GSK8612}/\mathrm{HTF})$ values. The comparison was restricted to the six singly localized, exact-site matches. The GARIN3 multiphosphorylated peptide and adjacent AK9 records were excluded from the single-site direction count. Results were summarized descriptively.

\subsection{Descriptive functional annotation}
\label{sec:supp_tbk1_function}

Candidate proteins were described using existing annotations for axonemal/dynein, cilia/flagella, fibrous-sheath and CatSper-associated functions. These categories were assembled after inspection of the candidate list and were not treated as prespecified enrichment tests.

\subsection{Computer-assisted sperm analysis}
\label{sec:supp_methods_casa}
\label{sec:supp_casa}

Each donor's purified sperm were divided into matched vehicle and inhibitor aliquots, using 0.1\% DMSO in vehicle controls. GSK8612 and U0126 were each used at 10~$\mu$M and tested in seven and five independent donors, respectively, at 1, 2 and 4~h. Drug-specific cohorts were non-overlapping and distinct from the omics cohort. Donors, rather than individual sperm tracks or fields, were the pairing units.

CASA used a Hamilton-Thorne CEROS system with a 20-$\mu$m-deep 2X-CEL chamber, stage temperature $37~^\circ$C, acquisition at 60~Hz and 30 frames. Seven fields and at least 200 sperm were evaluated per assay. Recorded parameters included total and progressive motility, VAP, VSL, VCL and ALH. HA was classified using VCL 150--1,000~$\mu$m/s, LIN 0--50\% and ALH 7--100~$\mu$m.

Two-sided paired $t$ tests compared treatment with vehicle separately for each endpoint and time. For each inhibitor, Holm correction was applied across 15 endpoint-by-time comparisons comprising HA, VCL, ALH, total motility and PR at 1, 2 and 4~h. The drug-specific comparison families were analysed separately; no between-drug treatment contrast was tested. Panel~j repeats the HA comparisons and is not an additional test.

U0126 CASA group and paired-difference plots display means $\pm$ SEM. The candidate-site response plots in Fig.~6d and Supplementary Fig.~\ref{fig:map2k2_u0126_casa}d display $\log_2$ ratios of condition-mean intensities without error bars. GSK8612 CASA summaries in Fig.~6 display means with pointwise 95\% confidence intervals. Supplementary Data 5 reports donor-paired mean differences, pointwise 95\% confidence intervals and Holm-adjusted $P$ values across 15 comparisons for each inhibitor. Paired intervals were calculated as $\bar d\pm t_{0.975,n-1}s_d/\sqrt n$ and are not simultaneous family-wise intervals.

\subsection{Statistical scope, source data and reproducibility}
\label{sec:supp_statistics}

Computational model-selection procedures are specified in Appendix~\ref{sec:supp_computational_methods}; Kinase Library and counterfactual analyses are described in Supplementary Notes~\ref{sec:supp_kinase_library} and~\ref{sec:supp_counterfactual}, respectively. Technical-replicate phosphoproteomics, candidate-site permutations and donor-level CASA have different inferential units. Protein-block resampling does not supply missing biological replication.

Processed phosphosite and protein matrices, excluded-site records and locked model rankings are provided in Supplementary Data 1--3; CASA donor-level values and statistics in Supplementary Data 4 and 5; enrichment summaries, running curves and leading-edge annotations in Supplementary Data 6--8; quality-control coordinates and correlations in Supplementary Data 9 and 10; and external phosphopeptide records and site mappings in Supplementary Data 11 and 12.

\end{appendices}

\end{document}